\documentclass[a4paper,fleqn]{cas-sc}

\usepackage[authoryear,longnamesfirst]{natbib}
\usepackage{float}
\usepackage[linesnumbered,ruled]{algorithm2e}
\usepackage{caption}
\usepackage[section]{placeins}

\usepackage{times}
\usepackage{soul}
\usepackage{url}
\usepackage[ruled,linesnumbered]{algorithm2e}
\SetKwInput{KwInput}{Input}
\SetKwInput{KwOutput}{Output}
\SetKw{Continue}{continue}
\SetKw{Break}{break}
\usepackage{multirow}
\usepackage{multicol}
\usepackage{rotating}
\usepackage[graphicx]{realboxes}
\usepackage{booktabs}
\usepackage{subfig}
\usepackage{caption}
\usepackage{xspace}
\usepackage[title]{appendix}

\usepackage{xcolor}

\def\tsc#1{\csdef{#1}{\textsc{\lowercase{#1}}\xspace}}
\tsc{WGM}
\tsc{QE}
\usepackage[english]{babel}
\usepackage{amsthm}

\theoremstyle{definition}

\begin{document}

\let\WriteBookmarks\relax
\def\floatpagepagefraction{1}
\def\textpagefraction{.001}

\captionsetup[figure]{labelfont={bf},name={Figure},labelsep=period}
\shorttitle{Solution-Hashing Search Based on Layout-Graph Transformation for Unequal Circle Packing} 

\shortauthors{Zhou et~al.}


\title [mode = title]{Solution-Hashing Search Based on Layout-Graph Transformation for Unequal Circle Packing} 

\author[1]{Jianrong Zhou}
\author[1]{Jiyao He}
\author[1]{Kun He$^*$}[orcid=0000-0001-7627-4604]

\affiliation[1]{organization={School of Computer Science and Technology, Huazhong University of Science and Technology},
            city={Wuhan},
            postcode={430074}, 
            country={China}}
    
\affiliation[]{organization={$^*$:Corresponding author. Email: brooklet60@hust.edu.cn}}

\begin{abstract}
The problem of packing unequal circles into a circular container stands as a classic and challenging optimization problem in computational geometry. 
This study introduces a suite of innovative and efficient methods to tackle this problem.
Firstly, we present a novel layout-graph transformation method that represents configurations as graphs, 
together with an inexact hash method facilitating fast comparison of configurations for isomorphism or similarity. 
Leveraging these advancements, we propose an Iterative Solution-Hashing Search algorithm 
adept at circumventing redundant exploration through efficient configuration recording. 
Additionally, we introduce several enhancements to refine the optimization and search processes,  
including an adaptive adjacency maintenance method, an efficient vacancy detection technique, and a Voronoi-based locating method. 
Through comprehensive computational experiments across various benchmark instances, our algorithm demonstrates superior performance over existing state-of-the-art methods, showcasing remarkable applicability and versatility.  
Notably, our algorithm surpasses the best-known results for 56 out of 179 benchmark instances
while achieving parity with the remaining instances.
\end{abstract}


\maketitle

\section{Introduction} \label{sec:intro}
The Circle Packing Problem (CPP) poses a fundamental challenge in computational geometry,  aiming to arrange circular items optimally within a specified space while preventing overlap. Typically, the objective is to achieve the maximum packing density~\citep{zeng2018adaptive,stoyan2020optimized,lai2023perturbation,amore2023efficient}.
CPP encompasses various variants, each introducing additional constraints or objectives, such as bin-packing~\citep{he2017greedy,he2021adaptive,yuan2022adaptive} and balanced constraints~\citep{he2013coarse,stetsyuk2016global,liu2016heuristic,wang2019stimulus,romanova2022balanced}.  
Despite its NP-hard complexity~\citep{fowler1981optimal,demaine2016circle}, significant efforts have been dedicated to addressing CPP, 
including the Packing Unequal Circles in a Circle (PUCC) problem, which holds diverse applications across industries and academia.

In this study, we focus on PUCC, which involves packing a set of $n$ circles $\{ c_1, c_2, ..., c_n \}$, with known radii $\{ r_1, r_2, ..., r_n \}$ ($r_1 \leq r_2 \leq ... \leq r_n$), into a circular container without overlap, 
while minimizing the radius of the circular container. Mathematically, PUCC can be formulated as a nonlinear programming problem in the two-dimensional Cartesian coordinate system: 
\begin{align}
    \min \quad & R \label{eq:tar} \\
    {\rm s.t.} \quad & \sqrt{{(x_{i} - x_{j})}^2 + {(y_{i} - y_{j})}^2} \geq r_{i} + r_{j}, \enskip 1 \leq i < j \leq n, \label{eq:pre1} \\ 
    \quad & \sqrt{x_{i}^{2} + y_{i}^{2}} + r_{i} \leq R, \enskip 1 \leq i \leq n, \label{eq:pre2}
\end{align}
where coordinates $(x_{i}, y_{i})$ and $(x_{j}, y_{j})$ denote the centers of circles $c_i$ and $c_{j}$, respectively, and $R$ denotes the radius of the circular container centered at the origin $(0, 0)$. 
Constraints (\ref{eq:pre1}) and (\ref{eq:pre2}) ensure that all circles are contained in the container with no overlap. 

PUCC finds a wide range of applications in industries, including circular cutting, container loading, cylinder packing, facility dispersion~\citep{castillo2008solving}, satellite packaging~\citep{wang2019stimulus}.  Its utility extends to data visualization and analysis tasks, demonstrating its versatility and importance in both practical and theoretical domains~\citep{wang2006visualization,murakami2015clonepacker,gortler2017bubble}. 
Notably, PUCC has garnered substantial interest in academia, evidenced by Al Zimmermann's Programming Contest 
(AZPC, see \href{http://recmath.com/contest/CirclePacking/index.php}{http://recmath.com/contest/CirclePacking/index.php}).
This contest, held in the autumn/winter of 2005, tasked participants with finding the densest packing arrangements for varying numbers of circles ($n = 5, 6, ..., 50$) with integer radii ($r_{i} = i, 1 \leq i \leq n$) within a circular container.  Drawing 155 participants, the contest spurred noteworthy studies and the development of 
various methodologies~\citep{addis2008efficiently,muller2009packing,schneider2009ultrametricity,carrabs2014tabu}.
 
Over the past decade, significant efforts have been devoted to tackling the PUCC problem~\citep{ye2013iterated,huang2013tabu,lopez2013packing,lopez2016formulation,zeng2016iterated}, resulting in impressive performances and 
the establishment of numerous computational records. 
Despite the extensive research and the multitude of proposed methods and algorithms, 
finding an optimal configuration remains highly challenging due to its NP-hard nature. Moreover, obtaining high-quality configurations for moderate-scale and large-scale instances is extremely time-consuming. Therefore, the development of advanced algorithms for PUCC is crucial. Not only can these algorithms facilitate the resolution of numerous real-world applications, but they also significantly contribute to the advancement of computer science and geometry research.

In this study, we propose the Iterative Solution-Hashing Search (I-SHS) algorithm, a novel stochastic optimization algorithm to address the PUCC problem. Rooted in a classic penalty model, the I-SHS algorithm is structured hierarchically, consisting of multiple optimization phases. Central to its functionality is a solution-hashing search heuristic, leveraging our novel layout-graph transformation and solution-hashing techniques. These innovations empower our algorithm to meticulously record explored solutions and circumvent duplicate explorations during the search, thereby facilitating the efficient discovery of high-quality configurations.

Moreover, we enhance the optimization phases of the I-SHS algorithm with several designs.
Firstly, we introduce the Adaptive Adjacency Maintenance (AAM) method, which focuses on maintaining the circle adjacency set throughout the layout optimization process. 
By dynamically calculating an adaptive maintenance feature utilizing deferring counters and deferring lengths, unnecessary maintenance is deferred, effectively reducing computational overhead. 
Secondly, we devise an efficient vacancy detection method employing continuous optimization techniques to locate and measure vacancies within configurations. 
Additionally, leveraging the characteristics of two-dimensional packing problems, we introduce a Voronoi-based locating method 
, complementing the vacancy detection process. 
The integration of these methods equips our algorithm to accurately and efficiently detect vacancies in any given configuration. 

To validate the efficacy of our proposed algorithm, we conduct extensive experiments across a range of benchmark instances with diverse radius distributions. Computational results demonstrate 
the superior performance of our algorithm over state-of-the-art methods, 
delivering a remarkable performance in solving the PUCC problem. 
Notably, our algorithm improves the best-known results for 56 out of 179 benchmark instances 
while achieving parity with the remainder. Furthermore, we conduct in-depth experiments to thoroughly analyze and showcase the efficiency, applicability, and generality of our proposed methodologies.

The main contributions and innovations of this study are summarized as follows:
\begin{itemize}
    \item \textit{Efficient Iterative Solution-Hashing Search (I-SHS) Algorithm:} We introduce the I-SHS algorithm, with remarkable applicability and generality across a wide range of instances with diverse radius distributions, for tackling PUCC. 
    \item \textit{Novel Methods for Configuration Comparison:} We propose two novel methods, the layout-graph transformation method and the solution hash method, to address the challenging task of determining 
    isomorphism or similarity between PUCC configurations.
    \item \textit{Efficient Vacancy Detection and Locating Methods: } We propose two novel methods, the vacancy detection method and Voronoi-based locating method, to accurately locate and measure vacancies within PUCC configurations. These methods significantly contribute to improving the algorithm's capability to identify optimal packing arrangements.
    \item \textit{Adaptive Adjacency Maintenance (AAM) Method: } We propose the AAM method, which dynamically maintains the circle adjacent set during the continuous optimization process of PUCC.  This adaptive approach effectively reduces computational overhead, enhancing the efficiency of the optimization process.
    \item \textit{Comprehensive and Excellent Experimental Validation: } Extensive experiments on various PUCC benchmarks yield abundant computational results, providing valuable insights for future comparison and analysis. Notably, our algorithm achieves many new best solutions, underscoring its effectiveness and competitiveness.  
\end{itemize}

The rest of this paper is organized as follows. 
In Section~\ref{sec:RW}, we provide a comprehensive review of related studies in the literature. 
Section~\ref{sec:model} presents the classic penalty model for solving PUCC that is adopted in our proposed algorithm.
In Section~\ref{sec:proalg},  we delve into the details of the proposed algorithm, providing a thorough explanation of its methodology. 
The experimental results and performance comparisons of the proposed algorithm are presented in Section~\ref{sec:exp}.
Finally, we summarize our findings in Section~\ref{sec:conclu}.

\section{Literature Review} \label{sec:RW}
Due to its popularity and the extensive research conducted in this field, PUCC has evolved into a well-explored and classic topic. As a result, there is a wealth of studies, methods, and computational results available.
A comprehensive review of CPP and PUCC studies prior to 2009 was provided by \citet{hifi2009literature}, offering a detailed overview of the topic. In this section, we provide a brief review of the most notable works for solving the PUCC problem, categorizing them into three main groups, namely constructive methods, penalty modeling methods, and mathematical programming and miscellaneous.  
Additionally, we provide a brief introduction to the best-known records and popular benchmarks for PUCC.

\subsection{Constructive Methods}
Constructive methods in the context of PUCC are based on an intuitive idea. 
It packs the given circles one by one into the container without overlapping, and a feasible configuration is obtained after all the circles have been packed.

\citet{huang2002learning} introduced a widely recognized evaluation metric, called Maximal Hole Degree (MHD), along with a circle placement rule evaluated by this criteria. They also employed a self-look-ahead strategy to solve the decision PUCC problem. 
\citet{huang2006new} then employed a dichotomous search method to minimize the container radius and solve the optimization PUCC problem, which has become a typical approach in this field. 
Since then, several follow-up studies adopted heuristics or metaheuristics based on the MHD metric and dichotomous search to solve the PUCC, such as Minimum Damage Heuristic~\citep{akeb2006hybrid}, Pruned–Enriched-Rosenbluth Method~\citep{lu2008perm}, Beam Search~\citep{akeb2009non,akeb2009beam,akeb2010adaptive}, etc.

In addition to the MHD metric, several significant studies focusing on constructive methods have explored alternative techniques. 
For instance, \citet{hifi2007dynamic} introduced a constructive method that involves packing circles one by one into the container while maintaining the positions of the packed circles and the radius of the circular container adaptively by solving a non-linear programming problem. 
Building upon their previous work, \citet{hifi2008adaptive} developed a more efficient three-phase method. This method consists of the constructive phase to construct a candidate layout by packing circles one by one into the container, the maintenance phase to dynamically maintain the position of packed circles and the size of the container, and the heuristic phase to search for a better layout. 
Furthermore, \citet{carrabs2014tabu} proposed a constructive method that applies a strength along a selected direction on each circle and shifts the circles to achieve a dense configuration.  To enhance the solution quality, they incorporated a Tabu Search heuristic as a restarting method, enabling the exploration of better solutions. 

\subsection{Penalty Modeling Methods}

In contrast to constructive methods, algorithms based on penalty modeling approach pack all circles into the container while allowing for overlapping among the circles and the container, and a penalty function is designed to evaluate the degree of overlaps. These algorithms aim to reduce the overlaps by minimizing the penalty function through continuous optimization techniques. Subsequently, post-processing approaches are employed to obtain a feasible configuration. 

\citet{huang1999two} introduced a classic penalty model, called the Quasi-Physical Quasi-Human (QPQH) method, also known as the elastic model. 
In this method, an objective function denoted as $E = \sum \sum d_{ij}^2 + \sum d_{i0}^2$ is introduced to quantify the overlapping degree.
Several subsequent studies have built upon the QPQH method and proposed various strategies to enhance the searching capability. For instance, \citet{wang2002improved} suggested a quasi-physical movement for basin hopping, \citet{huang2004short} introduced an early-escape strategy to accelerate the optimization process within the QPQH framework. 
Meanwhile, a significant number of studies have adopted heuristics or metaheuristics based on QPQH to solve the PUCC problem. 
Notable works include Simulated Annealing~\citep{zhang2004simulated,zhang2005personified,zhang2005effective}, Tabu Search~\citep{zhang2005effective,huang2012new,huang2013tabu,ye2013iterated,zeng2016iterated}, Extremal Optimization~\citep{huang2007extremal}, Energy Landscape Paving~\citep{liu2009improved,liu2010efficiently}, Population-Based Algorithm~\citep{huang2012packing}, Iterated Local Search~\citep{huang2012using}, Genetic Algorithm~\citep{flores2016evolutionary}, Differential Evolution~\citep{flores2016evolutionary}, etc. 
These studies leverage the QPQH method as a foundation and explore various techniques to enhance the packing quality.

Since 2009, several landmark studies and high-performance algorithms have emerged. 
\citet{muller2009packing} proposed an efficient Simulated Annealing algorithm to address the problem posed by Al Zimmermann’s Programming Contest
(AZPC). Their algorithm utilizes a novel hybrid penalty function and incorporates three adjustment operations, namely \textit{swap}, \textit{shift}, and \textit{jump}, in neighborhood construction. Their algorithm yields many better results compared to the existing AZPC records. 
In 2013, two significant studies~\citep{ye2013iterated,huang2013tabu} presented a similar concept, focusing on an iterated Tabu Search algorithm based on the QPQH method. Both studies exploited the \textit{swap} operation in neighborhood construction. Yet they differed in their diversification approaches. \citet{ye2013iterated} applied the \textit{shift} operation for perturbation and thereby improved the record held by \citet{muller2009packing}, while \citet{huang2013tabu} proposed a global perturbation method that improved the records of other benchmarks. 
In 2016, building upon the foundation of the QPQH method and iterated Tabu Search algorithms, \citet{zeng2016iterated} introduced a hybrid strategy that combines Variable Neighborhood Search and Tabu Search. Their algorithm improved the record held by \citet{ye2013iterated} and held many of the best records of other benchmarks. 

\subsection{Mathematical Programming and Miscellaneous}
In addition to the above two categories, several important studies have employed non-linear programming solvers to solve PUCC using mathematical programming techniques.  
\citet{addis2008efficiently} presented a powerful Population-Based algorithm, along with efficient neighborhood construction that involves swapping two similar circles. Their algorithm won AZPC in a racy final and emerged as a milestone in PUCC research. 
\citet{al2011packing} presented an adaptive algorithm incorporating a nested partitioning heuristic within the Tabu Search framework. 
\citet{lopez2013packing} introduced a formulation space search algorithm that leveraged the \textit{swap} operation as an improvement method.
Moreover, \citet{stoyan2020optimized} proposed a general methodology of hybrid methods and strategies for solving various packing problems, including the PUCC problem. Their approach involved a combination of techniques such as multi-start approaches, nonlinear programming techniques, greedy methods, and branch-and-bound algorithms. 

In addition to the aforementioned methods, there are also studies that explore alternative approaches to solve the PUCC problem. \citet{specht2015precise} introduced a precise algorithm that focused on detecting voids and they designed several jumping strategies and exchange heuristics to enhance a known configuration. \citet{ryu2020voropack} proposed an ingenious and efficient algorithm based on Voronoi diagrams to tackle the real-time PUCC problem.


\subsection{Computational Results and Benchmarks}

Over the years, a significant number of studies have been dedicated to the PUCC problem, resulting in a wealth of computational results and the creation of numerous benchmark instances. One notable resource is the Packomania website~\citep{Spechtweb}, which serves as a comprehensive repository for PUCC benchmark instances and their corresponding best-known results. 

The Packomania website serves as a valuable resource, providing a comprehensive and up-to-date history of PUCC benchmarks. It showcases the continuous improvement in the best-known results for the PUCC problem over the past two decades, despite the fact that many algorithms responsible for these advancements may not have been officially published. 
For instance, 
in the most famous benchmark of $r_{i} = i$ proposed by AZPC, the best-known results for $n \leq 23$ and $n = 25$ were achieved by the contestants who participated in AZPC, \citet{muller2009packing} attained several best-known results in the range of $24 \leq n \leq 38$, and \citet{zeng2016iterated} hold the best-known results for $39 \leq n \leq 49$. 
In the benchmark of $r_{i} = i^{-1/2}$ introduced by \citet{castillo2008solving}, \citet{zeng2016iterated} hold many best-known results within the range of $17 \leq n \leq 35$. 
Regarding the NR benchmark proposed by \citet{huang2006new}, \citet{ye2013iterated} and \citet{zeng2016iterated} hold most of the best-known results.

Based on our literature review and the updated history provided on the Packomania website, a clear observation emerges: heuristic methods based on the penalty function~\citep{muller2009packing,ye2013iterated,zeng2016iterated} exhibit significantly higher efficiency  than other methods employed in solving the PUCC problem. Furthermore, the continuous improvement in the best-known results over time indicates that there is still untapped potential for further advancements.


\section{The Penalty Model} \label{sec:model}

Given the intricate nature of addressing the constrained optimization problem posed by PUCC 
using local search methods or heuristics, the penalty model emerges as a relaxation approach to convert the problem into a series of unconstrained optimization problems. By fixing the radius of the circular container, algorithms leveraging the penalty model can obtain quasi-feasible solutions by solving a simpler subproblem of PUCC through unconstrained continuous optimization methods. Subsequently, a feasible solution can be obtained through post-processing methods or mathematical techniques.

\begin{figure}[!tb]
    \centering
    \begin{minipage}[b]{0.4\linewidth}
        \centering
        \subfloat[Overlapping depth $d_{ij}$ 
        ]{\includegraphics[width=1\linewidth]{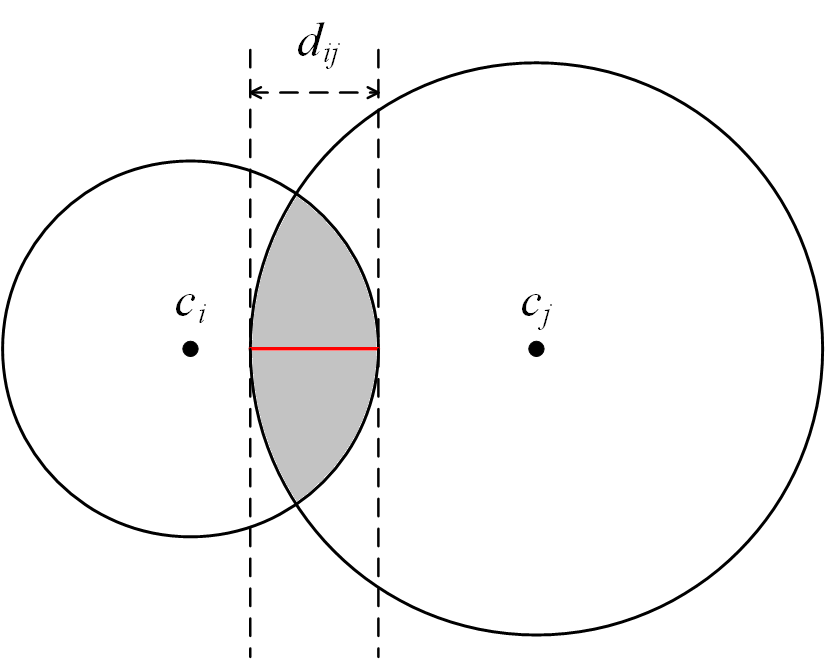}}
    \end{minipage}
    \begin{minipage}[b]{0.45\linewidth}
        \centering
        \subfloat[Overlapping depth $d_{i0}$ 
        ]{\includegraphics[width=1\linewidth]{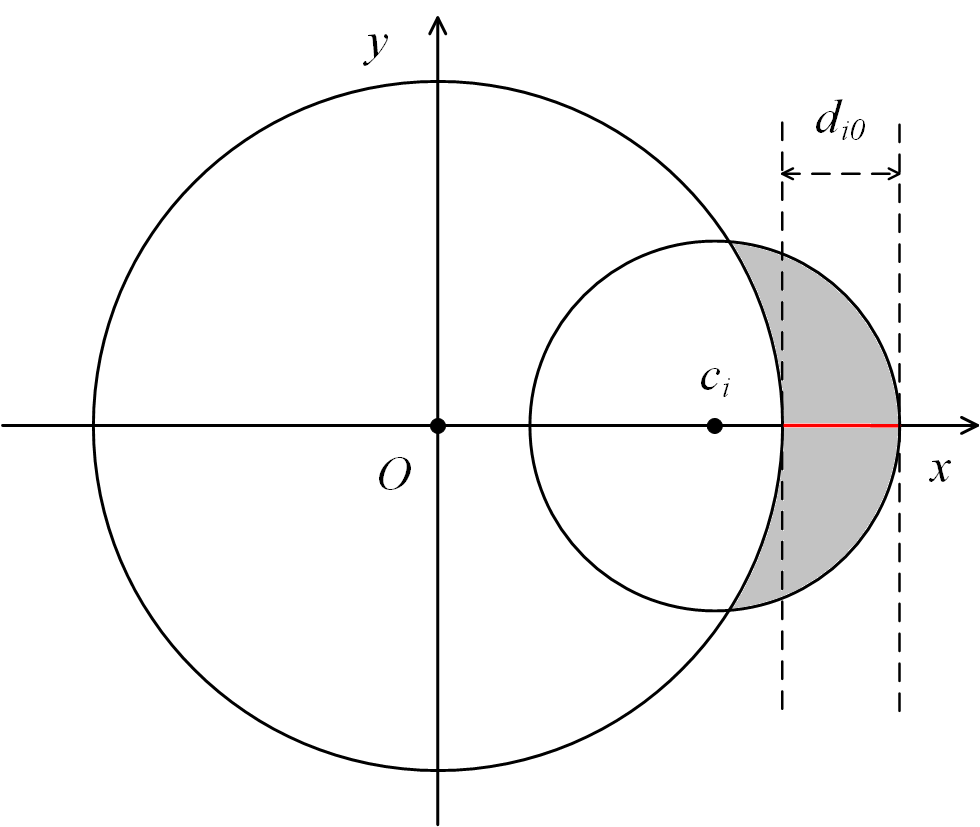}}
    \end{minipage}

    \caption{Illustration of two types of possible overlaps in PUCC configurations.}
    \label{fig:overlap}
\end{figure}

Following the landmark studies~\citep{huang1999two,ye2013iterated,zeng2016iterated},
we employ the classic penalty model introduced by ~\cite{huang1999two}, known as the quasi-physical and quasi human (QPQH) method, to tackle the PUCC problem. In this model, a certain level of overlap between circles or between circles and the container is permissible, and an objective function is designed to quantify the overlapping degree. The goal is to minimize the objective function, which is equivalent to minimizing the overlapping degree, for seeking 
a configuration of PUCC with minimal overlap. 
Consequently, the problem is transformed into an unconstrained optimization problem.

Formally, with a fixed radius $R$ for the container and a coordinate vector $X = ( x_1, y_1, x_2, y_2, ..., x_n, y_n )$, where $X \in \mathbb{R}^{2n}$ represents a packing configuration of $n$ circles with known radii $\{ r_1, r_2, ..., r_n \}$,  the objective function $E_R(X)$ can be formulated as follows:
\begin{align} 
    E_R(X) = \sum_{i=1}^{n-1} \sum_{j=i+1}^{n} d_{ij}^{2} + \sum_{i=1}^{n} d_{i0}^{2}, \label{eq:E}
\end{align}
with
\begin{align}
    & d_{ij} = \max \left( 0, r_i + r_j - \sqrt{{(x_i - x_j)}^2 + {(y_i - y_j)}^2} \right), \label{eq:d_ij} \\
    & d_{i0} = \max \left( 0, \sqrt{x_i^2 + y_i^2} + r_i - R \right), \label{eq:d_i0}
\end{align}
where $d_{ij}$ indicates the overlapping depth between two circles $c_i$ and $c_j$, $d_{i0}$ indicates the overlapping depth between circle $c_i$ and the circular container. Figure~\ref{fig:overlap} 
illustrates these two types of overlaps. 

In essence, $d_{ij}$ and $d_{i0}$ serve to quantify the violation of non-overlapping constraints in Formulas (\ref{eq:pre1}) and (\ref{eq:pre2}). The objective function $E_R(X)$ offers a measure of constraint violation in a given packing configuration. 
Consequently, a packing configuration $X$ is deemed feasible when condition $E_R(X) = 0$ holds, signifying  satisfaction of 
the two non-overlapping constraints; otherwise, it is considered infeasible. 
Algorithms leveraging this penalty model focus on solving the subproblem encapsulated by Equations~(\ref{eq:E})-(\ref{eq:d_i0}), aiming to minimize the objective function $E_R(X)$ to attain a feasible configuration for PUCC.



\section{The Proposed Algorithm} \label{sec:proalg}

\begin{figure}[!tb]
    \centering
    \includegraphics[width=0.8\linewidth]{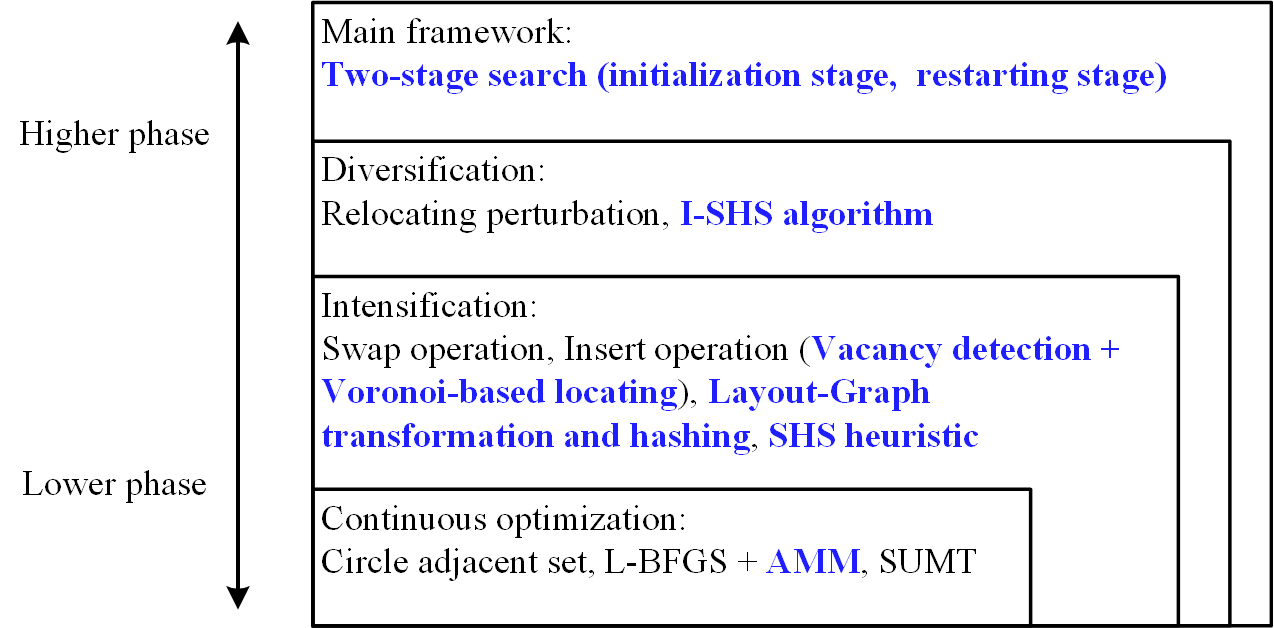} 
    \caption{The framework of our proposed algorithm.}
    \label{fig:framework}
\end{figure}

Our proposed algorithm is a stochastic optimization algorithm for solving PUCC, characterized by a hierarchical structure consisting of four phases: continuous optimization, intensification, diversification, and main framework. 
Each phase has different modules for achieving specific optimization objectives for solving PUCC, with modules from lower phases being invoked by those in higher phases. Figure~\ref{fig:framework} presents the four-phase hierarchical structure of the proposed algorithm, including the major methods and algorithms used in this study, where our proposed methods and algorithms are indicated in blue and bold. 
In this section, we introduce these phases in a bottom-up order, elucidating their roles and functionalities within the algorithmic framework.

\subsection{Continuous Optimization}


In the continuous optimization phase, 
there are two essential function modules, denoted as layout optimization and container optimization. The layout optimization module aims to find a local minimum solution with respect to the objective function $E_R(X)$ (defined in Equation (\ref{eq:E})) for a given solution, and the container optimization module aims to find a feasible solution for a given solution such that the container radius is minimized. Both function modules achieve their goals via continuous optimization methods.

\subsubsection{Layout Optimization} \label{sec:lay_opti}

\begin{figure}[tb]
    \centering
    \includegraphics[width=0.8\linewidth]{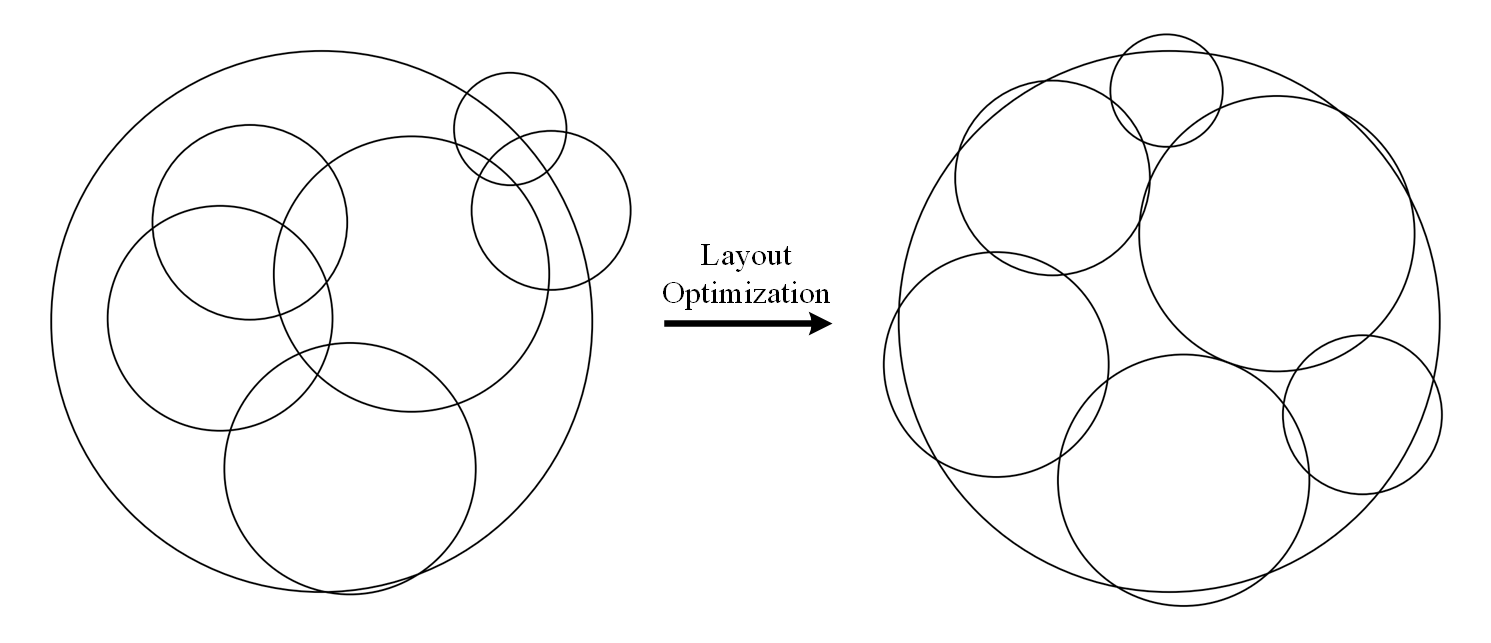}

    \caption{Illustration of the layout optimization process for a PUCC configuration.}
    \label{fig:layoutopt}
\end{figure}

Since we employ the penalty model to tackle PUCC, the layout optimization module aims to find a solution $X$ such that the objective function $E_R(X)$ is as minimal as possible by solving the problem consisting of Equations~(\ref{eq:E})-(\ref{eq:d_i0}). We employ an efficient quasi-Newton method, called Limited-memory Broyden-Fletcher-Goldfarb-Shanno (L-BFGS) method~\citep{liu1989limited}, as the basic continuous optimization method.
The call of the L-BFGS method can optimize solution $X$ to reach a local minimum with respect to the objective function $E_R(X)$.  
Figure~\ref{fig:layoutopt} illustrates an example of the layout optimization process that obtains a local minimum solution starting from a randomly initialized solution.

Before discussing the layout optimization module, we first introduce an important data structure, called the circle adjacency set, for accelerating the layout optimization process. The circle adjacency set is widely adopted for solving CPP, and its efficiency has been demonstrated in many studies~\citep{zeng2016iterated,zeng2018adaptive,he2018efficient,lai2022iterated,lai2023perturbation}.

First, we define that two circles $c_i$ and $c_j$ are considered adjacent if 
the distance between the centers of circles $c_i$ and $c_j$ is less than $\max \left( \frac{r_i + r_j}{2}, \frac{r_n}{4} \right) + r_i + r_j$, where $r_n$ is the largest radius of $n$ circles.
The circle adjacency set $\Gamma(i)$ of circle $c_i$, $\Gamma(i) \subseteq \{ c_1, c_2, ..., c_n\}$, consists of circles adjacent to circle $c_i$, defined as follows:
\begin{align}
    \Gamma(i) = \left\{ c_j \enskip \vline \enskip \forall j: \,  1 \leq j \leq n, \, i \neq j, \, \sqrt{(x_i - x_j)^2 + (y_i - y_j)^2} < \max \left( \frac{r_i + r_j}{2}, \frac{r_n}{4} \right) + r_i + r_j \right\}. \label{eq:gamma}
\end{align}

Consequently, the objective function $E_R(X)$ presented in Equation (\ref{eq:E}) can be calculated by enumerating the pairs of circles in the circle adjacency set $\Gamma$, reformulated as follows:
\begin{align}
    E_R(X) = \sum_{i=1}^{n} \sum_{j \in \Gamma(i)} d_{ij}^{2} [i < j] + \sum_{i=1}^{n} d_{i0}^{2}, \label{eq:E_gamma}
\end{align}
where ``$[~]$'' represents the Iverson bracket that $[P] = 1$ if statement $P$ holds, and $[P] = 0$ otherwise. The statement ``$i < j$'' ensures that each pair of circles is calculated only once. 


\begin{algorithm}[tb]
\DontPrintSemicolon
\caption{Layout Optimization} \label{alg:layopti}

\KwInput{Objective function $f$, input solution $X_0$}
\KwOutput{The local minimum solution $X$ with respect to $f(X)$}

\SetKwFunction{Flayout}{\textit{LayoutOptimization}}

\SetKwProg{Fn}{Function}{:}{\KwRet $X$}
\Fn{\Flayout($f$, $X_0$)}
{
    \tcc{AAM initialization}
    $cnt \leftarrow 0$, \enskip $len \leftarrow 1$ \;
    Construct a circle adjacency set $\Gamma$ of $X_0$ \;
    
    \tcc{L-BFGS procedure}
    $X \leftarrow X_0$ \;
    \While{$\lVert \nabla f(X) \rVert_2 > \epsilon_0$}    
    {
        Calculate a descent direction $d_k$ by two-loop recursion approach \;
        Calculate a step length $\alpha_k$ by a line search method \;
        $X \leftarrow X + \alpha_k d_k$ \tcc*{Update the solution}

        \tcc{AAM procedure}
        $cnt \leftarrow cnt + 1$ \;
        \If{$cnt \geq len$}
        {
            Reconstruct a new circle adjacency set $\Gamma'$ of $X$ \;
            \uIf{$\Gamma \neq \Gamma'$}
            {
                $cnt \leftarrow 0, \enskip len \leftarrow 1$ \;
                $\Gamma \leftarrow \Gamma'$ \tcc*{Update circle adjacency set}
            }
            \Else
            {
                $cnt \leftarrow 0, \enskip len \leftarrow 2 \times len$
            }
        }
    }
}
\end{algorithm}

To maintain the circle adjacency set during the layout optimization process, we propose an Adaptive Adjacency Maintenance (AAM) method. 
The layout optimization module (denoted by \textit{LayoutOptimization}) is depicted in Algorithm~\ref{alg:layopti}, which is the L-BFGS method coupled with our AAM method.
The call of \textit{LayoutOptimization}($E_R$, $X_0$) returns a local minimum solution $X$ with respect to $E_R(X)$.

In detail, starting from an objective function $f$ ($f = E_R$ for layout optimization) and an input solution $X_0$, \textit{LayoutOptimization} initializes two variables $cnt$ and $len$ of AAM as $cnt = 0, \,  len = 1$ and constructs a circle adjacency set $\Gamma$ by the input solution $X_0$ (lines 2-3). 
Then, \textit{LayoutOptimization} performs the L-BFGS procedure to find a local minimum solution (lines 4-19), returning it as the final result. The solution $X$ is regarded as reaching a local minimum when the norm of the gradient of the objective function $f(X)$ is tiny enough (i.e., $\lVert \nabla f(X) \rVert_2 \leq \epsilon_0, \  \epsilon_0 = 10^{-10}$ in this study).

At each iteration of L-BFGS, \textit{LayoutOptimization} applies the two-loop recursion approach, which is the core of L-BFGS, to calculate a descent direction $d_k$ and a line search method to obtain a descent step length $\alpha_k$. Then, the solution $X$ is updated through $d_k$ and $\alpha_k$ (lines 6-8). 
Subsequently, the AAM module is executed (lines 9-18). AAM first increases $cnt$ by 1 and compares $cnt$ with $len$. The maintenance process is triggered when $cnt$ reaches $len$, and AAM defers the maintenance process otherwise. 
In the maintenance process, a new circle adjacency set $\Gamma'$ is constructed and compared with the current circle adjacency set $\Gamma$. If $\Gamma'$ and $\Gamma$ are different, the current circle adjacency set $\Gamma$ is updated by $\Gamma'$, and the two variables are reset to $cnt = 0$ and $len = 1$. Otherwise, $cnt$ is set to 0, and $len$ is multiplied by 2.

In summary, AAM uses two variables, $cnt$ and $len$, to accomplish the adaptive maintenance feature. When the configuration is unstable during the layout optimization process, indicating the circle adjacency set constructed at each iteration is different, AAM constantly resets the two variables
, and the reconstruction is executed at each iteration. Otherwise, the variable $len$ grows exponentially, causing AAM to defer the maintenance process and reduce computational overhead. 

\subsubsection{Container Optimization} \label{sec:cont_opti}

The container optimization module is another essential component of the proposed algorithm for obtaining a feasible solution. It aims to slightly adjust the coordinates of $n$ circles and the container radius for a given packing configuration such that the resulting configuration becomes feasible while the container radius is locally minimized.

\begin{figure}[tb]
    \centering
    \includegraphics[width=0.8\linewidth]{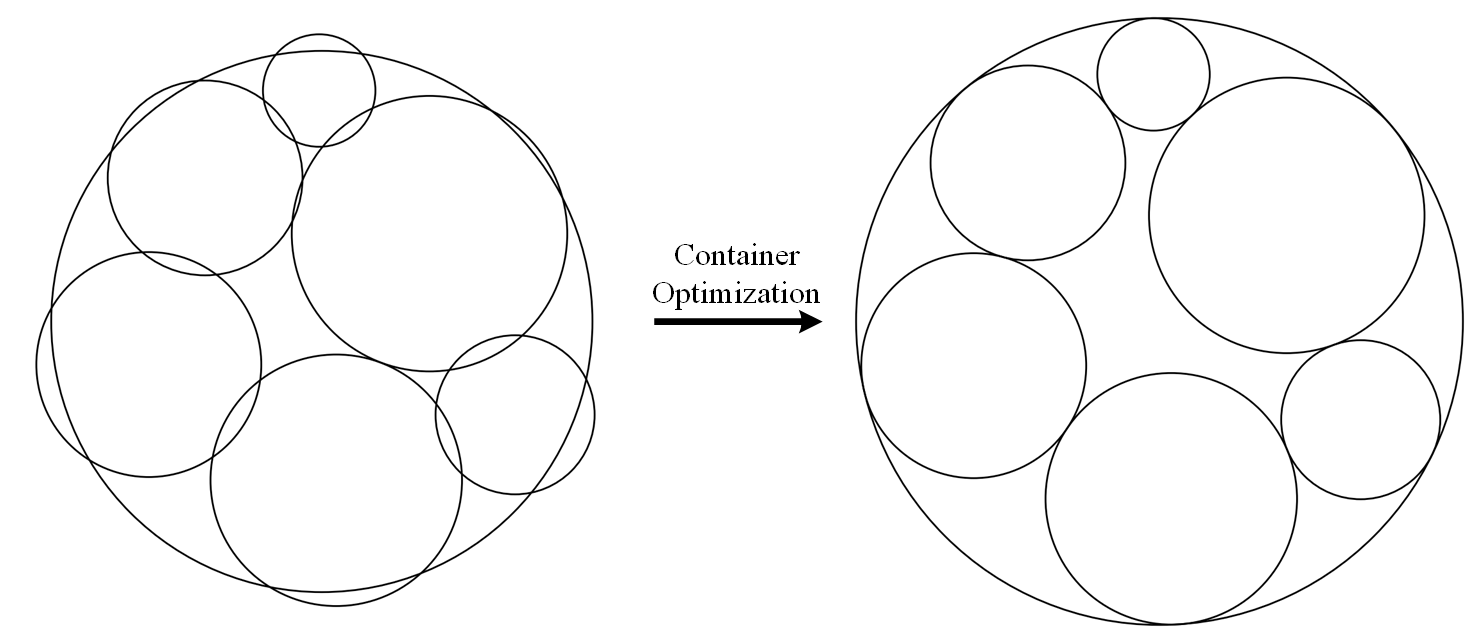}

    \caption{Illustration of the container optimization process for a PUCC configuration.}
    \label{fig:containeropt}
\end{figure}

Referring to existing circle packing studies~\citep{lai2022iterated,lai2023perturbation}, we adopt the sequential unconstrained minimization technique (SUMT)~\citep{fiacco1964computational} to handle this problem. The SUMT method converts the constrained optimization problem, consisting of Formulas~(\ref{eq:tar})-(\ref{eq:pre2}), to a sequence of unconstrained optimization problems and then solves them to obtain a feasible local minimum solution. 
Specifically, let a vector $Z = (x_0, y_0, x_1, y_1, ..., x_n, y_n, R)$, $Z \in \mathbb{R}^{2n+1}$, represents a packing configuration, consisting of the coordinates of $n$ circles and the value of $R$ for the container radius. A new objective function $U_{\rho}(Z)$ is formulated as follows:
\begin{equation} 
U_{\rho}(Z) = \sum_{i=1}^{n} \sum_{j \in \Gamma(i)} d_{ij}^{2} [i < j] + \sum_{i=1}^{n} d_{i0}^{2} + \rho R^{2} , \label{eq:U_rho}
\end{equation}
where $\rho R^2$ is a penalty term with $\rho$ as the penalty factor, $d_{ij}$ and $d_{i0}$ indicate the overlapping depths defined in Equations~(\ref{eq:d_ij}) and (\ref{eq:d_i0}). In short, the unconstrained optimization function $U_{\rho}(Z)$ consists of the penalty term $\rho R^2$ and $E_R(X)$ (formulated in Equation~(\ref{eq:E_gamma})), where the container radius $R$ becomes a variable.
The SUMT method sequentially minimizes several functions $U_{\rho}(Z)$ with decreasing penalty factors $\rho$ for an initial packing configuration $Z$ to obtain a resulting feasible configuration of which the container radius  $R$ reaches a local minimum.
Figure~\ref{fig:containeropt} illustrates an example of the container optimization process that obtains a feasible solution from an infeasible local minimum solution. 

\begin{algorithm}[tb]
\DontPrintSemicolon
\caption{Container Optimization} \label{alg:contopti}

\KwInput{Input solution $X_0$, input radius $R_0$}
\KwOutput{The feasible solution ($X$, $R$) where $R$ is locally minimized}

\SetKwFunction{Fcont}{\textit{ContainerOptimization}}

\SetKwProg{Fn}{Function}{:}{\KwRet {($X$, $R$)}}
\Fn{\Fcont($X_0$, $R_0$)}
{   
    $X \leftarrow X_0, \enskip R \leftarrow R_0, \enskip \rho \leftarrow 10^{-3}$ \;
    \While{$E_R(X) > \epsilon_1$}    
    {
        $Z \leftarrow$ ($X$, $R$) \;
        $Z' \leftarrow $ \textit{LayoutOptimization}($U_{\rho}$, $Z$) \;
        ($X$, $R$) $\leftarrow Z', \enskip \rho \leftarrow \rho / 2$ 
    }
}
\end{algorithm}

The container optimization module (denoted by \textit{ContainerOptimization}) is depicted in Algorithm~\ref{alg:contopti}. Starting from an input solution $X_0$ and an input radius $R_0$, \textit{ContainerOptimization} empirically initializes the penalty factor $\rho$ as $\rho = 10^{-3}$. Then, it executes several iterations until a feasible solution is obtained and returned as the final result, in which the radius of the feasible solution is locally minimized.
Note that a packing configuration is regarded as feasible when condition $E_R(X) \leq \epsilon_1$ holds ($\epsilon_1 = 10^{-25}$ in this study), i.e., the overlapping degree of the configuration is tiny enough.

At each iteration, the current solution $X$ and the current radius $R$ compose a new solution $Z$.
The \textit{LayoutOptimization} module is called to minimize the objective function $U_{\rho}(Z)$, 
obtaining an updated solution $(X, R)$ by decomposing the resulting solution $Z'$.
The penalty factor $\rho$ is halved for the next iteration. The iteration process terminates when the solution $(X, R)$ is feasible ($E_R(X) \leq \epsilon_1$).


\subsection{Intensification}

When solution $X$ gets stuck in a local minimum with respect to $E_R(X)$, several adjustment strategies are designed to adjust the positions of some circles within the solution, aiming to jump out of the local minimum and find a new solution with a smaller value of $E_R(X)$.
This iterative process, known as intensification, continually diminishes the value of $E_R(X)$, leading to the discovery of improved solutions.

Building upon seminal studies~\citep{addis2008efficiently,muller2009packing,ye2013iterated,zeng2016iterated}, we exploit two efficient adjustment operations: \textit{swap} and \textit{insert}. Meanwhile, we propose novel layout-graph transformation and hashing methods to tackle the problem of determining isomorphism or similarity between PUCC configurations. Leveraging these methodologies, we introduce the Solution-Hashing Search (SHS) algorithm as the intensification module for 
enhancing solution discovery. 

We first introduce the \textit{swap} and \textit{insert} operations, and then we present our proposed methods, including the vacancy detection method, the Voronoi-based locating method, and the layout-graph transformation and hashing methods. Finally, we introduce the SHS algorithm.


\subsubsection{Swap Operation} \label{sec:swap}

The \textit{swap} operation is a very efficient adjustment strategy in solving PUCC, which assists~\citet{addis2008efficiently} won AZPC in a racy final and is adopted in many landmark PUCC studies~\citep{muller2009packing,ye2013iterated,zeng2016iterated}.
Given a local minimum solution $X$, the \textit{swap} operation swaps the positions of two circles with different radii in solution $X$. Then, it continually optimizes the solution (i.e., call the \textit{LayoutOptimization} module presented in Algorithm~\ref{alg:layopti}) to obtain another local minimum solution $X'$, where $X'$ is called a \textit{swap} neighbor of $X$. The \textit{swap} neighborhood of $X$ consists of all \textit{swap} neighbors of $X$.

To reduce the \textit{swap} neighborhood size, 
a protocol is proposed that two circles $c_i$ and $c_j$ can be swapped if the rankings of their radii $r_i$ and $r_j$ are adjacent, i.e., $|ranking(r_i) - ranking(r_j)| = 1$ where $ranking(r_i)$ indicates the ranking of radius $r_i$ in the given $n$ radii $\{ r_1, r_2, ..., r_n \}$.
Specifically, a swapping list $L$ consists of the pairs of circles that can be swapped. The \textit{swap} neighborhood of $X$ is a set of candidate solutions obtained by performing the \textit{swap} operation on the pairs of circles in $L$. 
There is an example provided for facilitating understanding: let $n = 6$, a solution contains circles $\{ c_1, c_2, c_3, c_4, c_5, c_6 \}$ with their radii $\{ r_1, r_2, r_3, r_4, r_5, r_6 \} = \{ 2, 2, 3, 3, 5, 7 \}$; the ranking of the radii is $ranking = \{ 1, 1, 2, 2, 3, 4 \}$, so the swapping list is $L = \{ (c_1, c_3), (c_1, c_4), (c_2, c_3), (c_2, c_4), (c_3, c_5), (c_4, c_5), (c_5, c_6)\}$; circles $c_1$ and $c_2$ cannot be swapped because their radii are the same ($r_1 = r_2 = 2$), circles $c_1$ and $c_5$ cannot be swapped either because the rankings of their radii are not adjacent ($ranking(r_1) = 1$ and $ranking(r_5) = 3$).

\subsubsection{Insert Operation} \label{sec:insert}

The \textit{insert} operation is another important adjustment strategy that has been demonstrated to be complementary to the \textit{swap} operation in solving PUCC~\citep{zeng2016iterated}. 
The \textit{insert} operation picks a small circle from solution $X$ to place into a large vacancy. 
Then, similar to the \textit{swap} operation, it optimizes the solution to obtain a new local minimum solution $X'$, where $X'$ is called an \textit{insert} neighbor of $X$, and the \textit{insert} neighborhood of $X$ consists of the \textit{insert} neighbors of $X$.
However, the concept of vacancies is not intuitive, and the problem of detecting and measuring vacancies for PUCC is very challenging.
Several studies have proposed various methods to detect vacancies for solving CPP and PUCC~\citep{huang2010greedy,he2015action,specht2015precise,zeng2016iterated,ryu2020voropack}.

We propose a novel and efficient method to detect vacancies through a continuous optimization approach. 
Specifically, we use an additional circle with a changeable radius, called vacancy circle, to locate and measure a vacancy for a given configuration. 
Let a vector $u = (x_u, y_u, r_u)$, $u \in \mathbb{R}^{3}$, represent a vacancy circle 
where coordinate $(x_u, y_u)$ represents the center of the vacancy circle, and $r_u$ indicates its radius.

A objective function $P_{X,R,\rho}(u)$ is defined as follows:
\begin{align}
    P_{X,R,\rho}(u) =  \sum_{i \in \Gamma_u} d_{ui}^2 + d_{u0}^2 - \rho r_u , \label{eq:P}
\end{align}
with 
\begin{align}
    & d_{ui} = \max \left( 0, |r_u| + r_i - \sqrt{{(x_u - x_i)}^2 + {(y_u - y_i)}^2} \right), \label{eq:d_vi} \\
    & d_{u0} = \max \left( 0, \sqrt{x_u^2 + y_u^2} + |r_u| - R \right), \label{eq:d_v0} 
\end{align}
where $- \rho r_u$ is a penalty term with $\rho$ as the penalty factor, $d_{ui}$ indicates the overlapping depth between vacancy circle $u$ and circle $c_i$, $d_{u0}$ indicates the overlapping depth between vacancy circle $u$ and the circular container, and $\Gamma_u$ is the circle adjacency set of vacancy circle $u$ consisting of the packed circles adjacent to vacancy circle $u$ for accelerating the continuous optimization process (similar to $\Gamma(i)$ in Equation~(\ref{eq:gamma})).

Given a packing configuration $(X, R)$, starting from an arbitrary coordinate $(x_u, y_u)$ within the container with $r_u = 0$, the algorithm can use the SUMT method to sequentially minimize several objective functions $P_{X, R, \rho}(u)$ with decreasing penalty factors $\rho$ to reach the nearest local maximum vacancy circle $u^*$ of which radius $r_u$ is locally maximized. Note that the $n$ packed circles are fixed, and only vacancy circle $u$ changes during the vacancy detection process.  
Nevertheless, using the SUMT method is computationally expensive for detecting vacancies, which is demanded to solve a sequence of continuous optimization problems. 
To reduce the computational overhead, we only solve two stages of the problems with $\rho = 0.5$ and $0.1$ (i.e., minimize $P_{X,R,0.5}(u)$ first and then minimize $P_{X,R,0.1}(u)$) to reach an approximate local maximum vacancy circle $u'$ instead of using the SUMT method. 
In this way, the algorithm finds an approximate vacancy circle $u'$ very efficiently, and the approximate vacancy circle $u'$ is close enough to the precise vacancy circle $u^*$ found by the SUMT method, where the condition $\lVert u' - u^* \rVert_2 < 10^{-2}$ is usually satisfied in our experiments, indicating that the approximate vacancy circle $u'$ is sufficient to locate and measure a vacancy.

Obviously, the resulting vacancy circle is strongly affected by the initially selected coordinate. Several strategies are proposed for selecting the initial positions to detect vacancies, such as random spreading method~\citep{huang2010greedy}, action-space-based method~\citep{he2015action}, lattice covering method~\citep{zeng2016iterated}, etc. 
Leveraging the characteristics of two-dimensional packing problems, we propose a Voronoi-based locating method to select initial positions more properly for vacancy detection. 
The Voronoi diagram is a classic mathematical model that partitions a plane into regions based on a given set of sites~\citep{aurenhammer2000voronoi}. 
For a given configuration, we use the centers of $n$ circles as the set of sites to construct a Voronoi diagram by Fortune's algorithm~\citep{fortune1986sweepline}. 
Then, we select two types of coordinates based on the constructed Voronoi diagram as the initial positions. 

\begin{figure}[tb]
    \centering
    \begin{minipage}[b]{0.45\linewidth}
        \centering
        \subfloat[Voronoi diagram 
        ]{\includegraphics[width=1\linewidth]{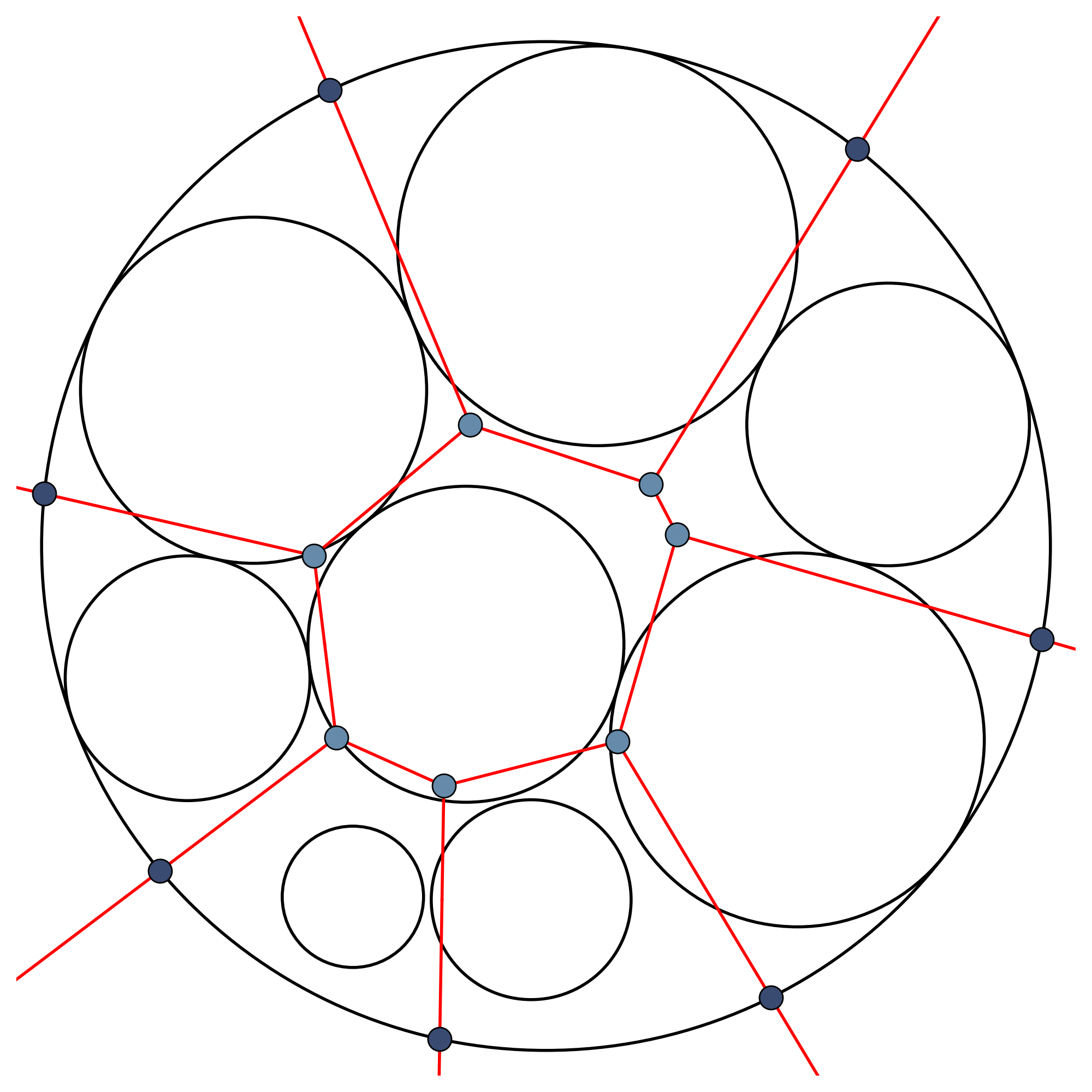}}
    \end{minipage}
    \begin{minipage}[b]{0.45\linewidth}
        \centering
        \subfloat[Vacancy detection result found by our methods]{\includegraphics[width=1\linewidth]{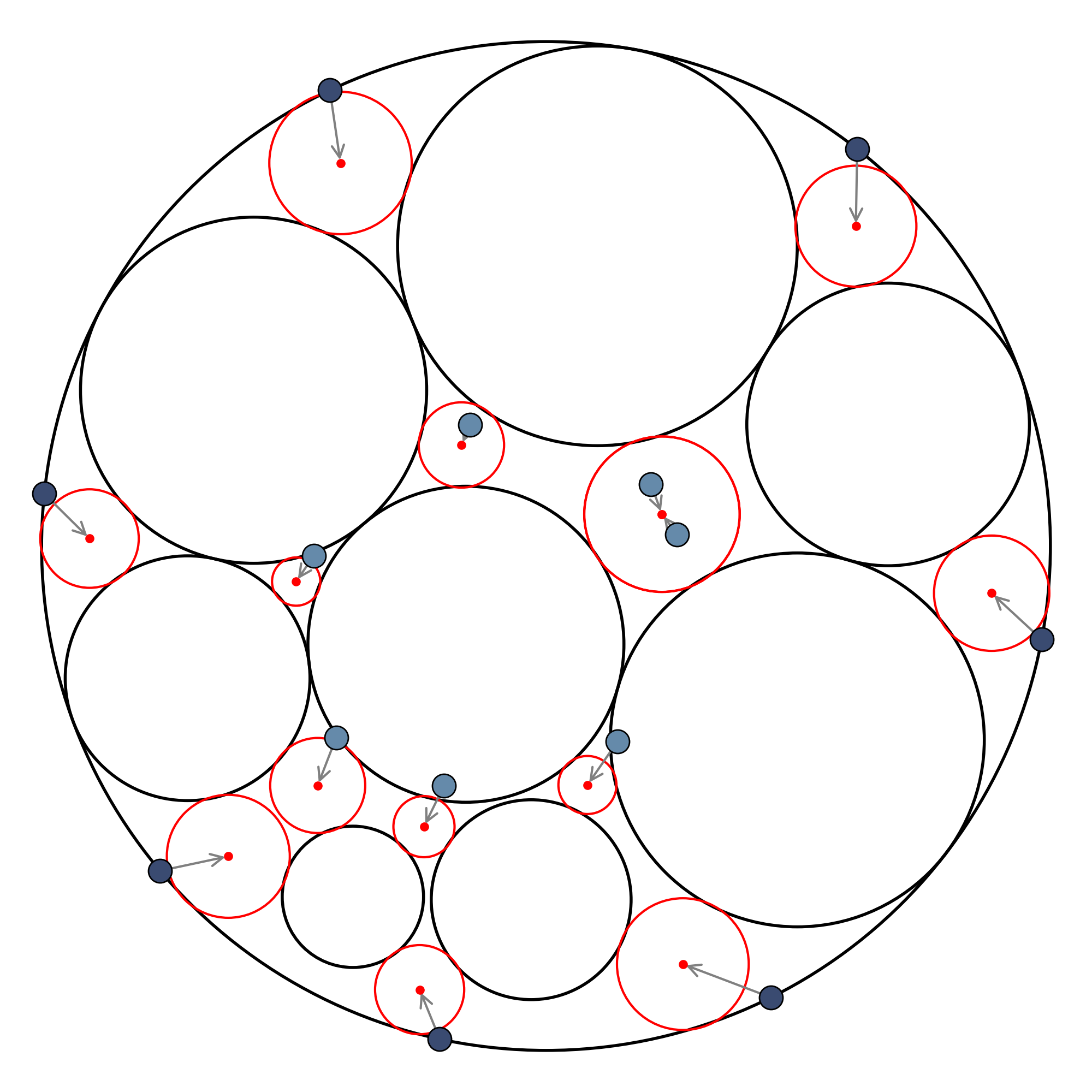}}
    \end{minipage}

    \caption{Illustration of the Voronoi diagram and vacancy detection method for a PUCC configuration.}
    \label{fig:vor_bubble}
\end{figure}

The first type of coordinates consists of the vertices of the Voronoi diagram. According to the definition of the Voronoi diagram, the distances from a vertex of Voronoi to certain three sites are equal. Therefore, the coordinate of vertices is usually close to the center of the vacancy circles that contact to three packed circles. Note that the vertices exceeding the container are excluded.
The second type of coordinates consists of intersections of the edges of  Voronoi and the circular container. The vacancy detection method initialized from the second type of coordinates usually finds the vacancy circles that contact to two packed circles and the circular container.
As a result, the algorithm solves the two problems that minimize $P_{X,R,0.5}(u)$ and $P_{X,R,0.1}(u)$ for each initial position to obtain a set of vacancy circles as the vacancy detection result. Figure~\ref{fig:vor_bubble} presents two graphical examples to showcase the Voronoi diagram with two types of coordinates and the vacancy detection result on a PUCC configuration.

After obtaining a set of vacancy circles, we use a simple strategy to construct the \textit{insert} neighborhood that places small circles into large vacancies. 
Specifically, a circle $c_i$ is regarded as small among $n$ circles with radii $r_1 \leq r_2 \leq ... \leq r_n$ when $i \leq \lceil \frac{n}{3} \rceil$. The set of vacancy circles is sorted in descending order based on the value of $r_u$ to obtain an ordered set $U = \{ u_1, u_2, ..., u_{|U|}\}$ with $r_{u,1} \geq r_{u,2} \geq ... \geq r_{u,|U|}$, vacancy circle $u_j$ is regarded as a large vacancy when $j \leq \lceil \frac{n}{3} \rceil$. 
Therefore, the \textit{insert} neighborhood of a solution $X$ consists of the solutions $X'$ that perform the \textit{insert} operation in an \textit{insert} operation set $S$ in solution $X$ where the \textit{insert} operation set is formulated as $S = \left\{ (c_i, u_j) \enskip \vline \enskip 1 \leq i, j \leq \lceil \frac{n}{3} \rceil \right\}$, the \textit{insert} operation $(c_i, u_j)$ indicates placing the center of circle $c_i$ to the center of vacancy circle $u_j$.

Notably, compared with the imprecise vacancy detection methods~\citep{huang2010greedy,he2015action,zeng2016iterated}, our proposed method can locate and measure vacancies more precisely and properly, especially via the SUMT method. Compared with the precise vacancy detection methods~\citep{specht2015precise,ryu2020voropack}, our proposed method is simpler with good generality, which can be easily adapted to solve other packing problems, such as the circle packing with containers in various shapes~\citep{lopez2013packing,stoyan2020optimized}, $d$-dimensional ($d \geq 3$) hypersphere packing~\citep{stoyan2020optimized,lai2023perturbation,hifi2023threshold}, non-Euclidean circle packing~\citep{lai2023iterated}, etc.

\subsubsection{Layout-Graph Transformation and Hashing} \label{sec:layhash}



\begin{figure}[tb]
    \centering
    \includegraphics[width=0.8\linewidth]{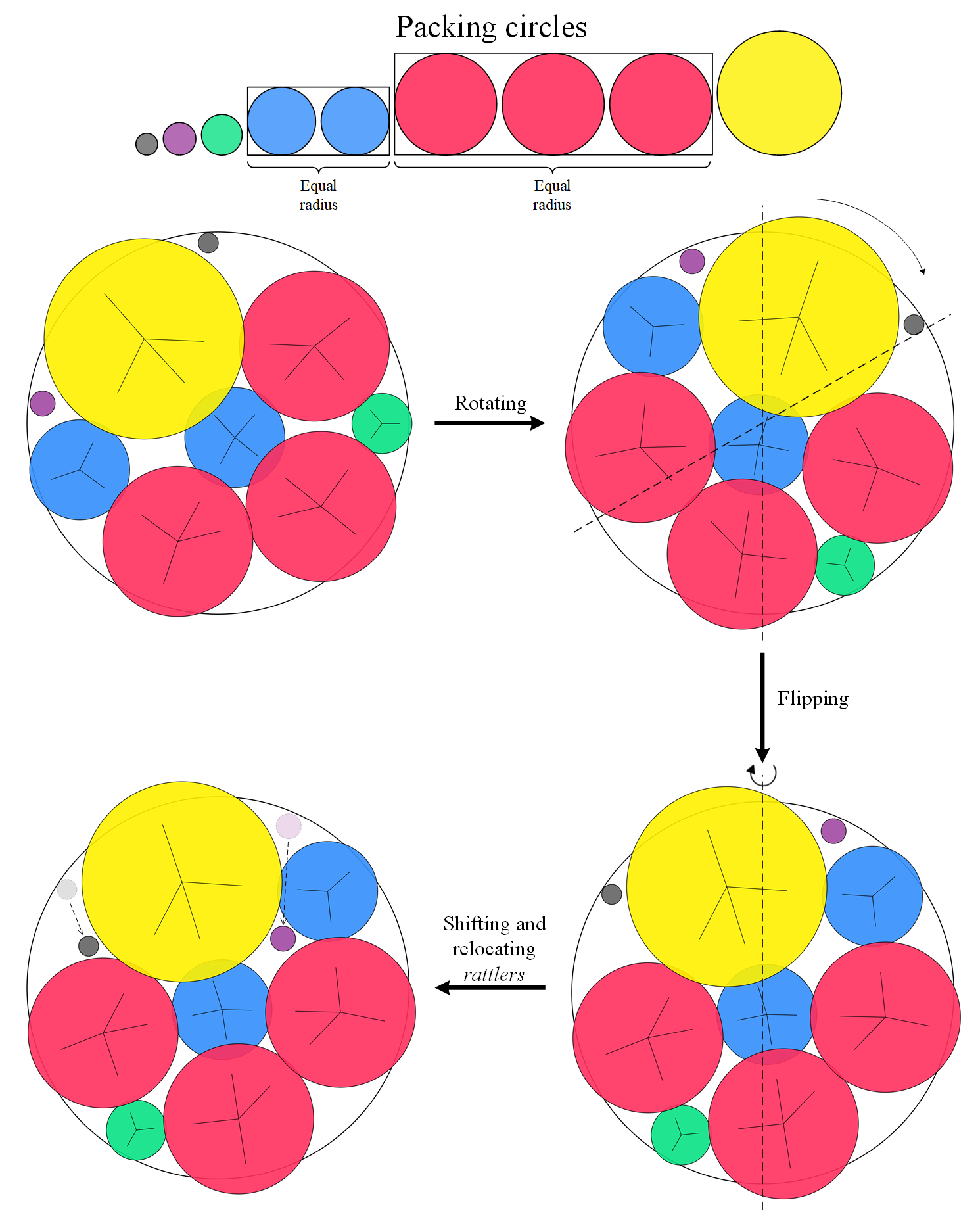}

    \caption{Illustration of three isomorphic or similar transformations of a PUCC configuration.}
    \label{fig:layouttrans}
\end{figure}

In the PUCC problem, to avoid duplicate exploration, it is essential to solve the problem of determining whether two configurations are isomorphic or similar. However, this problem is very challenging.
A configuration can be transformed into an isomorphic configuration by rotating or flipping, and it can be rotated at any angle or flipped at any symmetry axis passing through the center of the circular container. Thus, a configuration $X$ has uncountably infinite isomorphic configurations in space $\mathbb{R}^{2n}$. 
Besides, we call two configurations $X$ and $X'$  similar if $X'$ can be obtained by relocating or shifting \textit{rattlers} in $X$, where a circle is called \textit{rattler} if it still has degrees of freedom for movement inside the container. 
Obviously, a configuration $X$ also has uncountably infinite similar configurations in space $\mathbb{R}^{2n}$. 
Figure~\ref{fig:layouttrans} shows three isomorphic or similar transformations for a PUCC configuration.
To tackle this problem, we propose novel layout-graph transformation and hashing methods and apply these methods to solve PUCC.

\begin{figure}[tb]
    \centering
    \includegraphics[width=0.8\linewidth]{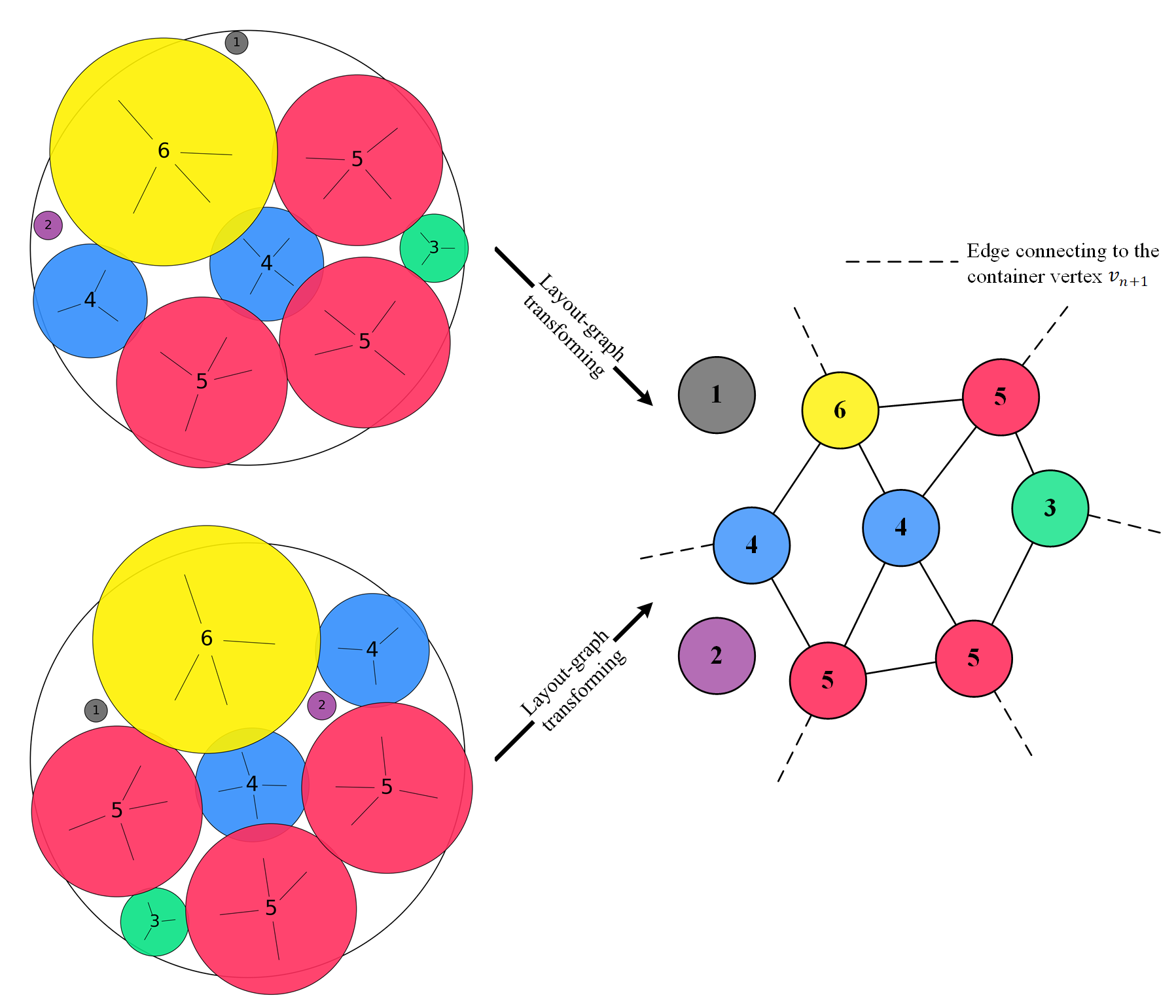}

    \caption{Illustration of transforming two configurations into an isomorphic graph with the vertices labeled.}
    \label{fig:layout2graph}
\end{figure}

Given a slightly compact and infeasible configuration $X$ ($E_R(X) > 0$) of which the value of $E_R(X)$ reaches a local minimum. The layout-graph transformation method transforms $X$ into a graph $G = (V, E)$. 
The vertex set $V$ of $G$ consists of $n+1$ vertices $\{ v_1, v_2, ..., v_{n+1} \}$, and vertex $v_i (1 \leq i \leq n)$ with label $l_i = ranking(r_i)$ represents circle $c_i$ of $X$. Generally, two circles $c_i$ and $c_j$ are regarded as the same when their radii are equal ($r_i = r_j$). Therefore, we use the ranking of radii to label the vertices, two vertices $v_i$ and $v_j$ are of the same type when their labels are the same $(l_i = l_j)$. The vertex $v_{n+1}$ with an unique label (for instance, $l_{n+1} = -1$) represents the circular container of $X$.
There is an undirected and unlabeled edge $(v_i, v_j)$  in the edge set $E$ of $G$ if two circles $c_i$ and $c_j$ overlap with each other in $X$. Similarly, an edge $(v_i, v_{n+1})$ adds to $E$ if circle $v_i$ overlaps with the container. As a result, the edge set $E$ consists of edges that represent the overlapping relationship of the configuration $X$. Note that two circles $c_i$ and $c_j$ are regarded as overlap if $d_{ij} > \epsilon_3$, and circle $c_i$ overlaps with the container if $d_{i0} > \epsilon_3$ ($\epsilon$ is a parameter and set as $\epsilon_3 = 10^{-8}$ in this study). Figure~\ref{fig:layout2graph} illustrates a graphic example that transforms two similar configurations into an isomorphic graph with the vertices labeled. 

Therefore, we can determine whether two configurations $X_1$ and $X_2$ are isomorphic or similar by transforming $X_1$ and $X_2$ to two graphs $G_1$ and $G_2$ and then solving the graph isomorphism problem of $G_1$ and $G_2$. In other words, we consider two configurations $X_1$ and $X_2$  isomorphic or similar if $G_1$ and $G_2$ are isomorphic. Unfortunately, the graph isomorphism problem is a well-known NP problem, but it is not known to belong to either P and NP-complete (if P $\neq$ NP)~\citep{kobler2012graph}. Many advanced exact graph isomorphism solvers~\citep{grohe2020improved,huang2021short,augaouglu2022isomorphism} run in sub-exponential or exponential time complexity. Therefore, using these solvers to tackle this problem is extremely computationally expensive and unacceptable in solving PUCC.

To tackle the graph isomorphism problem, we propose an inexact and efficient graph hash method. Let $A$ be the adjacency matrix of $G$, $A \in \{0, 1\}^{(n+1) \times (n+1)}$, where $A_{ij} = A_{ji} = 1$ if undirected edge $(v_i, v_j)$ occurs in $E$, and $A_{ij} = A_{ji} = 0$ otherwise. A graph hash function $H_{p,q,M}(A)$ is presented as follows:
\begin{align} 
    H_{p,q,M}(A) = \sum_{i=1}^{n} \sum_{j=i+1}^{n+1} A_{ij} p^i q^j \bmod M, \label{eq:hash}
\end{align}
where $p$ and $q$ are two small prime numbers, and $M$ is a large prime number as a modulus. The term $p^i q^j$ maps the overlapping relation of two circles $c_i$ and $c_j$ or circle $c_i$ and the container to an integer value. The function $H_{p,q,M}(A)$ is the sum of the overlapping relationship of $X$ and maps the sum to an integer value in interval $[0, M-1]$. $M$ is also called the size of the hash table.

Nevertheless, our algorithm explores a large number of configurations during the local search phase and records them in a hash table. To reduce the hash collision, we use two functions $H_{p_1,q_1,M_1}(A)$ and $H_{p_2,q_2,M_2}(A)$ to map a configuration $X$ to a pair $(h_1, h_2)$ of hash values, $(h_1, h_2) \in [0, M_1 - 1] \times [0, M_2 - 1]$, where $(p_1, q_1, M_1)$ and $(p_2, q_2, M_2)$ are different parameter settings (we set $(p_1, q_1, M_1) = (17, 193, 998244353)$ and $(p_2, q_2, M_2) = (97, 257, 1004535809)$ in this study). 
Consequently, two configurations are determined to be isomorphic or similar when their pairs of hash values are equal. Note that the size of the hash table is extremely large and sparse (the size is $M_1 M_2$). Some data structures can be applied to store the table, for example, the Binary Search Tree.

It is worth noting that circle $c_i$ is identified to be a \textit{rattler} when the degree of the corresponding vertex $v_i$ in $G$ is equal to zero, i.e., $\mathrm{deg}(v_i) = 0$.
Due to the characteristic of \textit{rattlers}, the neighborhoods of the \textit{swap} and \textit{insert} operations can be further simplified.
Obviously, performing the \textit{swap} operation on two \textit{rattlers} will not lead to a better configuration, so the pair $(c_i, c_j)$ of circles is removed from the swapping list $L$ when two circles $c_i$ and $c_j$ are \textit{rattlers}. 
Similarly, performing the \textit{insert} operation on a \textit{rattler} will not lead to a better configuration, so the \textit{insert} operations $(c_i, u_j)$ related to circle $c_i$ are removed from the \textit{insert} operation set $S$ when circle $c_i$ is a \textit{rattler}.

\subsubsection{Solution-Hashing Search} \label{sec:SHS}

\begin{algorithm}[tb]
\DontPrintSemicolon
\caption{Solution-Hashing Search Core} \label{alg:SHSCore}

\KwInput{Input solution $X_0$, fixed container radius $R$, recorded hash table $T_0$}
\KwOutput{The best solution $X^*$ found thus far, updated hash table $T$}

\SetKwFunction{Flayout}{\textit{SHSCore}}

\SetKwProg{Fn}{Function}{:}{\KwRet {$(X^*, T)$}}
\Fn{\Flayout($X_0$, $R$, $T_0$)}
{
    $unimproved \leftarrow 0$, \enskip $X^* \leftarrow X_0$, \enskip $T \leftarrow T_0$, \enskip $Q \leftarrow \emptyset$ \;
    
    $(h_1, h_2) \leftarrow LayoutGraphTransHashing(X_0, R)$ \;
    
    \If{$(h_1, h_2) \notin T$}
    {
        $Q \leftarrow Q \cup \{X_0\}$, \enskip $T \leftarrow T \cup \{(h_1, h_2)\}$ \tcc*{Add the solution to $Q$ and record its hash value}
    }

    \While{$unimproved < n \  \mathbf{and} \  Q \neq \emptyset$}    
    {   
        $X \leftarrow \underset{X' \in Q}{\arg\min} \, E_R(X')$ \tcc*{Select the best solution from $Q$}
        
        $Q \leftarrow Q \setminus \{X\}$ \tcc*{Remove the solution from $Q$}

        \uIf{$E_R(X) < E_R(X^*)$}
        {
            $unimproved \leftarrow 0$, \enskip $X^* \leftarrow X$ \tcc*{Update the current best solution}
        }
        \Else
        {
            $unimproved \leftarrow unimproved + 1$ \tcc*{Count the unimproved step}
        }

        \If{$E_R(X^*) \leq \epsilon_1$}
        {
            \Break
        }

        $N_S \leftarrow ConstructSwapNeighborhood(X)$

        \For{each $X_S \in N_S$} 
        {
            $(h_1, h_2) \leftarrow LayoutGraphTransHashing(X_S, R)$

            \If{$(h_1, h_2) \notin T$}
            {
                $Q \leftarrow Q \cup \{X_S\}$, \enskip $T \leftarrow T \cup \{(h_1, h_2)\}$ \tcc*{Add the solution to $Q$ and record its hash value}
            }
        }
    }
}
\end{algorithm}

The Solution-Hashing Search (SHS) heuristic is the intensification module of the proposed algorithm.
SHS uses the \textit{swap} and \textit{insert} operations (discussed in Sections~\ref{sec:swap} and \ref{sec:insert}) for a given configuration and a fixed container radius to seek a better configuration.
SHS adopts a hybrid local search strategy, consisting of the greedy search method based on the \textit{swap} and \textit{insert} neighborhoods and an iterative local search method based on the layout-graph transformation and hashing methods (discussed in Section~\ref{sec:layhash}), of which the iterative local search is called the SHS core.

The SHS core uses the \textit{swap} operation for a candidate solution to explore new solutions, and the explored solutions are recorded in a hash table by applying the layout-graph transformation and hashing methods. Through these methods, the SHS core can avoid duplicate exploration, i.e., avoid exploring isomorphic or similar solutions found before, and discover a better solution efficiently. The SHS core (denoted by \textit{SHSCore}) is depicted in Algorithm~\ref{alg:SHSCore}.

Given an input solution $X_0$ and a recorded hash table $T_0$. 
\textit{SHSCore} first obtains the hash value of $X_0$ and confirms whether the hash value occurs in the hash table $T_0$. If it exists, indicating the solution $X_0$ is found before, \textit{SHSCore} terminates immediately and returns solution $X_0$ and hash table $T_0$ as the result. Otherwise, \textit{SHSCore} adds $X_0$ into a candidate set $Q$ and records its hash value in $T_0$ as an updated hash table $T$ (lines 3-6). $X_0$ is recorded as the current best solution $X^*$. Subsequently, \textit{SHSCore} executes an iterative local search procedure to discover a better solution.

At each iteration, \textit{SHSCore} selects the best solution $X$ from the candidate set $Q$ (lines 8-9) and updates the current best solution $X^*$ (lines 10-14). Then, \textit{SHSCore} constructs the \textit{swap} neighborhood $N_S$ of $X$ and adds each solution $X_S$ of $N_S$ to the candidate set $Q$ if $X_S$ is not explored before, i.e., the hash value of $X_S$ does not occur in hash table $T$ (lines 18-24). Meanwhile, the hash table $T$ is updated for the next iteration. 
The iteration process terminates when a feasible solution is found ($E_R(X^*) \leq \epsilon_1$), the candidate set $Q$ is empty, or it does not find improvement for $n$ consecutive iteration steps. The best-found solution $X^*$ and the updated hash table $T$ are returned as the result. 

\begin{algorithm}[tb]
\DontPrintSemicolon
\caption{Solution-Hashing Search} \label{alg:SHS}

\KwInput{Input solution $X_0$, fixed container radius $R$, recorded hash table $T_0$}
\KwOutput{The best solution $X^*$ found thus far, updated hash table $T$}

\SetKwFunction{Flayout}{\textit{SHS}}

\SetKwProg{Fn}{Function}{:}{\KwRet {$(X^*, T)$}}
\Fn{\Flayout($X_0$, $R$, $T_0$)}
{
    $X^* \leftarrow X_0$, \enskip $T \leftarrow T_0$ \;
    \While{$E_R(X^*) > \epsilon_1$}
    {
        $N_S \leftarrow ConstructSwapNeighborhood(X^*)$ \;
        $X \leftarrow \underset{X' \in N_S}{\arg\min} \, E_R(X')$ \tcc*{Select the best solution from $N_S$}

        \If{$E_R(X) < E_R(X^*)$} 
        {
            $X^* \leftarrow X$ \tcc*{Update the current best solution}
            \Continue
        }

        $N_I \leftarrow ConstructInsertNeighborhood(X^*)$ \;
        $X \leftarrow \underset{X' \in N_I}{\arg\min} \, E_R(X')$ \tcc*{Select the best solution from $N_I$}

        \If{$E_R(X) < E_R(X^*)$} 
        {
            $X^* \leftarrow X$ \tcc*{Update the current best solution }
            \Continue
        }

        $(X, T) \leftarrow SHSCore(X^*, R, T)$ \tcc*{Call the \textit{SHSCore} module}

        \If{$E_R(X) < E_R(X^*)$} 
        {
            $X^* \leftarrow X$ \tcc*{Update the current best solution}
            \Continue
        }

        \Break \tcc*{No improvement found in \textit{SHSCore}, break the loop}
    }
}
\end{algorithm}

The SHS algorithm is the \textit{SHSCore} module coupled with the greedy search method, depicted in Algorithm~\ref{alg:SHS}. Given an input solution $X_0$, $X_0$  is first recorded as the current best solution $X^*$. Then, SHS improves the current best solution $X^*$ in the order of successively using the greedy search method based on the \textit{swap} neighborhood, the greedy search method based on the \textit{insert} neighborhood, and the \textit{SHSCore} method. The calling order is sorted according to the efficiency of these methods. In experiments, we observe that the \textit{swap} operation is much more efficient than the \textit{insert} operation, and the greedy search method is more efficient than the iterative local search method (i.e., \textit{SHSCore}).

Specifically, SHS selects the best solution $X$ from the \textit{swap} neighborhood of $X^*$ and improves $X^*$ repeatedly until no improvement is found (lines 4-9). Then, SHS turns to use the \textit{insert} neighborhood to improve $X^*$. As soon as an improvement is found by exploiting the \textit{insert} neighborhood, SHS switches back to using the more efficient \textit{swap} neighborhood (lines 10-15). Otherwise, it indicates that $X^*$ cannot obtain an improvement by exploiting \textit{swap} and \textit{insert} neighborhoods, SHS calls the \textit{SHSCore} module to further seek an improvement, and it switches back to exploit the \textit{swap} neighborhood when a better solution is found (lines 16-20). Finally, SHS returns the best solution $X^*$ as the result when a feasible solution is found or no improvement has occurred in the \textit{SHSCore} module.

\subsection{Diversification} \label{sec:I-SHS}

Unlike the intensification operation that aims at seeking improvement, diversification operations focus on partially changing the current solution, thereby guiding the local search method to explore nearby promising areas. 
Various perturbation operations and strategies have been proposed to diversify the exploration for solving PUCC and its variants. These include the quasi-physical method~\citep{huang1999two}, relocating method~\citep{zeng2016iterated}, shifting method~\citep{muller2009packing,ye2013iterated}, global perturbation method~\citep{huang2013tabu}, genetic algorithm~\citep{zeng2018adaptive}, etc. 

\begin{algorithm}[tb]
\DontPrintSemicolon
\caption{Iterative Solution-Hashing Search} \label{alg:I-SHS}

\KwInput{Best radius $R_0^*$ found before, radius shrinking ratio $\alpha$, recorded hash table $T_0$}
\KwOutput{The best feasible solution $(X^*, R^*)$ found thus far, updated hash table $T$}

\SetKwFunction{Flayout}{\textit{I-SHS}}

\SetKwProg{Fn}{Function}{:}{\KwRet {$(X^*, R^*, T)$}}
\Fn{\Flayout($R_0, \alpha, T_0$)}
{
    $unimproved \leftarrow 0$, \enskip $T \leftarrow T_0$ \;
    $X \leftarrow RandomLayout(R_0)$ \tcc*{Generate a random solution}
    $X \leftarrow LayoutOptimization(E_{R_0}, X)$ \tcc*{Optimize the solution}
    $(X^*, R^*) \leftarrow ContainerOptimization(X, R_0)$ \tcc*{Obtain an initial feasible solution}

    $R' \leftarrow (1 - \alpha) \min(R_0, R^*)$ \tcc*{Obtain a compact radius}

    $X \leftarrow LayoutOptimization(E_{R'}, X)$ \tcc*{Obtain a compact solution}

    $X' \leftarrow X$ \tcc*{Recorded the best compact solution}
    
    \While{$unimproved < Maxiter$}
    {
        $(X, T) \leftarrow SHS(X, R', T)$ \tcc*{Call \textit{SHS} to improve the solution}

        \uIf{$E_{R'}(X) < E_{R'}(X')$}
        {
            $unimproved \leftarrow 0$, \enskip $X' \leftarrow X$ \;
            $(X_t^*, R_t^*) \leftarrow ContainerOptimization(X, R')$ \;

            \If{$R_t^* < R^*$}
            {
                $(X^*, R^*) \leftarrow (X_t^*, R_t^*)$ \tcc*{Update the best feasible solution}
                $R' \leftarrow (1 - \alpha) \min(R_0, R^*)$ \tcc*{Update the compact radius}
                $X \leftarrow LayoutOptimization(E_{R'}, X)$ \;
                $X' \leftarrow X$ \tcc*{Update the best compact solution}
            }
        }
        \Else
        {
            $unimproved \leftarrow unimproved + 1$
        }

        $X \leftarrow PerturbLayout(X')$ \tcc*{Perturb the solution to obtain offspring}
        $X \leftarrow LayoutOptimization(E_{R'}, X)$ \tcc*{Optimize the offspring}
    }
}
\end{algorithm}

We use the simple and efficient relocating method as the perturbation operation of the proposed algorithm. This method randomly selects $m$ circles from a configuration and randomly places them into the circular container with possible overlaps, where $m$ is also called the perturbation strength. Following the landmark studies~\citep{ye2013iterated,zeng2016iterated} that gain the state-of-the-art results, we set the perturbation strength $m$ according to the strategy that randomly chooses a value from interval $[1, \lceil n / 6 \rceil]$.

The proposed Iterative Solution-Hashing Search (I-SHS) algorithm aims to find a feasible configuration with the container radius as minimum as possible, which iteratively calls the SHS algorithm (presented in Algorithm~\ref{alg:SHS}) for seeking improvement and applies the relocating method for diversifying the exploration. The I-SHS procedure is described in Algorithm~\ref{alg:I-SHS}. 

I-SHS first generates a solution that randomly packs the $n$ circles into the container with any possible overlaps for a given radius, and then the \textit{LayoutOptimization} and \textit{ContainerOptimization} modules (presented in Algorithms~\ref{alg:layopti} and \ref{alg:contopti}) are adopted to obtain an initial feasible solution $(X^*, R^*)$ quickly (lines 3-5).
To seek a better feasible solution and construct the overlapping relationship for the layout-graph transformation method, I-SHS slightly shrinks the current best-found radius as a shrunk radius $R'$ and obtains a compact solution $X$, which is recorded as the best compact solution $X'$ (lines 6-8). Subsequently, I-SHS performs an iterative search procedure to seek a better solution (lines 9-25). 
At each iteration, I-SHS calls the SHS algorithm to improve the compact solution $X$ (line 10) and applies the perturbation method to obtain offspring for the next iteration (lines 23-24). If a better compact solution $X$ is found ($E_{R'}(X) < E_{R'}(X')$), indicating that $X$ may lead to a better feasible solution, I-SHS immediately updates the best compact solution $X'$ and uses the \textit{ContainerOptimization} module for $X$ to seek a potential feasible improvement (lines 12-19). I-SHS terminates when the unimproved iteration step reaches the maximum unimproved limit (\textit{Maxiter}), and returns the best-found feasible solution $(X^*, R^*)$ as the result.

\subsection{Main Framework} \label{sec:mainframe}

\begin{algorithm}[tb]
\DontPrintSemicolon
\caption{Main Framework of the I-SHS Algorithm} \label{alg:MF}

\KwInput{Input $n$ circles with their radii}
\KwOutput{The best found feasible configuration $(X^*, R^*)$}

\tcc{Initialization stage}

$\alpha \leftarrow 10^{-2}$, \enskip $\beta \leftarrow 0.2$, \enskip $T \leftarrow \emptyset$ \;

$R \leftarrow \sqrt{ \frac{ \sum_{i=1}^{n} r_i^2  }{ 0.9 } }$ \tcc*{Estimate a dense radius}
$X \leftarrow RandomLayout(R)$ \tcc*{Generate a random initial solution}
$X \leftarrow LayoutOptimization(E_R, X)$ \;
$(X, T) \leftarrow SHS(X, R, T)$ \;
$(X^*, R^*) \leftarrow ContainerOptimization(X, R)$ \;

\While{$\mathbf{true}$}
{
    $R \leftarrow (1 - \alpha) \times R^*$ \tcc*{Shrink the best radius}
    $X \leftarrow RandomLayout(R)$ \tcc*{Generate a random solution}
    $X \leftarrow LayoutOptimization(E_R, X)$ \;
    $(X, T) \leftarrow SHS(X, R, T)$ \;
    $(X^*_t, R^*_t) \leftarrow ContainerOptimization(X, R)$ \;
    \uIf{$R^*_t < R^*$}
    {
        $(X^*, R^*) \leftarrow (X^*_t, R^*_t)$
    }
    \Else
    {
        \Break
    }
}

\tcc{Restarting stage}

\While{$time() < t_{max}$}
{
    $(X^*_t, R^*_t, T) \leftarrow I\text{-}SHS(R^*, \alpha, T) $ \;
    \If{$R^*_t < R^*$}
    {
        $(X^*, R^*) \leftarrow (X^*_t, R^*_t)$
    }
    $\alpha \leftarrow \max(10^{-4}, \alpha \times \beta)$ \tcc*{Decay the shrinking ratio}
}

\Return $(X^*, R^*)$

\end{algorithm}

The main framework of the proposed I-SHS algorithm consists of the initialization stage and the restarting stage, described in Algorithm~\ref{alg:MF}.

The initialization stage aims to produce a dense and feasible configuration rapidly as a good start for the following search process (i.e., I-SHS). Since we adopt the penalty model (discussed in Section~\ref{sec:model}), where the algorithm aims to seek a feasible configuration in a container with a fixed radius. Starting from a proper and tight container radius can drive the search method to find dense and high-quality feasible configurations efficiently. 

To achieve this purpose, the algorithm sets a tight radius estimated by an upper bound packing density $\rho$, and the packing density is formulated as $\rho = \frac{\sum_{i=1}^{n} \pi r_i^2}{\pi R^2}$ where $r_i$ is the radius of circle $c_i$ and $R$ is the radius of the circular container.
Hence, a radius can be estimated by a packing density, formulated as $R = \sqrt{\frac{\sum_{i=1}^{n} r_i^2}{\rho}}$. 
According to the density of best-known records of various PUCC instances at the Packomania website~\citep{Spechtweb}, we use $\rho_0 = 0.9$ as an initial upper bound density to estimate the initial radius (line 2). 

The algorithm performs a rapid search procedure to obtain a feasible solution starting from the initial radius. In detail, the algorithm generates a random solution $X$ and then optimizes it by the \textit{LayoutOptimization} module (presented in Algorithm~\ref{alg:layopti}). The SHS algorithm (presented in Algorithm~\ref{alg:SHS}) is adopted to further improve the solution $X$, and a dense and feasible solution $(X^*, R^*)$ is obtained by calling the \textit{ContainerOptimization} module (presented in Algorithm~\ref{alg:contopti}) (lines 2-6).
Then, the algorithm shrinks radius $R^*$ as a new initial radius and iteratively performs the rapid search procedure until no further improvement can be found (lines 7-18). The resulting feasible solution $(X^*, R^*)$ is a good start for the I-SHS algorithm (discussed in Section~\ref{sec:I-SHS}).

In the restarting stage, the algorithm uses the current best-found radius $R^*$ as the input radius for I-SHS and continuously calls the I-SHS algorithm to discover a denser feasible solution until the cut-off time ($t_{max}$) is reached (lines 19-25). The radius shrinking ratio $\alpha$ used in I-SHS is iteratively reduced from $10^{-2}$ to $10^{-4}$. This gradual reduction facilitates I-SHS to easily find significant improvements in the early search stage and discover subtle enhancements in the later search stage. The current best-found solution is updated when I-SHS returns a better solution. Finally, the algorithm outputs the best-found feasible solution as its result.

\section{Experiments and Evaluations} \label{sec:exp}

In this section, we present extensive computational experiments to evaluate our proposed algorithm for solving the PUCC problem. The evaluation is based on the well-known benchmark instances and comparisons with the best-known results at the Packomania website~\citep{Spechtweb} and the landmark studies in the literature. Furthermore, we conduct experiments to provide an insight analysis and exhibit the efficiency of our proposed methods.

\subsection{Experimental Settings and Benchmarks}

Our algorithm was implemented in the C++ programming language and compiled using g++ 9.4.0. Experiments were performed on a computer with AMD EPYC\texttrademark~7H12 CPU and 256 GBytes RAM, running under a Linux OS. 
Because of the randomness, the algorithm was independently performed multiple times with different random seeds (CPU timestamps) for each tested instance to evaluate the overall performance of the algorithm. 

Experiments were based on various PUCC benchmarks to evaluate the applicability, generality, and overall performance of the algorithm on instances with various radius distributions. These benchmarks are described as follows.

\begin{itemize}
    \item $r_i = i$ benchmark, proposed by AZPC, is the famous PUCC benchmark. 
    \item NR benchmark, proposed by \citet{huang2006new}, has 24 irregular instances with the number of circles in the range of $10 \leq n \leq 60$, which is a popular handmade PUCC benchmark. The detailed information of these instances is referred to~\citet{huang2006new}.
    \item $r_i = i^{-1/2}$ benchmark, proposed by \citet{castillo2008solving}, is a popular PUCC benchmark.
    \item $r_i = i^{1/2}$ benchmark, recorded at the Packomania website~\citep{Spechtweb}. 
    \item $r_i = i^{-2/3}$ benchmark, recorded at the Packomania website~\citep{Spechtweb}. 
    \item $r_i = i^{-1/5}$ benchmark, recorded at the Packomania website~\citep{Spechtweb}. 
\end{itemize}

Notably, for the contest instances ($r_i = i$), the contestants ran their algorithm without any limitation on the computational resources to obtain their best results, and many studies devoted a large number of resources to solving these instances. For instance, \citet{addis2008efficiently} and \citet{muller2009packing} as the contestants obtained their best results without any limitation on the computational resource, \citet{ye2013iterated} set the cut-off time to 24 hours for each tested instance and did not limit the number of runs to obtain their best results, \citet{zeng2016iterated} obtained their best results without any limitation on the cut-off time and the number of runs. As a result, the best-known results of these benchmark instances are extremely difficult to reach and improve.
Because of the difficulty, we evaluated our algorithm on these benchmark instances with 20 runs for each tested instance and a cut-off time of 10,000 seconds for each run. 

For the NR and $r_i = i^{-1/2}$ benchmarks, we performed our algorithm 10 runs under a cut-off time of 10,000 seconds for each tested instance, which is consistent with the state-of-the-art method~\citep{zeng2016iterated} for the comparable reason.
To ensure consistency, the settings for the number of runs and the cut-off time for the rest of the benchmarks ($r_i = i^{1/2}$, $r_i = i^{-2/3}$, $r_i = i^{-1/5}$) are the same as the NR and $r_i = i^{-1/2}$ benchmarks, i.e., 10 runs for each tested instance and 10,000 seconds for each run.

\begin{table}[tb]
\centering
\caption{
Settings of parameters.
}
\label{tb:para}

\setlength{\tabcolsep}{2mm}

\begin{tabular}{llll}
\toprule
Parameter               & Section & Description                                           & Value                       \\ \midrule
$\epsilon_0$ & \ref{sec:lay_opti}   & Precision control parameter of layout optimization    & $10^{-10}$ \\
$\epsilon_1$ & \ref{sec:cont_opti}   & Precision control parameter of container optimization & $10^{-25}$ \\
$\epsilon_2$ & \ref{sec:layhash} & Precision control parameter of overlapping            & $10^{-8}$ \\
$(p_1, q_1, M_1)$  & \ref{sec:layhash}   & Parameters used in hash function                      & $(17, 193, 998244353)$      \\
$(p_2, q_2, M_2)$  & \ref{sec:layhash}   & Parameters used in hash function                      & $(97, 257, 1004535809) $     \\
$m$                       & \ref{sec:I-SHS}     & Perturbation strength of   diversification            & $[1, \lceil n / 6 \rceil]$                \\
\textit{Maxiter}                 & \ref{sec:I-SHS}     & Maximum unimproved limit of I-SHS                     & $200$                         \\
$\rho_0$                     & \ref{sec:mainframe}     & Packing density of initial estimation                 & $0.9$                         \\
$\alpha_0$                   & \ref{sec:mainframe}     & Initial container radius shrinking ratio         & $10^{-2}$        \\
$\beta$                    & \ref{sec:mainframe}     & Decay coefficient of container radius shrinking ratio       & $0.2$                         \\ \bottomrule
\end{tabular}

\end{table}

The default settings of other parameters used in the algorithm are given in Table~\ref{tb:para}. These default settings were empirically determined via preliminary experiments. The detailed computational results and comparison are provided in Appendix~\ref{append:detail}.

\subsection{Computational Results and Comparisons}

Tables~\ref{tb:ins1}-\ref{tb:sum} summarize the computational results and comparisons of our algorithm (I-SHS) with the best-known results at the Packomania website~\citep{Spechtweb} and the landmark studies in the literature.

\begin{table}[tb]
\centering
\caption{
Computational results and comparison on the $r_i = i$ benchmark instances in the range of $5 \leq n \leq 35$.
}
\label{tb:ins1}

\scalebox{0.9}{
\begin{tabular}{llllllrr}
\toprule
   &                      &              &              &          & \multicolumn{3}{l}{I-SHS (this study)}               \\ \cline{6-8} 
$n$  & $R^*$ & FSS          & ITS-PUCC     & ITS-VND  & $R_{best}$               & SR    & $time(s)$ \\ \midrule
5  & 9.00139774           & 9.00139775   & 9.00139774   & matched  & \textbf{9.00139774}   & 20/20 & 0.00    \\
6  & 11.05704039          & 11.05704040  & 11.05704039  & matched  & \textbf{11.05704039}  & 20/20 & 0.01    \\
7  & 13.46211067          & 13.46211068  & 13.46211067  & matched  & \textbf{13.46211067}  & 20/20 & 0.01    \\
8  & 16.22174667          & 16.22174668  & 16.22174667  & matched  & \textbf{16.22174667}  & 20/20 & 0.02    \\
9  & 19.23319390          & 19.23319391  & 19.23319390  & matched  & \textbf{19.23319390}  & 20/20 & 0.04    \\
10 & 22.00019301          & 22.00019301  & 22.00019301  & matched  & \textbf{22.00019301}  & 20/20 & 0.02    \\
11 & 24.96063428          & 24.96063429  & 24.96063428  & matched  & \textbf{24.96063428}  & 20/20 & 0.10    \\
12 & 28.37138943          & 28.37138944  & 28.37138943  & matched  & \textbf{28.37138943}  & 20/20 & 0.19    \\
13 & 31.54586701          & 31.54586702  & 31.54586701  & matched  & \textbf{31.54586701}  & 20/20 & 0.23    \\
14 & 35.09564714          & 35.09564714  & 35.09564714  & matched  & \textbf{35.09564714}  & 20/20 & 0.76    \\
15 & 38.83799550          & 38.83799682  & 38.83799550  & matched  & \textbf{38.83799550}  & 20/20 & 1.93    \\
16 & 42.45811643          & 42.45811644  & 42.45811643  & matched  & \textbf{42.45811643}  & 20/20 & 13.43   \\
17 & 46.29134211          & 46.34518193  & 46.29134211  & matched  & \textbf{46.29134211}  & 20/20 & 12.27   \\
18 & 50.11976262          & 50.20889346  & 50.11976262  & matched  & \textbf{50.11976262}  & 20/20 & 5.37    \\
19 & 54.24029359          & 54.36009421  & 54.24029359  & matched  & \textbf{54.24029359}  & 20/20 & 30.70   \\
20 & 58.40056747          & 58.48047359  & 58.40056747  & matched  & \textbf{58.40056747}  & 20/20 & 2081.80 \\
21 & 62.55887709          & 63.00078332  & 62.55887709  & matched  & \textbf{62.55887709}  & 20/20 & 56.72   \\
22 & 66.76028624          & 66.96471591  & 66.76028624  & matched  & \textbf{66.76028624}  & 20/20 & 127.52  \\
23 & 71.19946160          & 71.69822657  & 71.19946160  & matched  & \textbf{71.19946160}  & 20/20 & 1900.06 \\
24 & 75.74914258          & 76.12311970  & 75.74914258  & matched  & \textbf{75.74914258}  & 20/20 & 643.55  \\
25 & 80.28586443          & 80.81682360  & 80.28586443  & matched  & \textbf{80.28586443}  & 20/20 & 1596.54 \\
26 & 84.97819106          & 85.48743800  & 84.97819106  & 84.9782  & \textbf{84.97819106}  & 20/20 & 1417.22 \\
27 & 89.75096268          & 90.93173506  & 89.75096268  & 89.7510  & \textbf{89.75096268}  & 19/20 & 3903.17 \\
28 & 94.52587710          & 95.64064140  & 94.52587710  & 94.5259  & \textbf{94.52587710}  & 20/20 & 796.63  \\
29 & 99.48311156          & 100.72003130 & 99.48311156  & 99.4832  & \textbf{99.48311156}  & 16/20 & 4791.38 \\
30 & 104.54036376         & 105.88817223 & 104.54036376 & 104.5404 & \textbf{104.54036376} & 15/20 & 4148.65 \\
31 & 109.62924066         & 111.07712597 & 109.62924066 & 109.6292 & \textbf{109.62924066} & 6/20  & 4273.86 \\
32 & 114.79981466         & 116.61226677 & 114.79981466 & 114.7999 & \textbf{114.79981466} & 2/20  & 5014.04 \\
33 & 120.06565963         & N/A          & 120.06565963 & 120.0657 & \textbf{120.06565963} & 4/20  & 5293.43 \\
34 & 125.36693920         & N/A          & 125.44355791 & 125.3670 & \textbf{125.36693920} & 1/20  & 4257.74 \\
35 & 130.84907875         & N/A          & 130.84907874 & 130.8491 & \textbf{130.84907875} & 2/20  & 5186.00 \\ \bottomrule
\end{tabular}
}

\end{table}

Table~\ref{tb:ins1} presents the computational results and comparisons for the $r_i = i$ benchmark instances within the range of $5 \leq n \leq 35$. 
The first column denotes the instance sizes ($n$). 
The second column displays the best-known results ($R^*$) sourced from the Packomania website. 
Columns 3-5 showcase the best outcomes from three state-of-the-art methods: FSS~\citep{lopez2013packing}, ITS-PUCC~\citep{ye2013iterated}, and ITS-VND~\citep{zeng2016iterated}. Note that \citet{lopez2013packing} (FSS) provide results for $5 \leq n \leq 32$, while \citet{zeng2016iterated} (ITS-VND) present results starting from $n = 26$ and claim to have matched the best-known results for smaller instances. 
Columns 6-8 show the results of our algorithm, including the best result ($R_{best}$) over 20 independent runs, the success rate (SR) of achieving the best result, and the average running time ($time(s)$) in seconds for each run to obtain the final outcome. For $R_{best}$, 
results that match the best-known ones are highlighted in bold.

Table~\ref{tb:ins1} shows that our algorithm matches all the best-known results across the 31 tested instances. In comparison to previous methods that require massive computational resources, our algorithm achieves these best results within a reasonable running time (about 55.5 hours in total for 20 runs) and with constrained computational resources. Additionally, the success rate of instances with $n \leq 28$ except for $n = 27$ is 100\% (19/20 for $n = 27$). This underscores the robust solving capability of our algorithm for PUCC.

\begin{table}[tb]
\centering
\caption{
Computational results and comparison on the 24 NR benchmark instances.
}
\label{tb:NR}
\scalebox{0.9}{
\begin{tabular}{llllllrrr}
\toprule
           &                      &          &          &          & \multicolumn{4}{l}{I-SHS (this study)}                                             \\ \cline{6-9} 
Instance   & $R^*$ & GP-TS    & ITS-PUCC & ITS-VND  & $R_{best}$               & \multicolumn{1}{l}{$R_{best} - R^*$} & SR    & $time(s)$ \\ \midrule
NR10\_1    & 99.88507689          & 99.8851  & 99.8851  & 99.8851  & 99.88507689           & 0                           & 10/10 & 0.04    \\
NR11\_1    & 60.70996138          & 60.7100  & 60.7100  & 60.7100  & 60.70996138           & 0                           & 10/10 & 0.67    \\
NR12\_1    & 65.02442246          & 65.0245  & 65.0244  & 65.0245  & 65.02442246           & 0                           & 10/10 & 0.70    \\
NR14\_1    & 113.55876291         & 113.5588 & 113.5588 & 113.5588 & 113.55876291          & 0                           & 10/10 & 11.20   \\
NR15\_1    & 38.91138666          & 38.9158  & 38.9114  & 38.9114  & 38.91138666           & 0                           & 10/10 & 2.29    \\
NR15\_2    & 38.83799550          & 38.8380  & 38.8380  & 38.8380  & 38.83799550           & 0                           & 10/10 & 2.63    \\
NR16\_1    & 143.37978108         & 143.3798 & 143.3798 & 143.3798 & 143.37978108          & 0                           & 10/10 & 71.72   \\
NR16\_2    & 127.69782537         & 127.7174 & 127.6978 & 127.6978 & 127.69782537          & 0                           & 10/10 & 31.71   \\
NR17\_1    & 49.18730653          & 49.1874  & 49.1873  & 49.1873  & 49.18730653           & 0                           & 10/10 & 45.51   \\
NR18\_1    & 196.98262400         & 197.0367 & 196.9826 & 196.9826 & 196.98262400          & 0                           & 10/10 & 113.15  \\
NR20\_1    & 125.11775418         & 125.1178 & 125.1178 & 125.1178 & 125.11775418          & 0                           & 10/10 & 16.25   \\
NR20\_2    & 121.78871660         & 121.9944 & 121.7887 & 121.7887 & 121.78871660          & 0                           & 10/10 & 223.98  \\
NR21\_1    & 148.09678792         & 148.3373 & 148.0968 & 148.0968 & 148.09678792          & 0                           & 10/10 & 293.77  \\
NR23\_1    & 174.34254220         & 174.8524 & 174.3425 & 174.3426 & 174.34254220          & 0                           & 5/10  & 4528.68 \\
NR24\_1    & 137.75905206         & 138.0044 & 137.7591 & 137.7591 & 137.75905206          & 0                           & 10/10 & 2023.08 \\
NR25\_1    & 188.71878994         & 189.3736 & 188.8314 & 188.7188 & 188.71878994          & 0                           & 6/10  & 6458.62 \\
NR26\_1    & 244.57428028         & 246.0853 & 244.5743 & 244.5743 & 244.57428028          & 0                           & 10/10 & 1360.38 \\
NR26\_2    & 300.26307937         & 302.0687 & 300.2631 & 300.2631 & 300.26307937          & 0                           & 10/10 & 1728.43 \\
NR27\_1    & 220.65960596         & 221.4882 & 220.9393 & 220.8172 & 220.65960596          & 0                           & 1/10  & 5215.47 \\
NR30\_1    & 177.25866811         & 178.0093 & 177.5125 & 177.3852 & \textbf{177.17846105} & -8.02E-02                   & 1/10  & 4454.86 \\
NR30\_2    & 172.65018482         & 173.1641 & 172.9665 & 172.6502 & \textbf{172.51354788} & -1.37E-01                   & 1/10  & 5281.77 \\
NR40\_1    & 352.40262684         & 355.1307 & 352.4517 & 352.4027 & \textbf{352.23082161} & -1.72E-01                   & 1/10  & 5795.36 \\
NR50\_1    & 376.80638007         & 377.9105 & 377.9080 & 377.0317 & \textbf{375.88541573} & -9.21E-01                   & 1/10  & 4726.23 \\
NR60\_1    & 514.83631921         & 519.4515 & 518.6792 & 517.1091 & \textbf{514.09716814} & -7.39E-01                   & 1/10  & 7815.28 \\ \midrule
\# Improved &                      &          &          &          & 5                     & \multicolumn{1}{l}{}        &       &         \\
\# Matched  &                      &          &          &          & 19                    & \multicolumn{1}{l}{}        &       &         \\
\# Worse    &                      &          &          &          & 0                     & \multicolumn{1}{l}{}        &       &         \\ \bottomrule
\end{tabular}
}
\end{table}

Table~\ref{tb:NR} presents the computational results and comparisons for the 24 NR benchmark instances. 
The first column denotes the instance name, while the second column displays the best-known results ($R^*$) sourced from the Packomania website.
Columns 3-5 present the best outcomes from three state-of-the-art methods: GP-TS~\citep{huang2013tabu}, ITS-PUCC~\citep{ye2013iterated}, and ITS-VND~\citep{zeng2016iterated}. 
Columns 6-9 exhibit the results of our algorithm, including the best result ($R_{best}$) over 10 independent runs, the difference ($R_{best} - R^*$) between the best result and the best-known result presented in scientific notation, the success rate (SR) of achieving the best result, and the average running time ($time(s)$) in seconds for each run to obtain the final outcome. 
Improved results are highlighted in bold compared to the best-known result in terms of $R_{best}$, and a negative value of the difference indicates an improvement.
Additionally, the last three rows of the table summarize the number of instances where our algorithm obtained improved, matched, or worse results compared to the best-known outcome.

\begin{table}[tb]
\centering
\caption{
Computational results and comparison on the $r_i = i^{-1/2}$ benchmark instances in the range of $5 \leq n \leq 35$.
}
\label{tb:ins3}

\scalebox{0.9}{
\begin{tabular}{llllllrrr}
\toprule
           &                      &            &        &         & \multicolumn{4}{l}{I-SHS (this study)}                                           \\ \cline{6-9} 
$n$          & $R^*$ & FSS        & GP-TS  & ITS-VND & $R_{best}$             & \multicolumn{1}{l}{$R_{best} - R^*$} & SR    & $time(s)$ \\ \midrule
5          & 1.75155245           & 1.75155246 & 1.7516 & 1.7516  & 1.75155245          & 0                           & 10/10 & 0.01    \\
6          & 1.81007693           & 1.81007694 & 1.8101 & 1.8101  & 1.81007693          & 0                           & 10/10 & 0.01    \\
7          & 1.83872406           & 1.83872407 & 1.8388 & 1.8388  & 1.83872406          & 0                           & 10/10 & 0.02    \\
8          & 1.85840095           & 1.85840095 & 1.8585 & 1.8585  & 1.85840095          & 0                           & 10/10 & 0.01    \\
9          & 1.87881275           & 1.87881276 & 1.8789 & 1.8789  & 1.87881275          & 0                           & 10/10 & 0.06    \\
10         & 1.91343551           & 1.91343552 & 1.9135 & 1.9135  & 1.91343551          & 0                           & 10/10 & 0.57    \\
11         & 1.92918775           & N/A          & N/A      & N/A       & 1.92918775          & 0                           & 10/10 & 0.57    \\
12         & 1.94982343           & 1.95197444 & 1.9499 & 1.9499  & 1.94982343          & 0                           & 10/10 & 1.38    \\
13         & 1.96523681           & N/A          & N/A      & N/A       & 1.96523681          & 0                           & 10/10 & 1.92    \\
14         & 1.98024874           & 1.98606765 & 1.9863 & 1.9803  & 1.98024874          & 0                           & 10/10 & 15.55   \\
15         & 1.99270927           & N/A          & N/A      & N/A       & 1.99270927          & 0                           & 10/10 & 35.51   \\
16         & 2.00458577           & 2.02213023 & 2.0084 & 2.0046  & 2.00458577          & 0                           & 10/10 & 27.59   \\
17         & 2.01525778           & N/A          & N/A      & N/A       & 2.01525778          & 0                           & 10/10 & 30.57   \\
18         & 2.02814858           & 2.05613259 & 2.0397 & 2.0281  & 2.02814858          & 0                           & 10/10 & 63.38   \\
19         & 2.04199731           & N/A          & N/A      & N/A       & 2.04199731          & 0                           & 10/10 & 541.85  \\
20         & 2.05144226           & 2.08237082 & 2.0716 & 2.0515  & \textbf{2.05144134} & -9.20E-07                   & 10/10 & 724.41  \\
21         & 2.06233110           & N/A          & N/A      & N/A       & 2.06233110          & 0                           & 10/10 & 1568.79 \\
22         & 2.06796317           & N/A          & N/A      & N/A       & 2.06796317          & 0                           & 10/10 & 622.06  \\
23         & 2.07977236           & N/A          & N/A      & N/A       & 2.07977236          & 0                           & 8/10  & 4152.34 \\
24         & 2.09038316           & N/A          & N/A      & N/A       & 2.09038316          & 0                           & 7/10  & 6347.90 \\
25         & 2.09752783           & 2.13929005 & 2.1236 & 2.0976  & 2.09752783          & 0                           & 5/10  & 4346.33 \\
26         & 2.10763017           & N/A          & N/A      & N/A       & \textbf{2.10702271} & -6.07E-04                   & 5/10  & 4847.16 \\
27         & 2.11589049           & N/A          & N/A      & N/A       & \textbf{2.11317010} & -2.72E-03                   & 2/10  & 4929.92 \\
28         & 2.12284917           & N/A          & N/A      & N/A       & \textbf{2.12147566} & -1.37E-03                   & 1/10  & 4369.85 \\
29         & 2.13197913           & N/A          & N/A      & N/A       & \textbf{2.12807595} & -3.90E-03                   & 1/10  & 6545.35 \\
30         & 2.13748392           & 2.18190380 & 2.1679 & 2.1400  & \textbf{2.13624146} & -1.24E-03                   & 1/10  & 6033.12 \\
31         & 2.14898330           & N/A          & N/A      & N/A       & \textbf{2.14356970} & -5.41E-03                   & 1/10  & 5688.82 \\
32         & 2.15535415           & N/A          & N/A      & N/A       & \textbf{2.15130900} & -4.05E-03                   & 1/10  & 5824.71 \\
33         & 2.16580368           & N/A          & N/A      & N/A       & \textbf{2.15890144} & -6.90E-03                   & 1/10  & 4993.72 \\
34         & 2.17436411           & N/A          & N/A      & N/A       & \textbf{2.16293568} & -1.14E-02                   & 1/10  & 6388.91 \\
35         & 2.17116210           & 2.21958858 & 2.2037 & 2.1719  & \textbf{2.16988756} & -1.27E-03                   & 1/10  & 5243.74 \\ \midrule
\# Improved &                      &            &        &         & 11                  & \multicolumn{1}{l}{}        &       &         \\
\# Matched  &                      &            &        &         & 20                  & \multicolumn{1}{l}{}        &       &         \\
\# Worse    &                      &            &        &         & 0                   & \multicolumn{1}{l}{}        &       &         \\ \bottomrule
\end{tabular}
}
\end{table}

Table~\ref{tb:ins3} presents the computational results and comparisons for the $r_i = i^{-1/2}$ benchmark instances in the range of $5 \leq n \leq 35$. 
The first column indicates the instance sizes ($n$), while the second column displays the best-known results ($R^*$) sourced from the Packomania website.
Columns 3-5 showcase the best outcomes from three state-of-the-art methods: FSS~\citep{lopez2013packing}, GP-TS~\citep{huang2013tabu}, and ITS-VND~\citep{zeng2016iterated}. 
Note that the evaluations of these three methods were conducted on only 14 instances of this benchmark, and the best results from all these instances are included in the table.
Columns 6-9 exhibit the outcomes of our algorithm, including the best result ($R_{best}$) over 10 independent runs, the difference ($R_{best} - R^*$) between the best result and the best-known result presented in scientific notation, the success rate (SR) of achieving the best result, and the average running time ($time(s)$) in seconds for each run to obtain the final outcome. 
Improved results are highlighted in bold compared to the best-known outcome in terms of $R_{best}$, and a negative value of the difference indicates an improvement.
Additionally, the last three rows of the table summarize the number of instances where our algorithm obtained improved, matched, or worse results compared with the best-known outcome.



Tables~\ref{tb:NR} and \ref{tb:ins3} show that our algorithm performs excellently in solving these benchmark instances. Our algorithm improves many of the best-known results on the larger instances of the two benchmarks and matches many of the best-known results on the smaller instances with a success rate of 100\%. Besides, our algorithm obtains many better results than the state-of-the-art methods, especially in larger instances. 
It demonstrates that our algorithm significantly outperforms the state-of-the-art methods and exhibits strong capability for solving PUCC.

\begin{table}[tb]
\centering
\caption{
Computational results and comparison on the instances of three benchmarks in the range of $5 \leq n \leq 35$.
}
\label{tb:ins245}
\scalebox{0.9}{
\begin{tabular}{lllllllll}
\toprule
           & \multicolumn{2}{l}{$r_i = i^{1/2}$ benchmark} &  & \multicolumn{2}{l}{$r_i = i^{-2/3}$ benchmark} &  & \multicolumn{2}{l}{$r_i = i^{-1/5}$ benchmark} \\ \cline{2-3} \cline{5-6} \cline{8-9} 
$n$          & $R^*$         & $R_{best}$                      &  & $R^*$          & $R_{best}$                      &  & $R^*$          & $R_{best}$                      \\ \midrule
5          & 4.52148027                   & 4.52148027                   &  & 1.62996052                    & 1.62996052                   &  & 2.24461584                    & 2.24461584                   \\
6          & 5.35096299                   & 5.35096299                   &  & 1.62996052                    & 1.62996052                   &  & 2.38798638                    & 2.38798638                   \\
7          & 6.04937848                   & 6.04937848                   &  & 1.62997277                    & 1.62997277                   &  & 2.42262334                    & 2.42262334                   \\
8          & 6.77426665                   & 6.77426665                   &  & 1.63148407                    & 1.63148407                   &  & 2.52382090                    & \textbf{2.52381953}          \\
9          & 7.55900237                   & 7.55900237                   &  & 1.63786399                    & 1.63786399                   &  & 2.63002657                    & 2.63002657                   \\
10         & 8.30346812                   & 8.30346812                   &  & 1.64695723                    & 1.64695723                   &  & 2.71578482                    & 2.71578482                   \\
11         & 9.07212587                   & 9.07212587                   &  & 1.65031382                    & 1.65031382                   &  & 2.76746630                    & 2.76746630                   \\
12         & 9.86532030                   & 9.86532030                   &  & 1.65676025                    & 1.65676025                   &  & 2.82901840                    & 2.82901840                   \\
13         & 10.58832628                  & 10.58832628                  &  & 1.66277803                    & 1.66277803                   &  & 2.92391778                    & 2.92391778                   \\
14         & 11.36497759                  & 11.36497759                  &  & 1.67018352                    & 1.67018352                   &  & 2.98478396                    & 2.98478396                   \\
15         & 12.06692333                  & 12.06692333                  &  & 1.67300105                    & 1.67300105                   &  & 3.04049205                    & 3.04049205                   \\
16         & 12.81931152                  & 12.81931152                  &  & 1.67963181                    & 1.67963181                   &  & 3.09641443                    & 3.09641443                   \\
17         & 13.56954137                  & 13.56954137                  &  & 1.68384097                    & 1.68384097                   &  & 3.15212915                    & 3.15212915                   \\
18         & 14.32166883                  & 14.32166883                  &  & 1.68602524                    & 1.68602524                   &  & 3.21482378                    & 3.21482378                   \\
19         & 15.03535243                  & 15.03535243                  &  & 1.68953260                    & 1.68953260                   &  & 3.25933278                    & 3.25933278                   \\
20         & 15.79127663                  & 15.79127663                  &  & 1.69644772                    & \textbf{1.69279776}          &  & 3.31183101                    & 3.31183101                   \\
21         & 16.53963351                  & 16.53963351                  &  & 1.69894302                    & \textbf{1.69473728}          &  & 3.36159380                    & 3.36159380                   \\
22         & 17.28558985                  & 17.28558985                  &  & 1.70066803                    & \textbf{1.69838000}          &  & 3.41145125                    & 3.41145125                   \\
23         & 18.04998520                  & 18.04998520                  &  & 1.70393378                    & \textbf{1.70223864}          &  & 3.45013735                    & 3.45013735                   \\
24         & 18.78807360                  & 18.78807360                  &  & 1.70538770                    & \textbf{1.70420700}          &  & 3.50251027                    & \textbf{3.49921018}          \\
25         & 19.54468228                  & \textbf{19.52687708}         &  & 1.71233042                    & \textbf{1.70631982}          &  & 3.54524557                    & \textbf{3.54088547}          \\
26         & 20.29026179                  & \textbf{20.27823046}         &  & 1.71459329                    & \textbf{1.70969312}          &  & 3.59113378                    & \textbf{3.58686982}          \\
27         & 21.04950183                  & \textbf{21.02971196}         &  & 1.71898563                    & \textbf{1.71211430}          &  & 3.63268719                    & \textbf{3.62413712}          \\
28         & 21.79712526                  & \textbf{21.76057988}         &  & 1.72190904                    & \textbf{1.71437832}          &  & 3.67394734                    & \textbf{3.66722675}          \\
29         & 22.54735751                  & \textbf{22.49836633}         &  & 1.72242777                    & \textbf{1.71658642}          &  & 3.70961474                    & \textbf{3.70434027}          \\
30         & 23.25868018                  & \textbf{23.24258773}         &  & 1.72504075                    & \textbf{1.71816387}          &  & 3.75302053                    & \textbf{3.74670554}          \\
31         & 24.03727192                  & \textbf{23.99738366}         &  & 1.72537598                    & \textbf{1.72046887}          &  & 3.79463689                    & \textbf{3.78299544}          \\
32         & 24.78200903                  & \textbf{24.74151096}         &  & 1.72708467                    & \textbf{1.72253372}          &  & 3.82636068                    & \textbf{3.81794567}          \\
33         & 25.54486051                  & \textbf{25.47851512}         &  & 1.73189080                    & \textbf{1.72382222}          &  & 3.86602448                    & \textbf{3.85326849}          \\
34         & 26.30459206                  & \textbf{26.23123612}         &  & 1.73298460                    & \textbf{1.72519586}          &  & 3.90172754                    & \textbf{3.88952299}          \\
35         & 27.03382244                  & \textbf{26.97763983}         &  & 1.73378808                    & \textbf{1.72867585}          &  & 3.93668876                    & \textbf{3.92408008}          \\ \midrule
\# Improved &                              & 11                           &  &                               & 16                           &  &                               & 13                           \\
\# Matched  &                              & 20                           &  &                               & 15                           &  &                               & 18                           \\
\# Worse    &                              & 0                            &  &                               & 0                            &  &                               & 0                            \\ \bottomrule
\end{tabular}
}
\end{table}

Table~\ref{tb:ins245} summarizes the computational results and comparisons of the three benchmark instances in the range of $5 \leq n \leq 35$. 
The first column denotes the instance sizes ($n$). 
Columns 2-3, columns 4-5, and columns 6-7 display the results for the $r_i = i^{1/2}$, $r_i = i^{-2/3}$, and $r_i = i^{-1/5}$ benchmark instances. Columns 2, 4, and 6 present the best-known result $R^*$ sourced from the Packomania website for each benchmark, while columns 3, 5, and 7 present the best result $R_{best}$ over 10 independent runs obtained by our algorithm for each benchmark. 
Compared with the current best-known results, improved results are highlighted in bold in terms of $R_{best}$.
The last three rows of the table summarize the number of instances where our algorithm obtained improved, matched, or worse results compared to the best-known results.

From Table~\ref{tb:ins245}, it is evident that our algorithm exhibits excellent performance in solving PUCC. Our algorithm totally improves the best-known results for 40 out of the 93 tested instances while it matches the best-known results for the remaining instances. Notably, the improved results primarily pertain to larger instances. 
This highlights the efficiency, applicability, and generality of our algorithm across instances with various radius distributions.
Table~\ref{tb:sum} provides a comprehensive summary of the computational results for all benchmark instances, of which our algorithm improves and matches the best-known results for 56 and 123 out of the total 179 instances, respectively.

\begin{table}[tb]
\centering
\caption{
Summary of computational results.
}
\label{tb:sum}
\scalebox{0.9}{
\begin{tabular}{lllllllll}
\toprule
Benchmark        &  & No. instance & & No. improved & & No. matched & & No. worse \\ \midrule
$r_i = i$        &  & 31           & & 0            & & 31          & & 0         \\
$r_i = i^{-1/2}$ &  & 31           & & 11           & & 20          & & 0         \\
$r_i = i^{1/2}$  &  & 31           & & 11           & & 20          & & 0         \\
$r_i = i^{-2/3}$ &  & 31           & & 16           & & 15          & & 0         \\
$r_i = i^{-1/5}$ &  & 31           & & 13           & & 18          & & 0         \\
NR               &  & 24           & & 5            & & 19          & & 0         \\ \midrule
Total            &  & 179          & & 56           & & 123         & & 0         \\ \bottomrule
\end{tabular}
}

\end{table}

\begin{figure}[tb]
    \centering
    \begin{minipage}[b]{0.38\linewidth}
        \centering
        \subfloat[Previous best solution for $r_i = i^{-1/2}$ with $n=20$]
        {\includegraphics[width=1\linewidth]{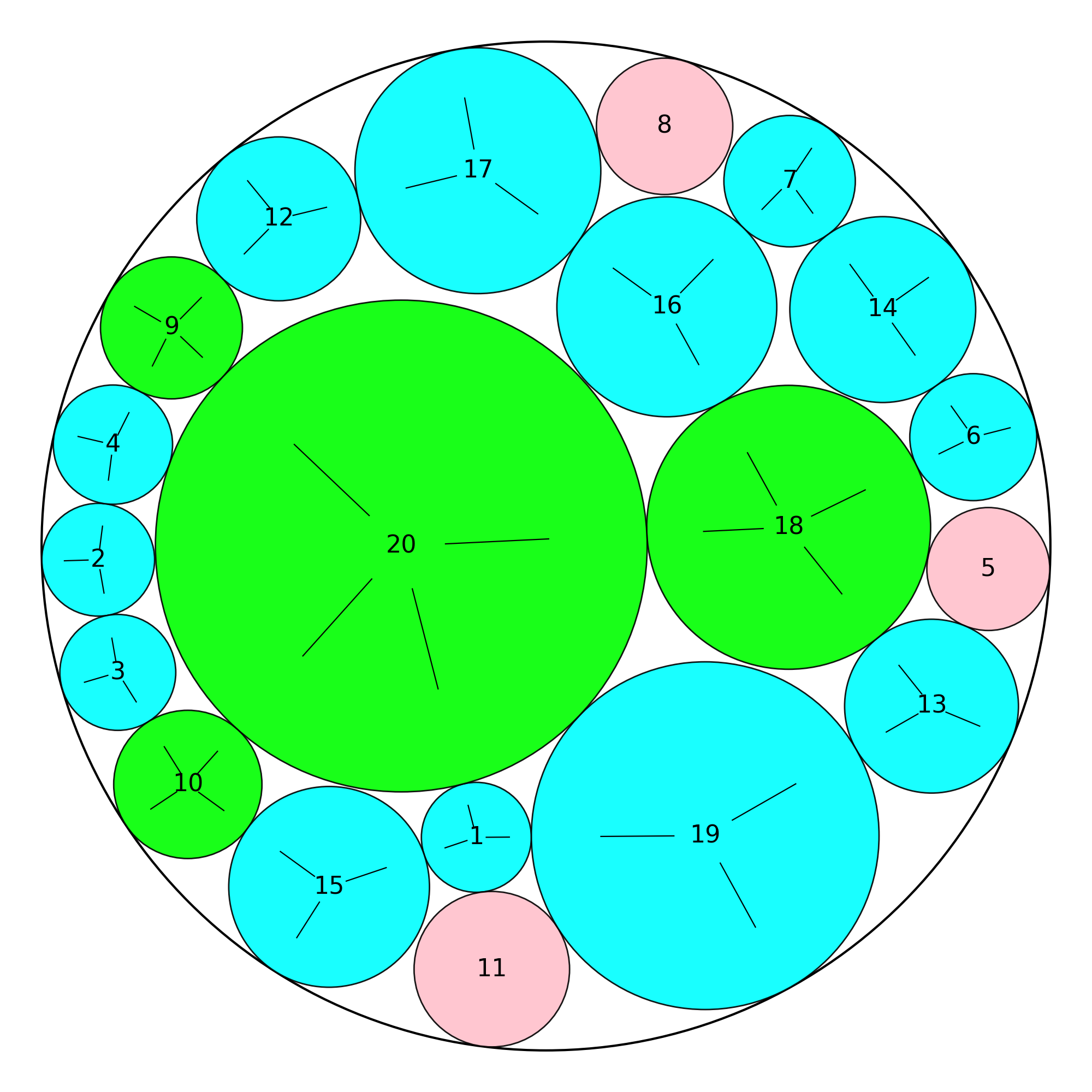}}
    \end{minipage}
    \begin{minipage}[b]{0.38\linewidth}
        \centering
        \subfloat[Improved best solution for $r_i = i^{-1/2}$ with $n=20$]
        {\includegraphics[width=1\linewidth]{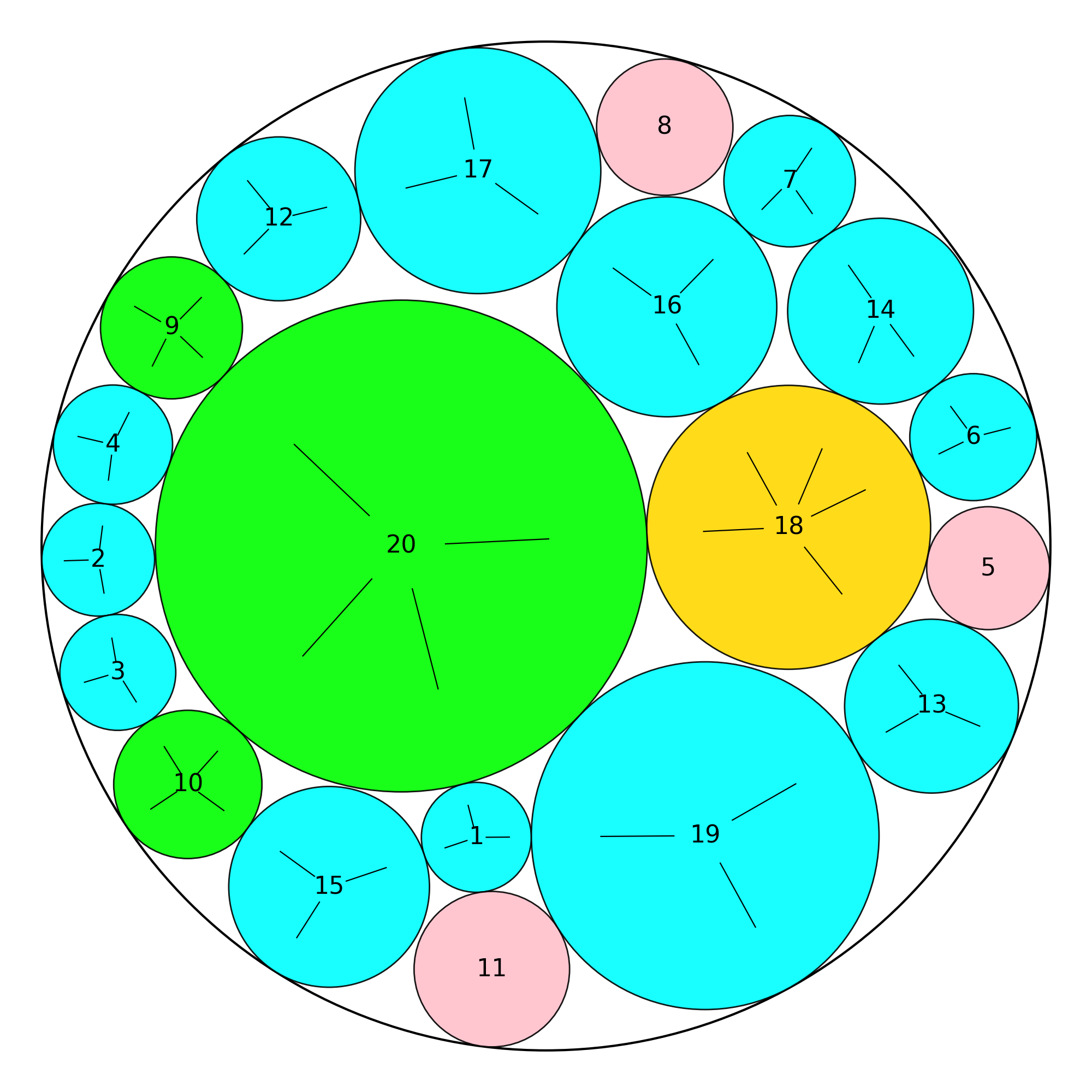}}
    \end{minipage}
    \begin{minipage}[b]{0.38\linewidth}
        \centering
        \subfloat[Previous best solution for $r_i = i^{-1/5}$ with $n=8$]
        {\includegraphics[width=1\linewidth]{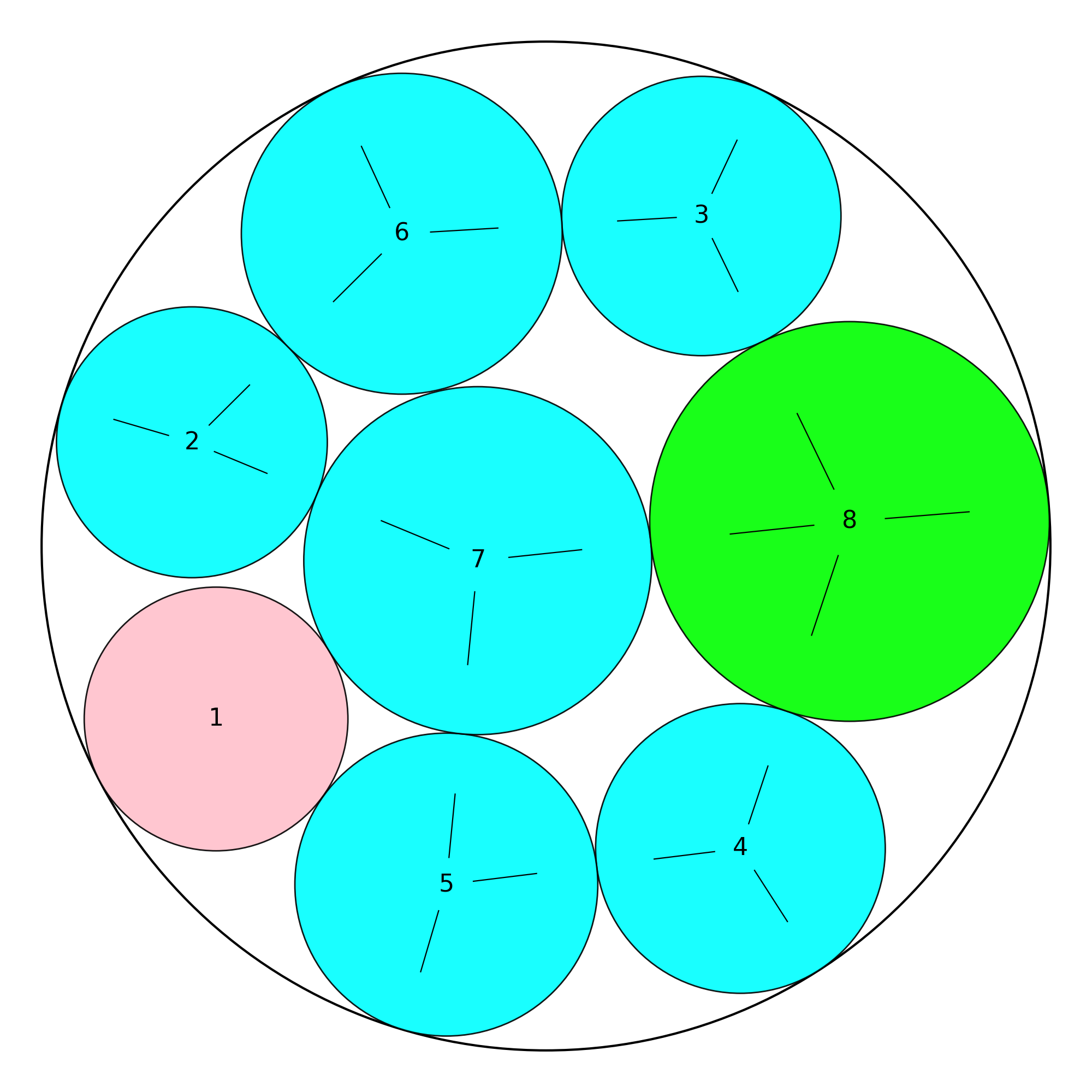}}
    \end{minipage}
    \begin{minipage}[b]{0.38\linewidth}
        \centering
        \subfloat[Improved best solution for $r_i = i^{-1/5}$ with $n=8$]
        {\includegraphics[width=1\linewidth]{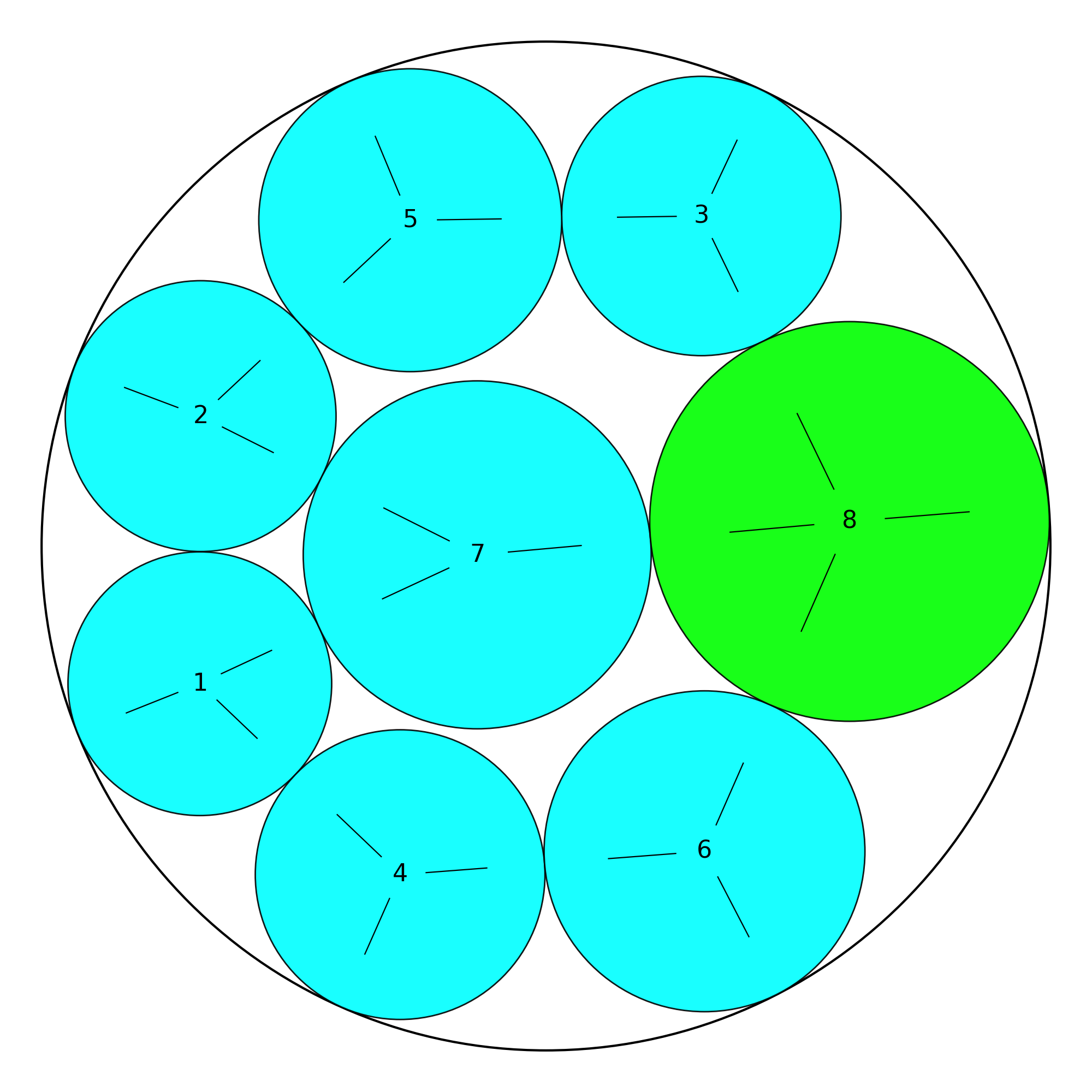}}
    \end{minipage}

    \caption{Comparisons between the previous best-known solutions and improved solutions for the two small-scale instances.}
    \label{fig:improved_sample_cmp}
\end{figure}

\begin{figure}[tb]
    \centering
    \begin{minipage}[b]{0.32\linewidth}
        \centering
        \subfloat[$r_i = i^{-1/2}$ for $n = 29$]
        {\includegraphics[width=1\linewidth]{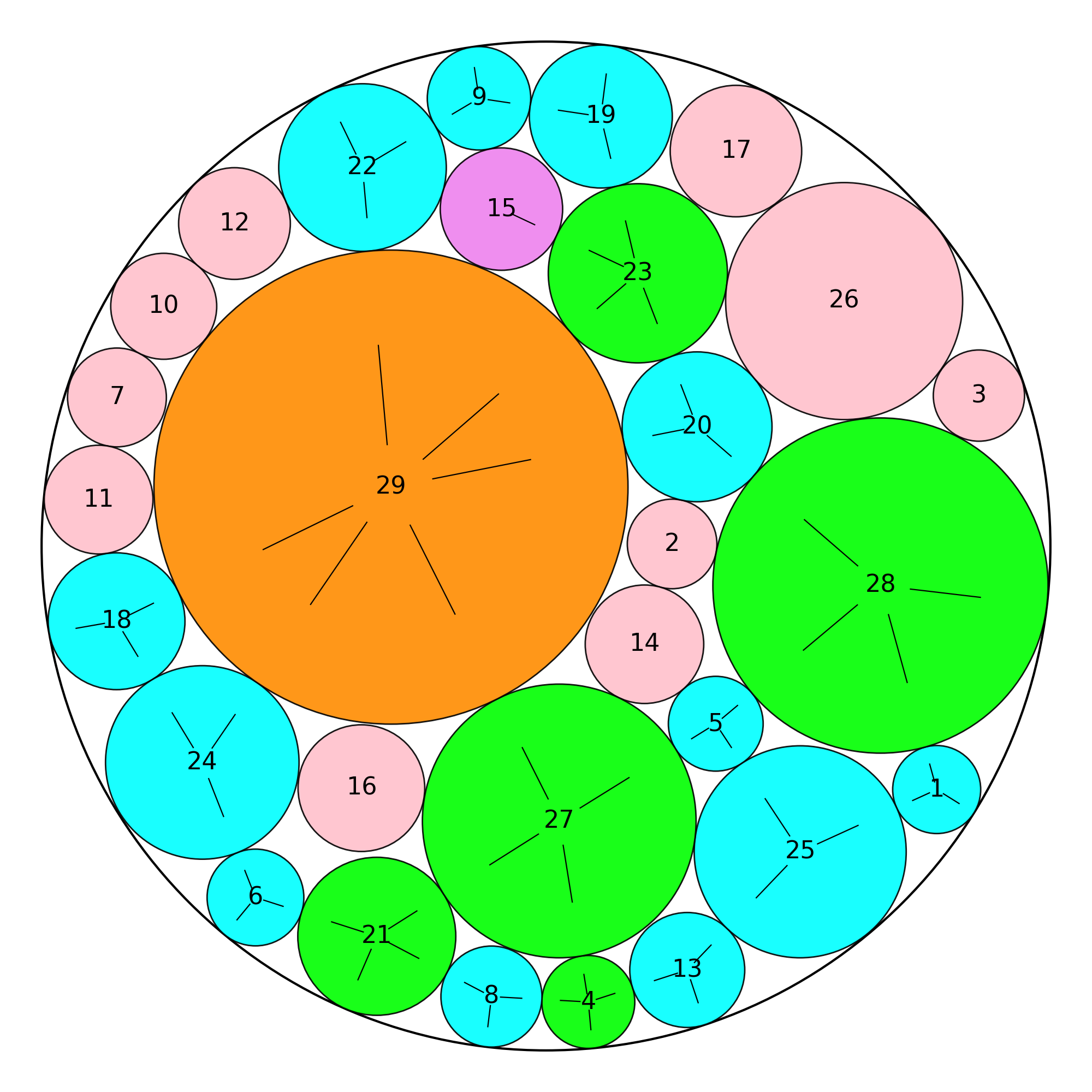}}
    \end{minipage}
    \begin{minipage}[b]{0.32\linewidth}
        \centering
        \subfloat[$r_i = i^{-1/2}$ for $n = 32$]
        {\includegraphics[width=1\linewidth]{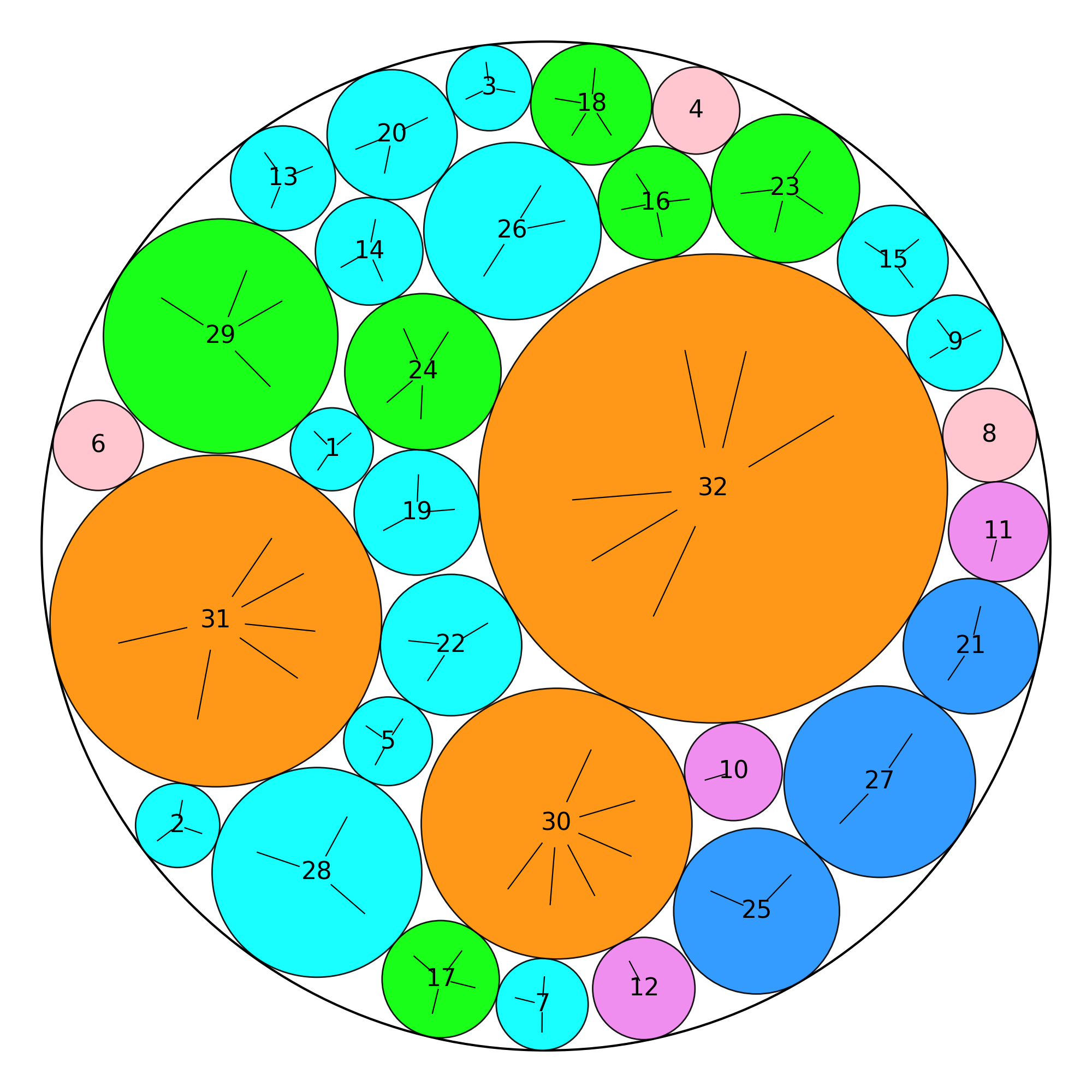}}
    \end{minipage}
    \begin{minipage}[b]{0.32\linewidth}
        \centering
        \subfloat[$r_i = i^{-1/2}$ for $n = 35$]
        {\includegraphics[width=1\linewidth]{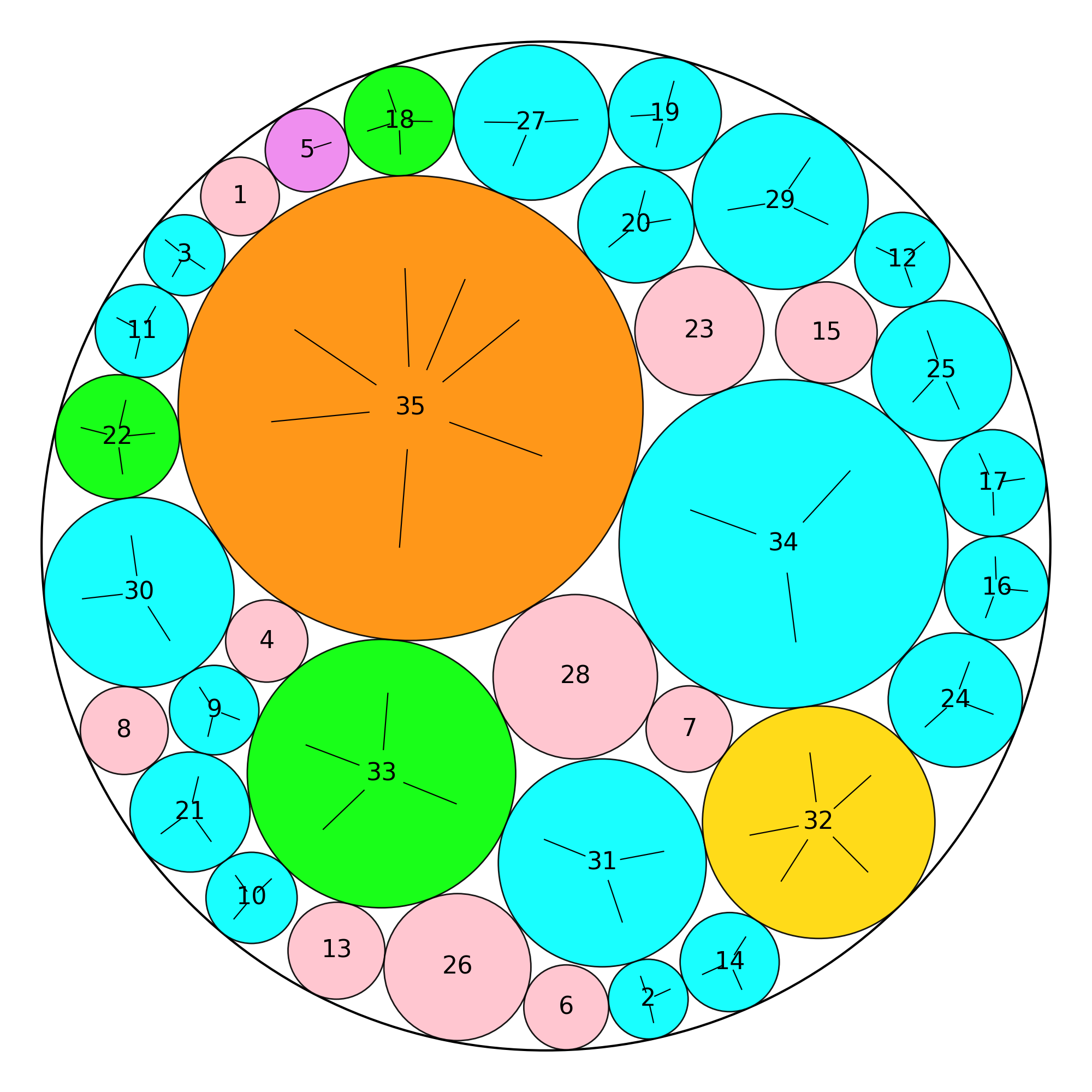}}
    \end{minipage}
    \begin{minipage}[b]{0.32\linewidth}
        \centering
        \subfloat[NR40\_1]
        {\includegraphics[width=1\linewidth]{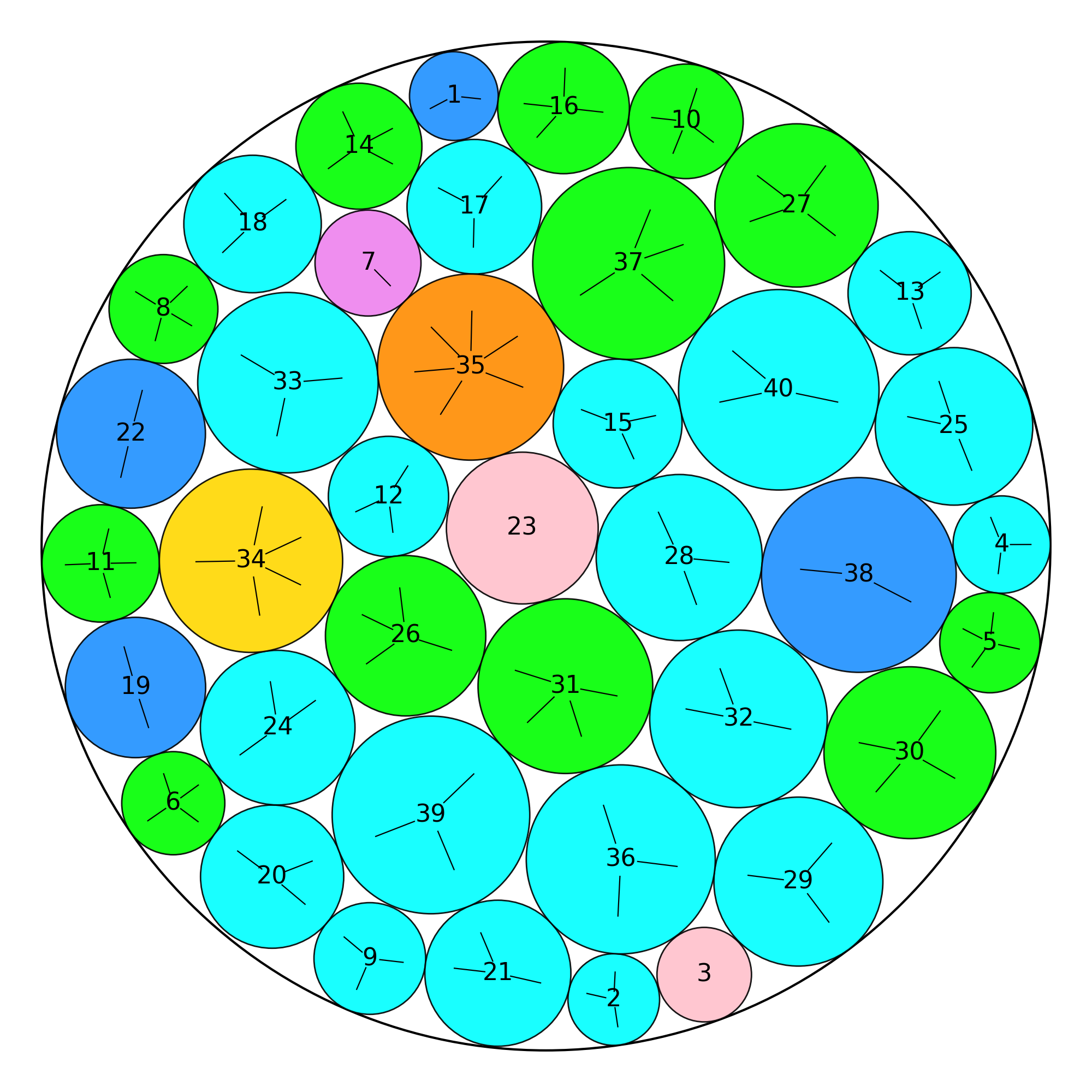}}
    \end{minipage}
    \begin{minipage}[b]{0.32\linewidth}
        \centering
        \subfloat[NR50\_1]
        {\includegraphics[width=1\linewidth]{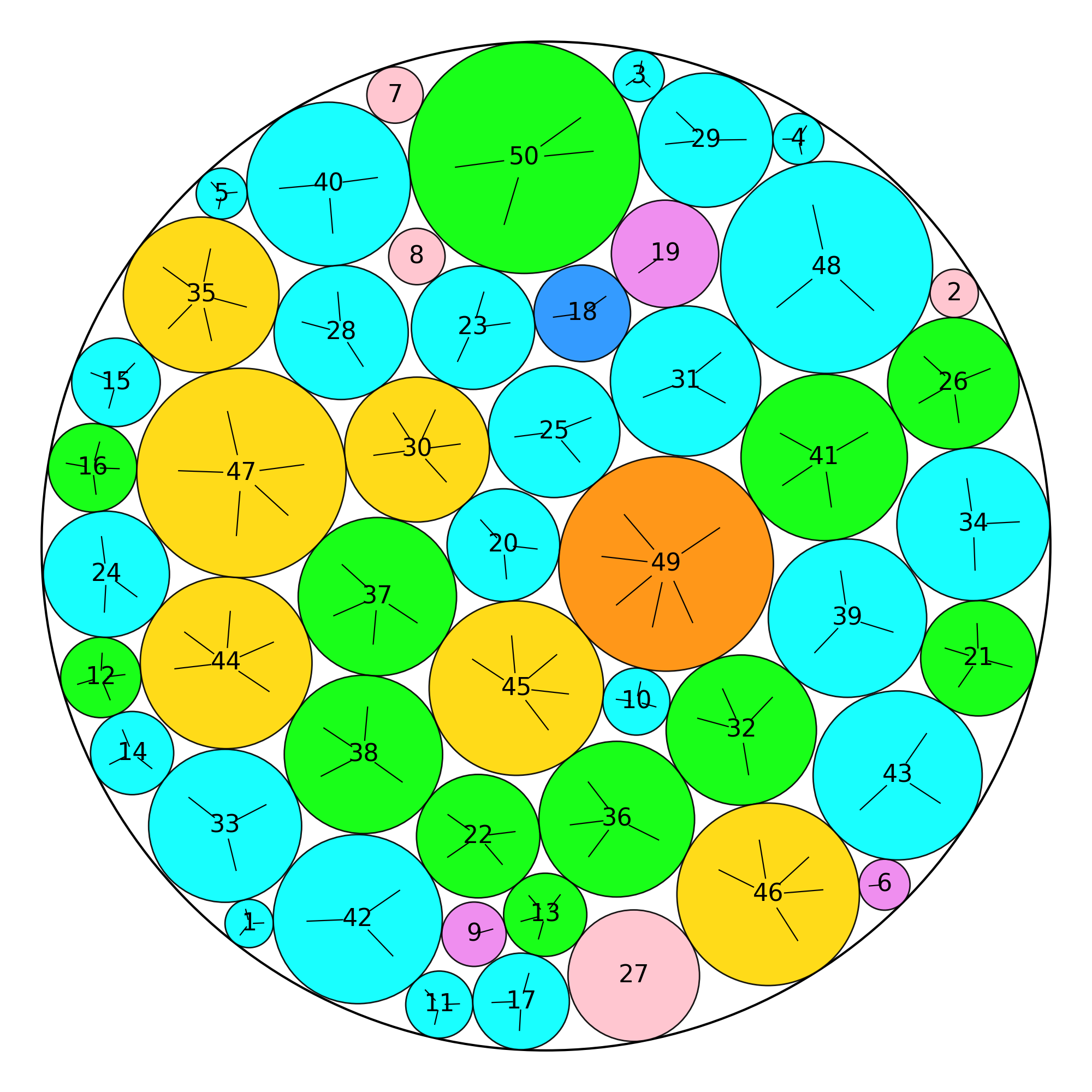}}
    \end{minipage}
    \begin{minipage}[b]{0.32\linewidth}
        \centering
        \subfloat[NR60\_1]
        {\includegraphics[width=1\linewidth]{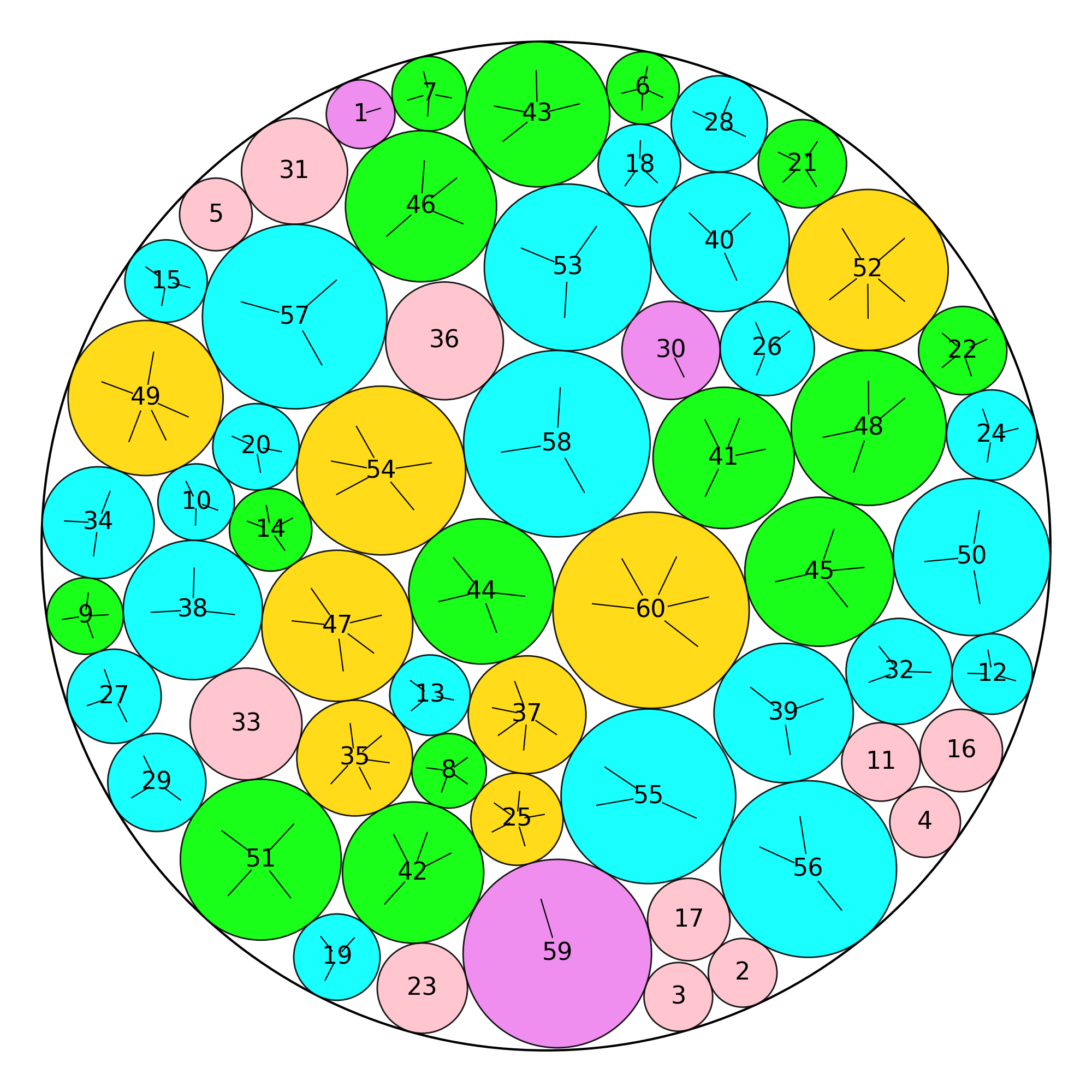}}
    \end{minipage}
    \begin{minipage}[b]{0.32\linewidth}
        \centering
        \subfloat[$r_i = i^{1/2}$ for $n = 27$]
        {\includegraphics[width=1\linewidth]{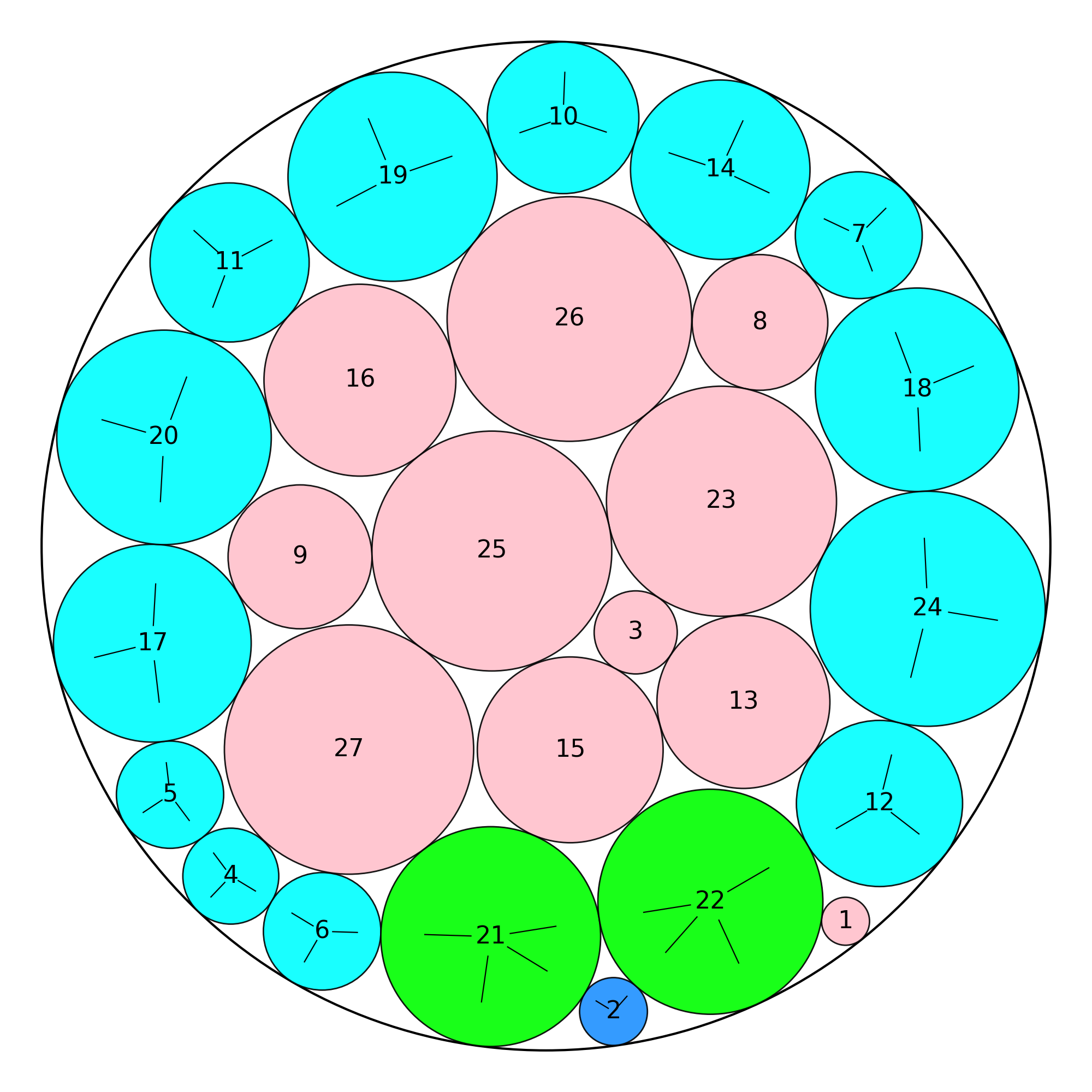}}
    \end{minipage}
    \begin{minipage}[b]{0.32\linewidth}
        \centering
        \subfloat[$r_i = i^{1/2}$ for $n = 30$]
        {\includegraphics[width=1\linewidth]{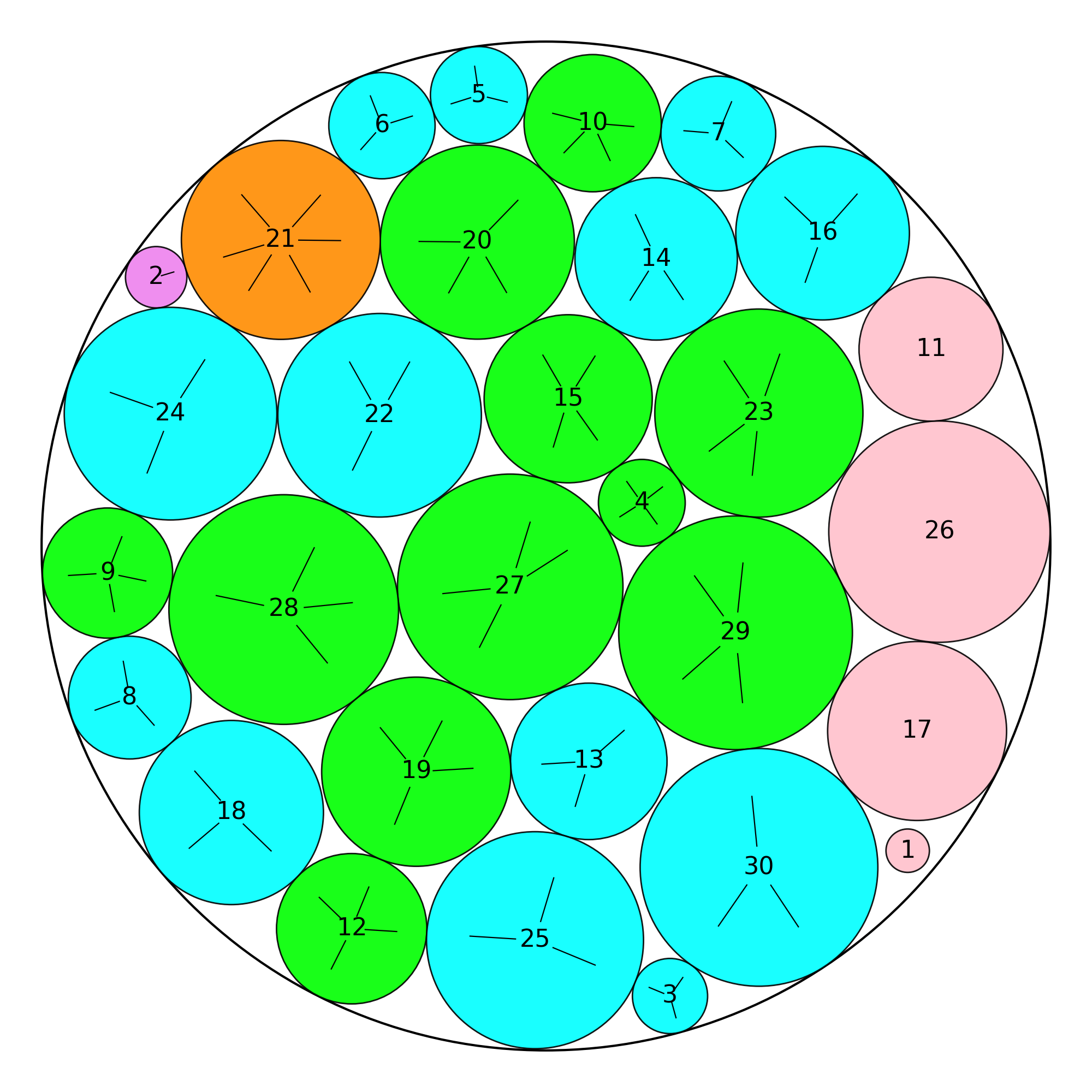}}
    \end{minipage}
    \begin{minipage}[b]{0.32\linewidth}
        \centering
        \subfloat[$r_i = i^{1/2}$ for $n = 34$]
        {\includegraphics[width=1\linewidth]{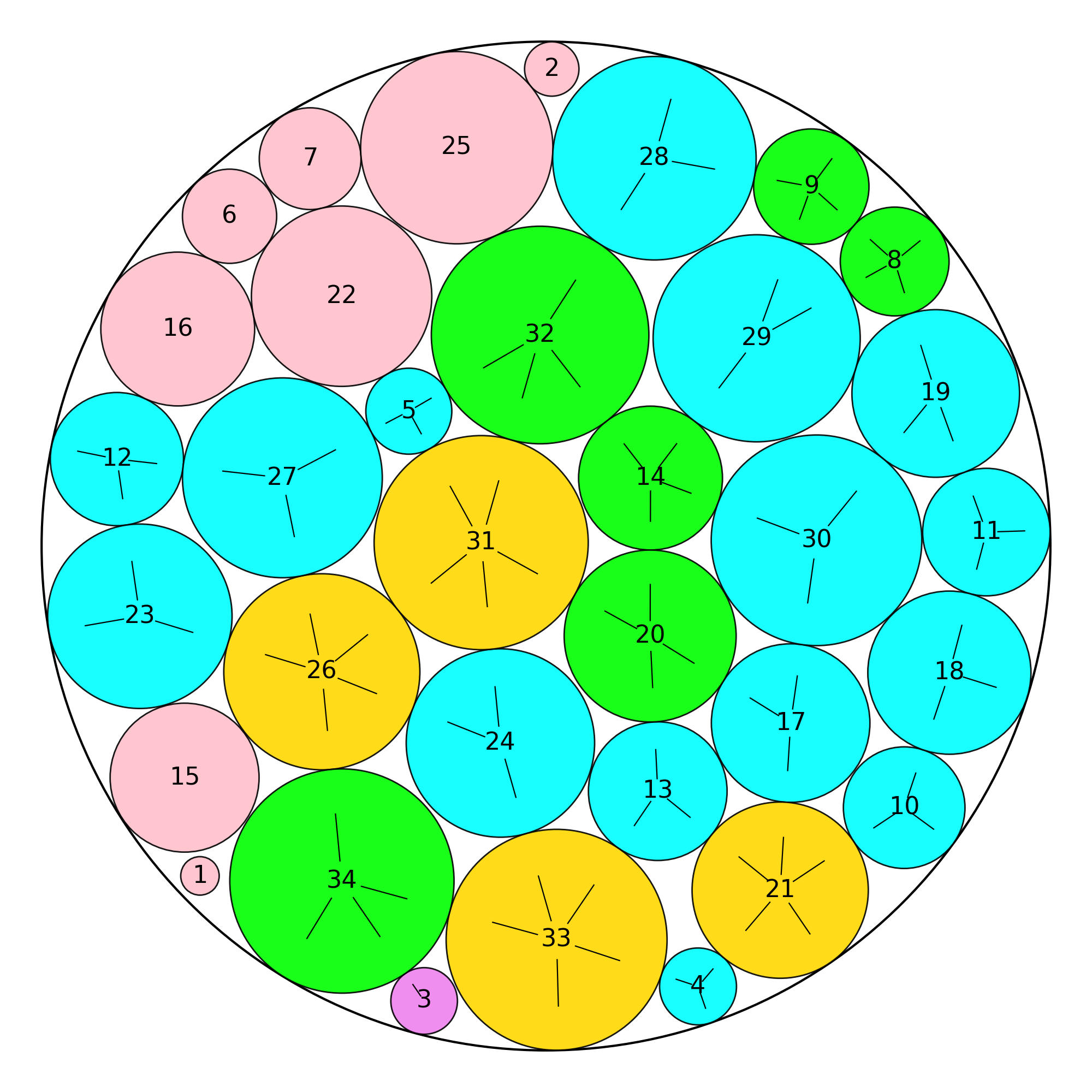}}
    \end{minipage}

    \caption{Improved solutions found in this study for nine instances selected from each benchmark.}
    \label{fig:improved_sample_1}
\end{figure}

\begin{figure}[tb]
    \centering
    \begin{minipage}[b]{0.32\linewidth}
        \centering
        \subfloat[$r_i = i^{-2/3}$ for $n = 27$]
        {\includegraphics[width=1\linewidth]{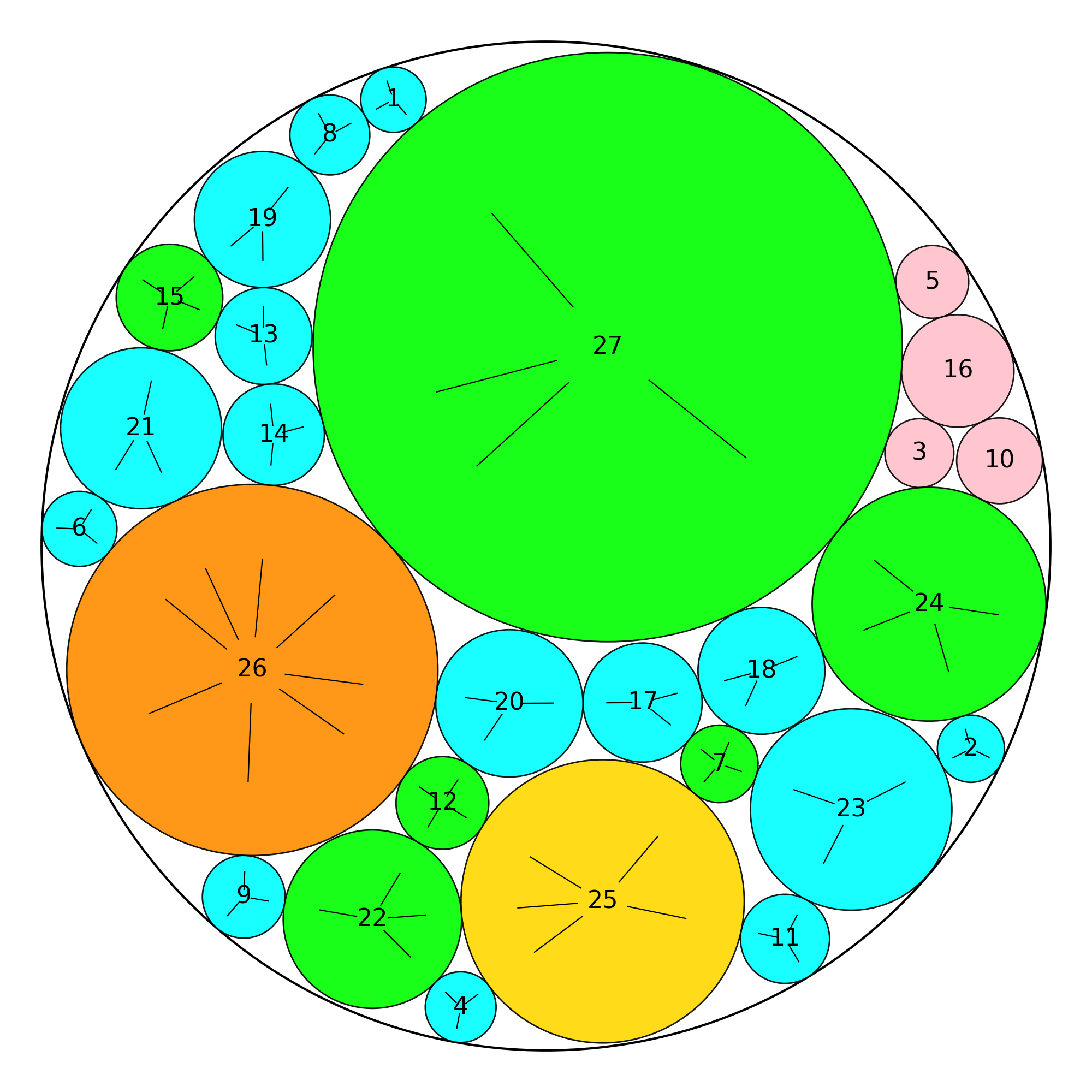}}
    \end{minipage}
    \begin{minipage}[b]{0.32\linewidth}
        \centering
        \subfloat[$r_i = i^{-2/3}$ for $n = 33$]
        {\includegraphics[width=1\linewidth]{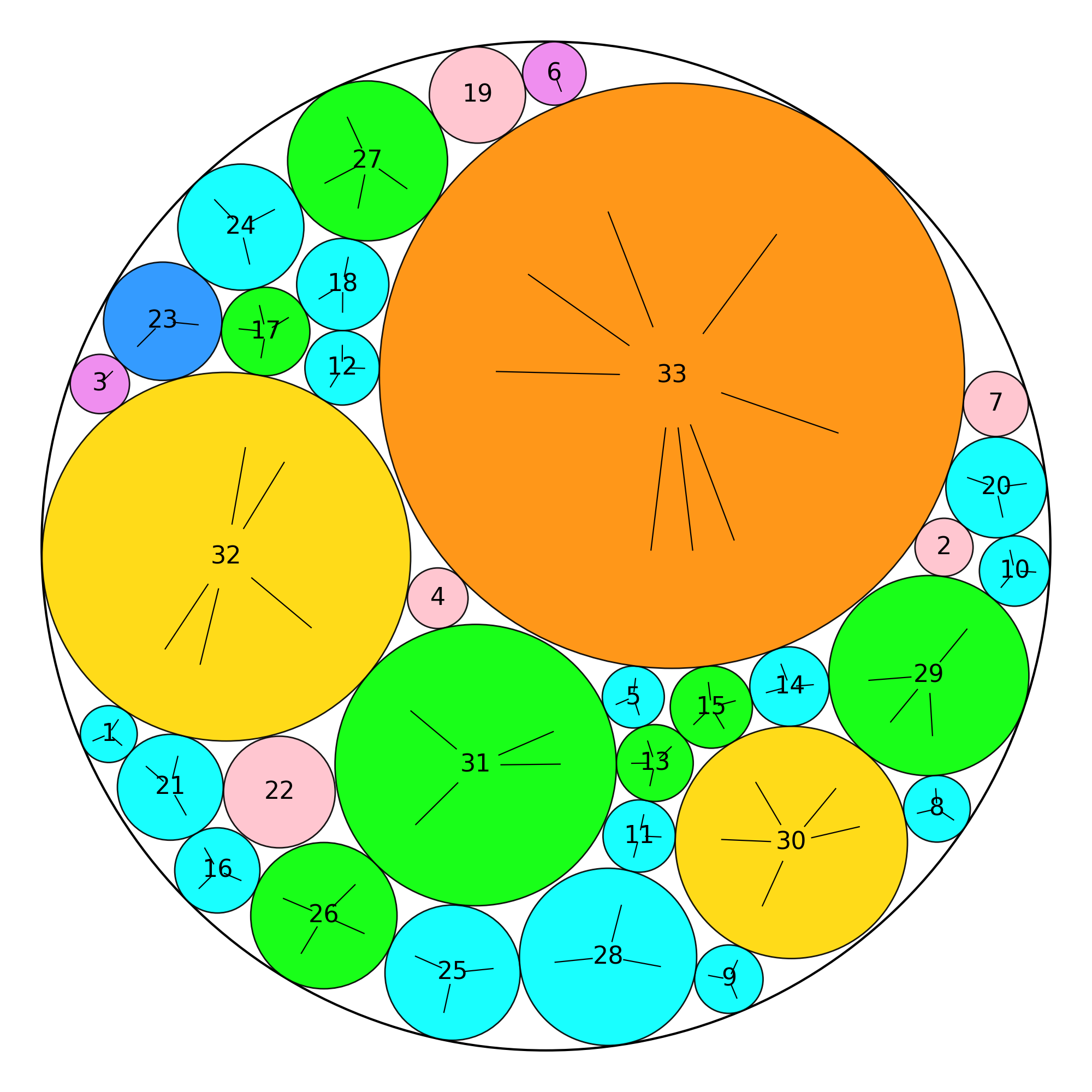}}
    \end{minipage}
    \begin{minipage}[b]{0.32\linewidth}
        \centering
        \subfloat[$r_i = i^{-2/3}$ for $n = 35$]
        {\includegraphics[width=1\linewidth]{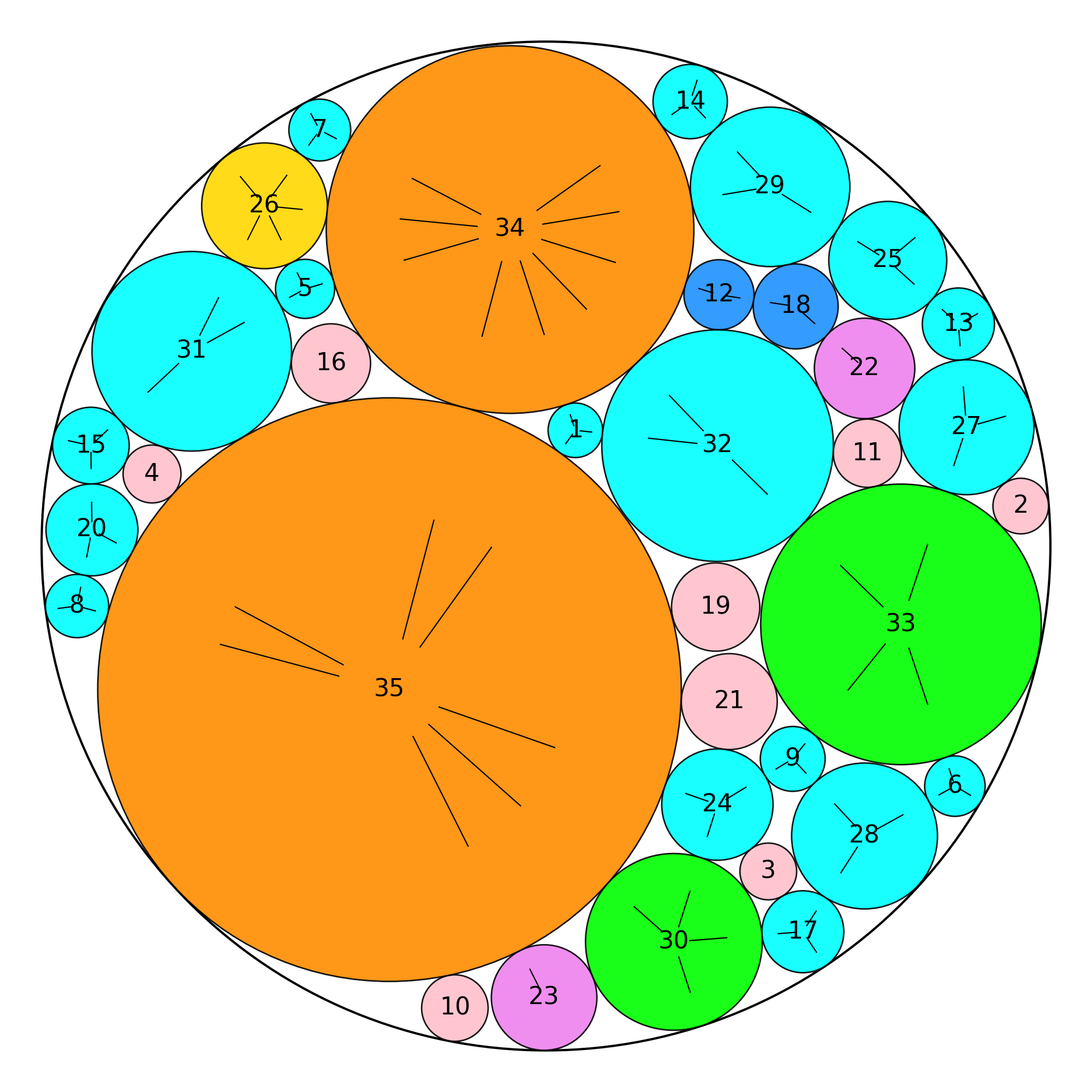}}
    \end{minipage}
    \begin{minipage}[b]{0.32\linewidth}
        \centering
        \subfloat[$r_i = i^{-1/5}$ for $n = 26$]
        {\includegraphics[width=1\linewidth]{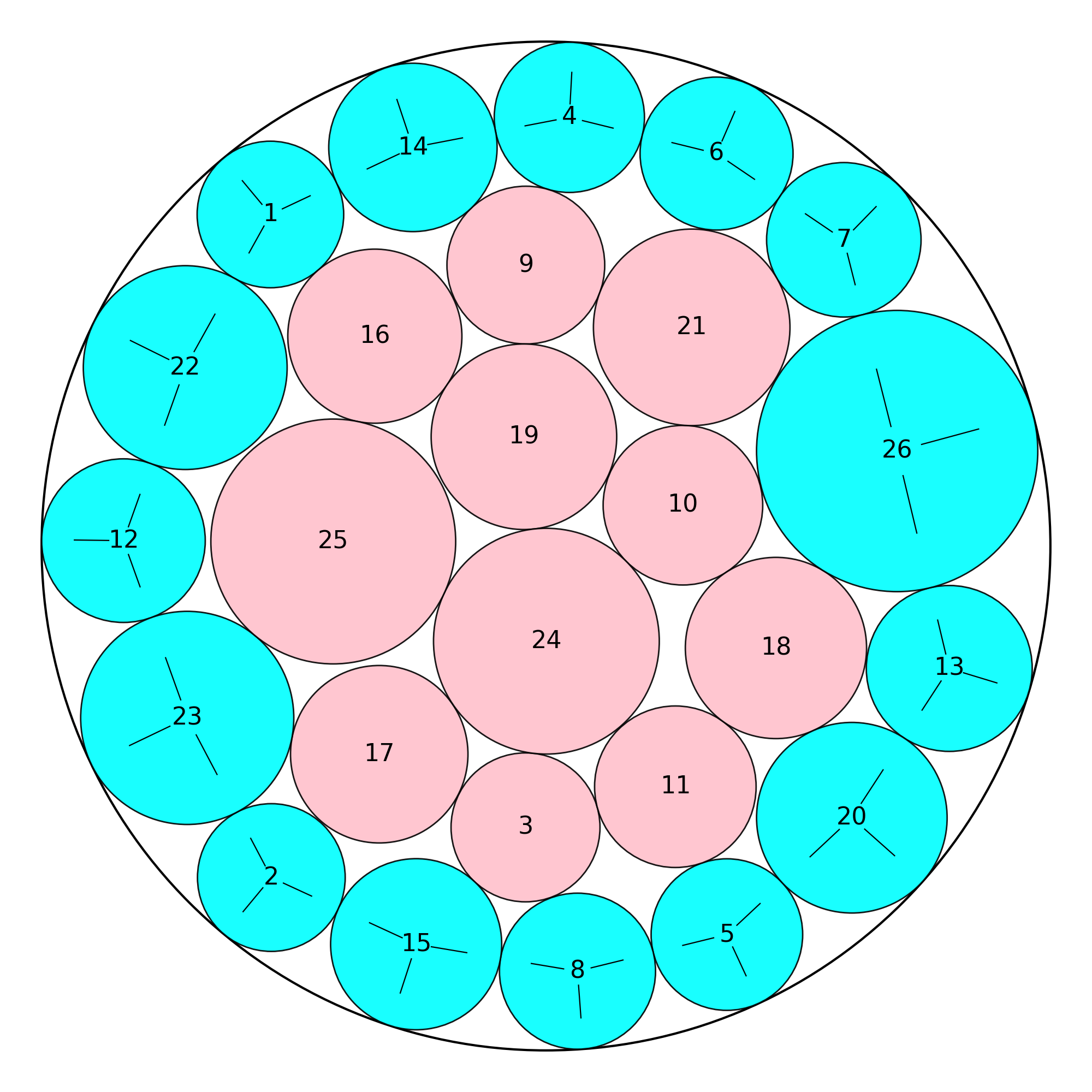}}
    \end{minipage}
    \begin{minipage}[b]{0.32\linewidth}
        \centering
        \subfloat[$r_i = i^{-1/5}$ for $n = 29$]
        {\includegraphics[width=1\linewidth]{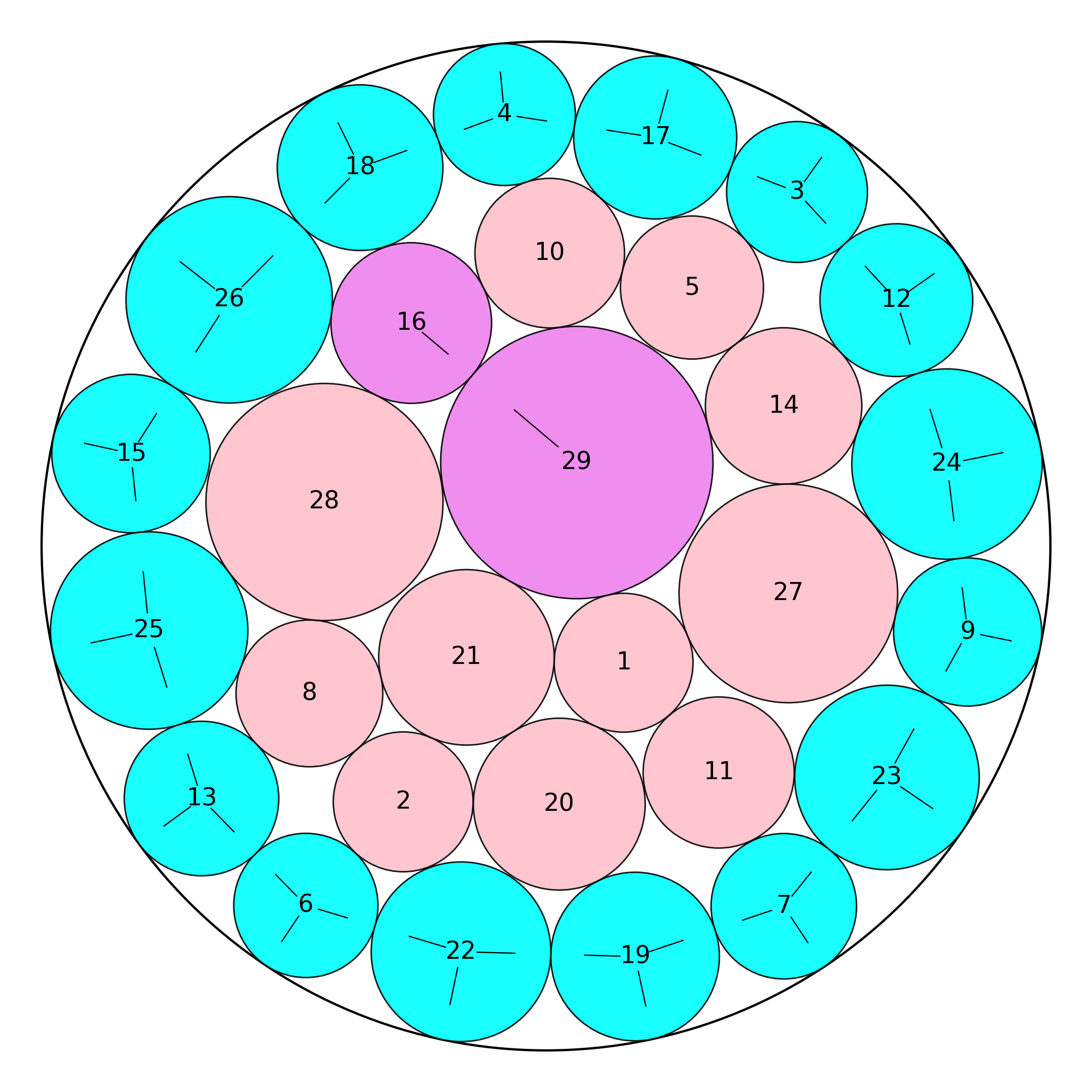}}
    \end{minipage}
    \begin{minipage}[b]{0.32\linewidth}
        \centering
        \subfloat[$r_i = i^{-1/5}$ for $n = 33$]
        {\includegraphics[width=1\linewidth]{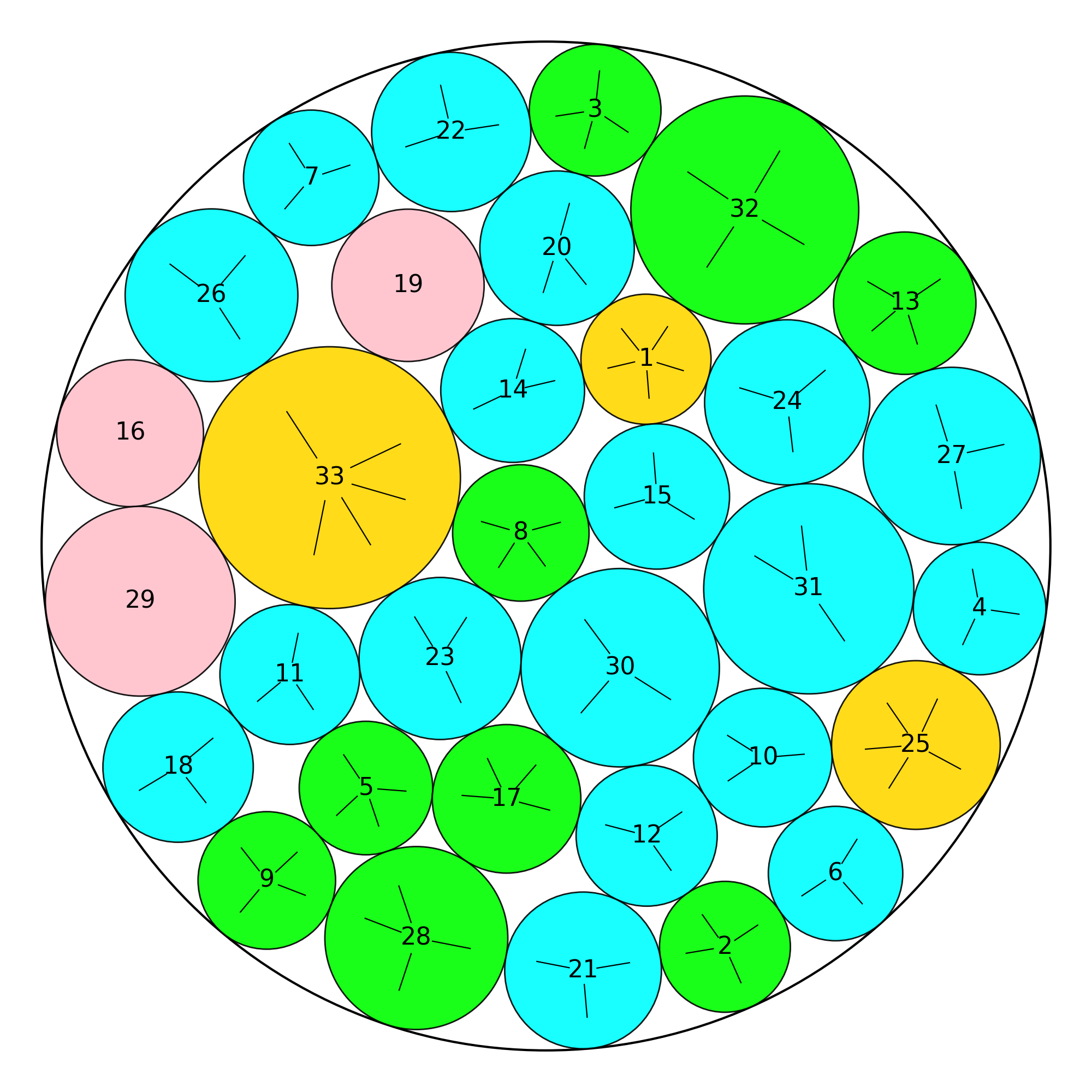}}
    \end{minipage}

    \caption{Improved solutions found in this study for six instances selected from each benchmark.}
    \label{fig:improved_sample_2}
\end{figure}

In experiments, our algorithm obtained two slight improvements for the instances $r_i = i^{-1/2}$ with $n = 20$ ($R_{best} - R^* = $ -9.20E-07 presented in Table~\ref{tb:ins3}) and $r_i = i^{-1/5}$ with $n = 8$ ($R_{best} - R^* = $ -1.37E-06 presented in Table~\ref{tb:ins245}). 
For the intuitive purpose of presenting and analyzing these improvements, we provide the illustrations of the comparisons between the previous best configurations sourced from the Packomania website~\citep{Spechtweb} and the improved configuration found in this study, shown in Figure~\ref{fig:improved_sample_cmp}. 
Moreover, we provide the illustrations of several improved and representative configurations found in this study selected from each benchmark, presented in Figures~\ref{fig:improved_sample_1} and~\ref{fig:improved_sample_2}.

From Figure~\ref{fig:improved_sample_cmp}, we observe that even a slight improvement could lead to different configurations. Specifically, on the instance $r_i = i^{-1/2}$ with $n = 20$, the previous best configuration and the improved configuration are very similar, but circle $c_{14}$ contact to circles $c_6$, $c_7$ and the circular container in the previous best configuration, circle $c_{14}$ contact to circles $c_6$, $c_7$ and $c_{18}$ in the improved configuration. As a result, such a difference leads to slight improvement. 
On the instance $r_i = i^{-1/5}$ with $n = 8$, the positions of circles $c_4$, $c_5$ and $c_6$ in the improved configuration are different from the previous best configuration, leading to a slight improvement. 
These experimental results demonstrate the superior performance of our algorithm, revealing a slight improvement that exceeds the expectations of previous methods.


\subsection{Efficiency of the Proposed Methods}

The novel search ingredient of our proposed algorithm is the SHS core (discussed in Section~\ref{sec:SHS}), which is based on the proposed novel layout-graph transformation and solution hash methods (discussed in Section~\ref{sec:layhash}). To assess the efficiency and performance of our proposed methods, we conducted extra experiments and comparisons of our algorithm with two algorithmic variants. 
The first algorithmic variant, denoted by I-Greedy, removes the SHS core module from the proposed algorithm (i.e., remove the codes in Algorithm~\ref{alg:SHS} lines 16-20), which degenerates into only using the greedy methods as the local search strategy in the intensification phase.
The second algorithmic variant, denoted by I-TS, replaces the SHS core with the Tabu Search subroutine of the state-of-the-art ITS-VND algorithm~\citep{zeng2016iterated} (i.e., improve the solution by the Tabu Search instead of the SHS core in Algorithm~\ref{alg:SHS} line 16). We ensure the implementation of the Tabu Search subroutine and the parameter settings used in the subroutine are the same as ITS-VND.
It is worth noting that the I-TS variant is very similar to the ITS-VND algorithm in the intensification and diversification phases.

\begin{figure}[tb]
    \centering
    \begin{minipage}[b]{0.32\linewidth}
        \centering
        \subfloat[$r_i = i$ for $n = 30$]{\includegraphics[width=1\linewidth]{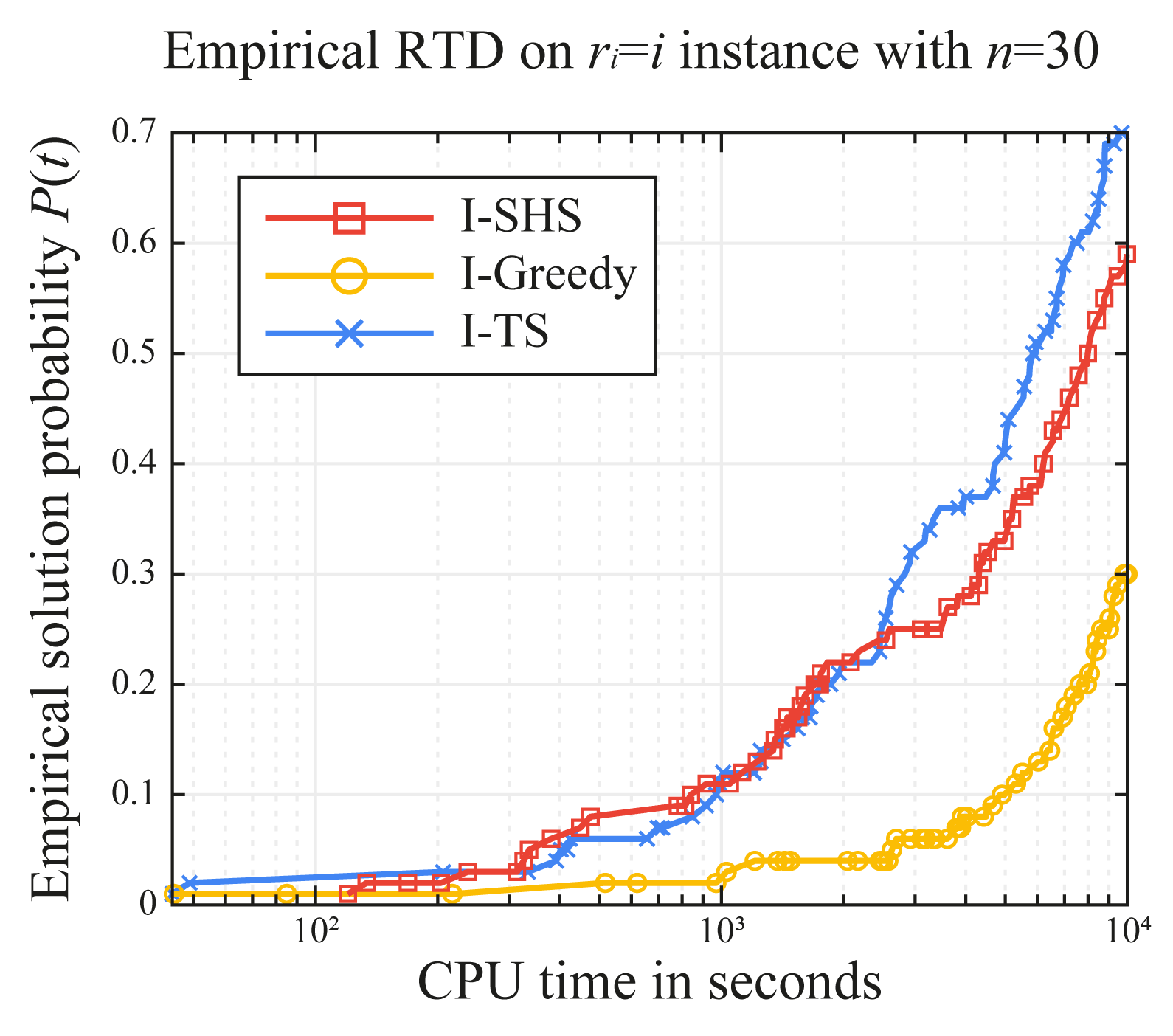}}
    \end{minipage}
    \begin{minipage}[b]{0.32\linewidth}
        \centering
        \subfloat[NR25\_1]
        {\includegraphics[width=1\linewidth]{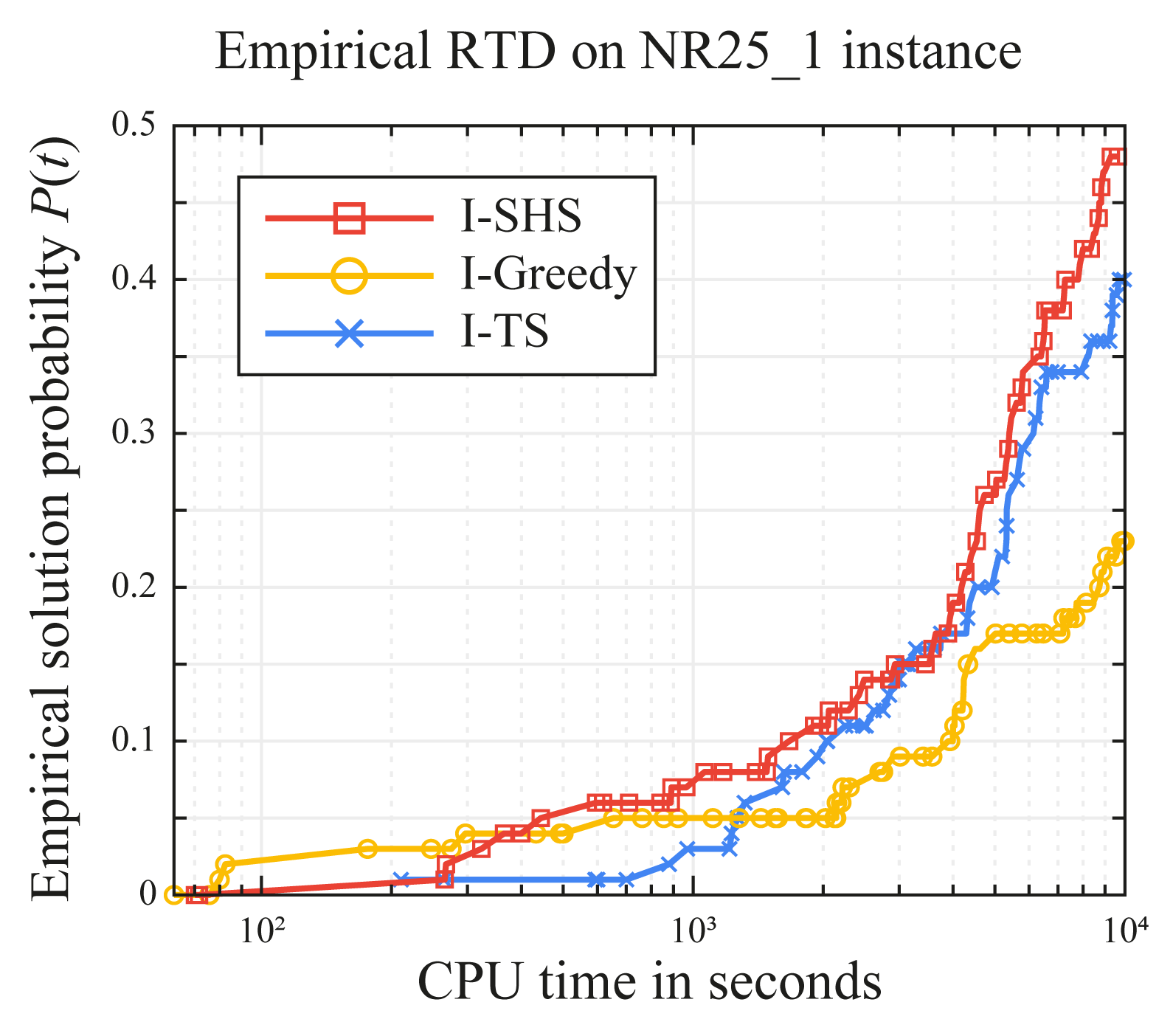}}
    \end{minipage}
    \begin{minipage}[b]{0.32\linewidth}
        \centering
        \subfloat[$r_i = i^{-1/2}$ for $n = 23$]{\includegraphics[width=1\linewidth]{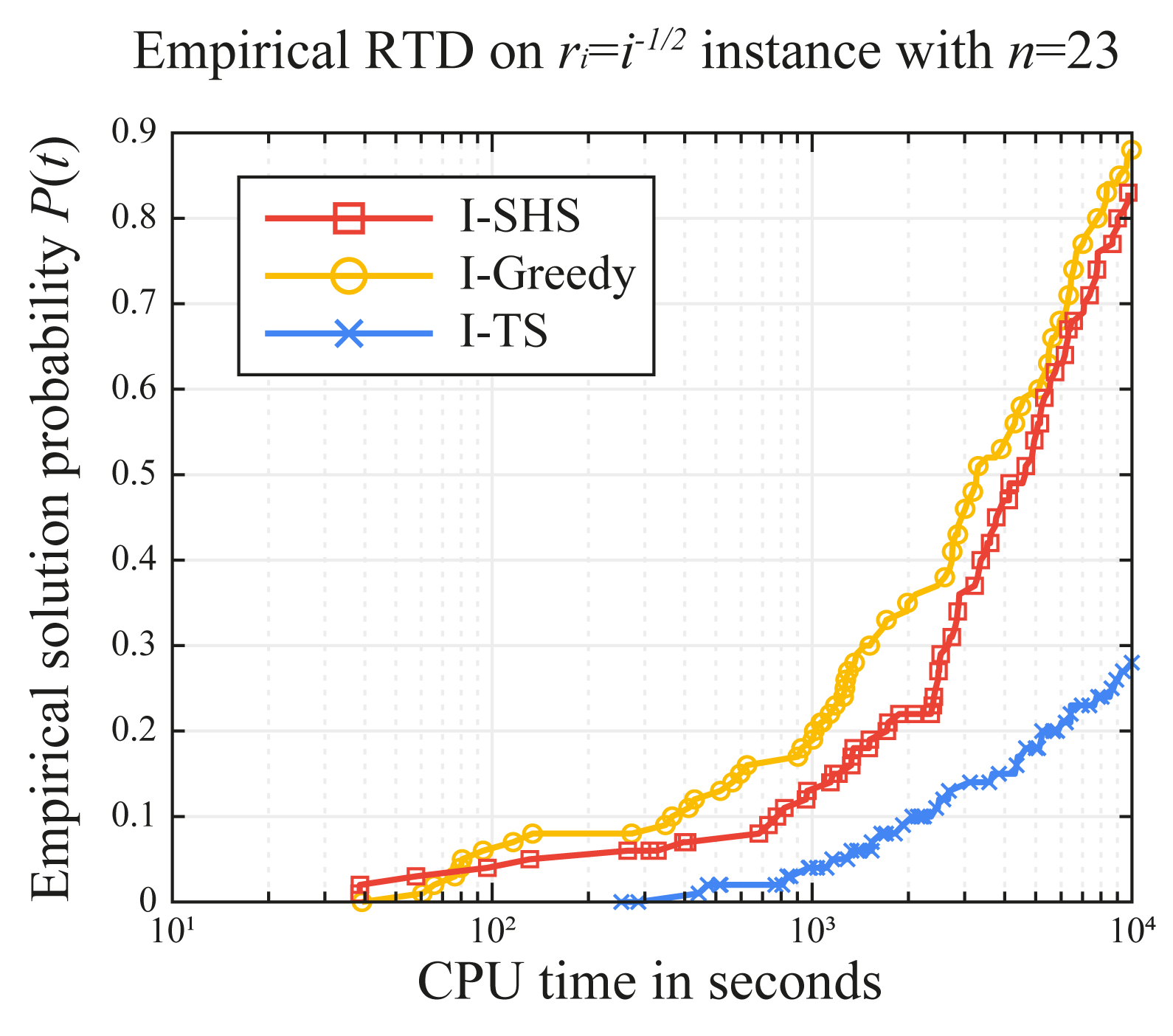}}
    \end{minipage}
    \begin{minipage}[b]{0.32\linewidth}
        \centering
        \subfloat[$r_i = i^{1/2}$ for $n = 26$]{\includegraphics[width=1\linewidth]{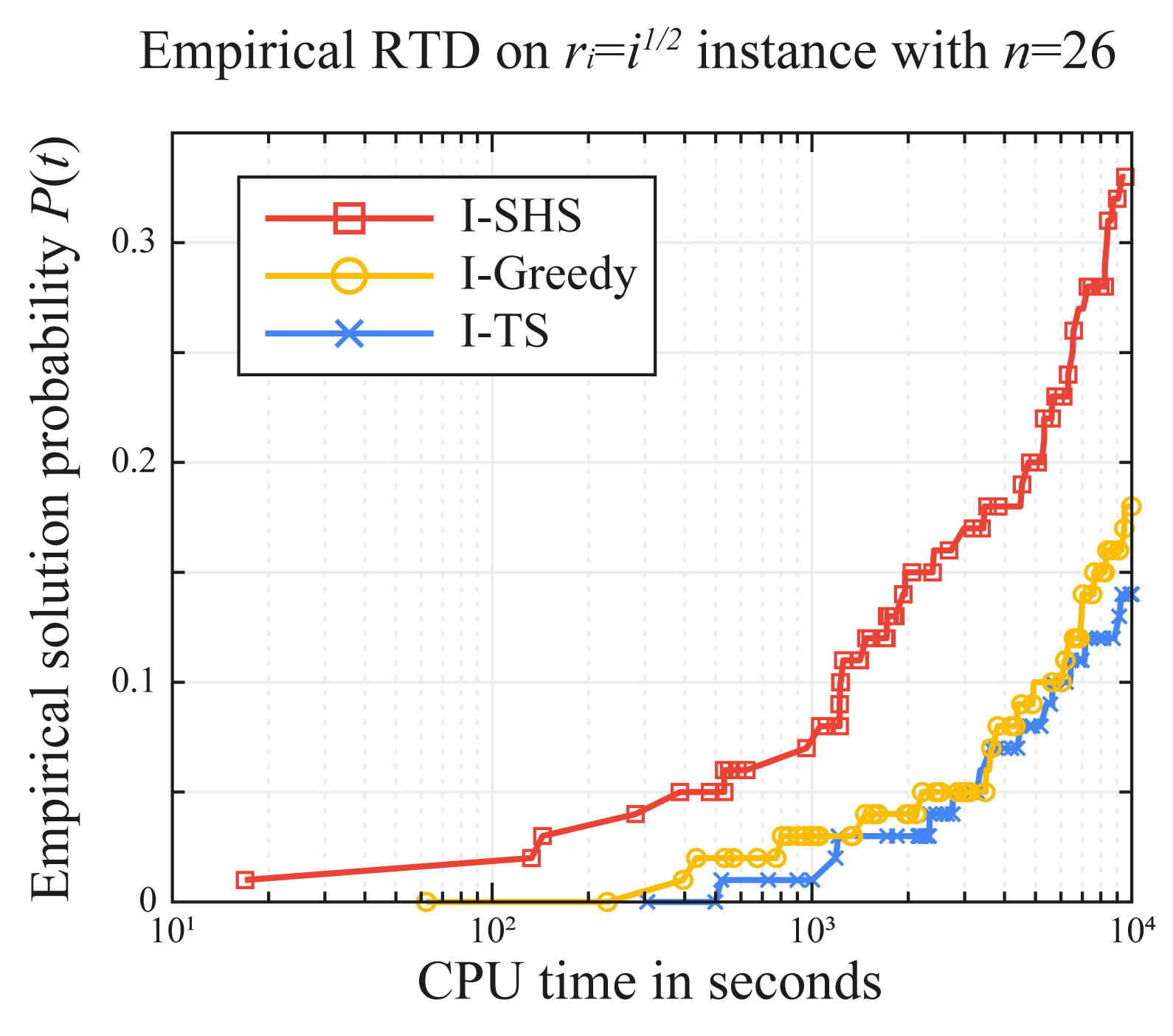}}
    \end{minipage}
    \begin{minipage}[b]{0.32\linewidth}
        \centering
        \subfloat[$r_i = i^{-2/3}$ for $n = 24$]{\includegraphics[width=1\linewidth]{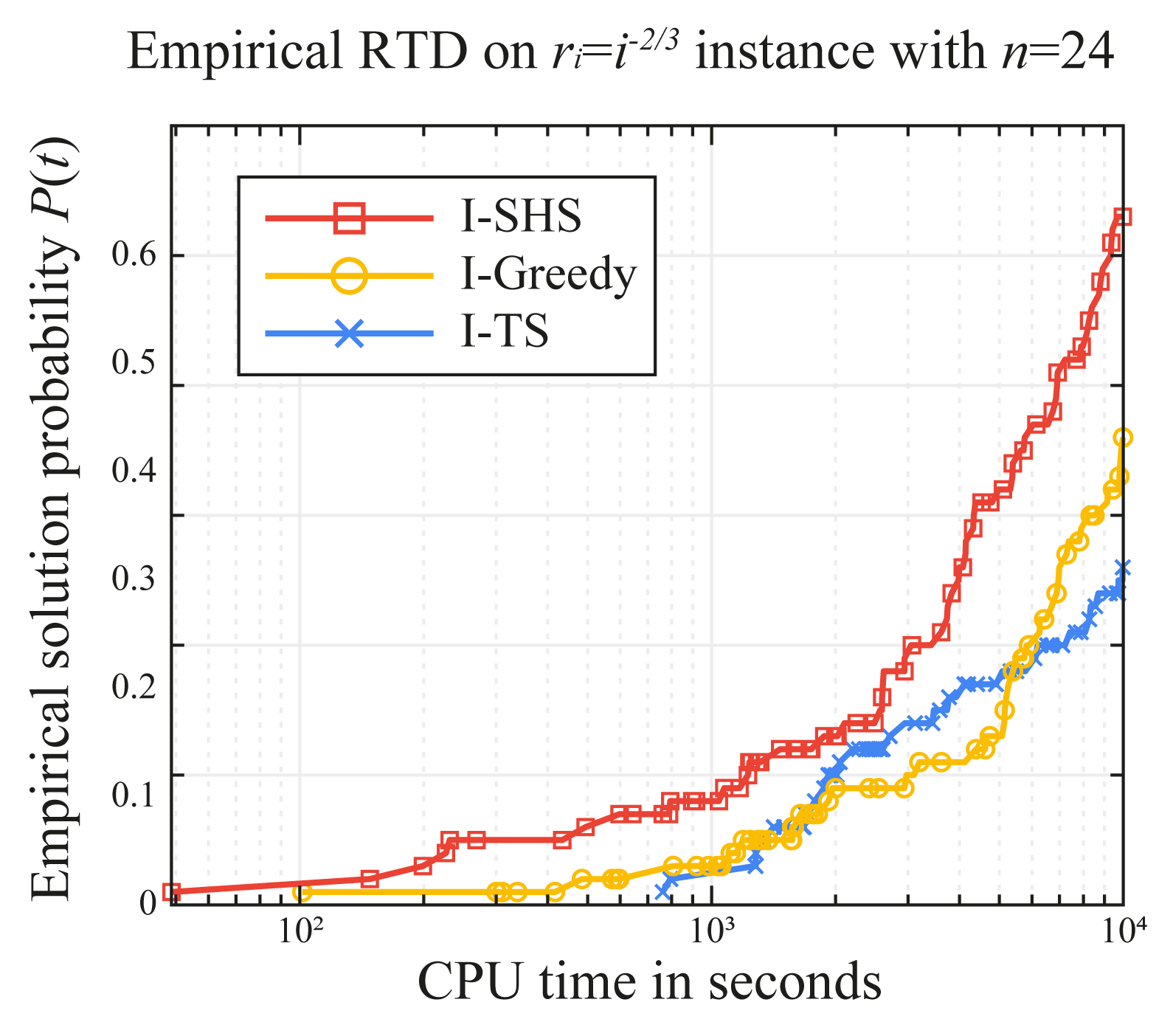}}
    \end{minipage}
    \begin{minipage}[b]{0.32\linewidth}
        \centering
        \subfloat[$r_i = i^{-1/5}$ for $n = 28$]{\includegraphics[width=1\linewidth]{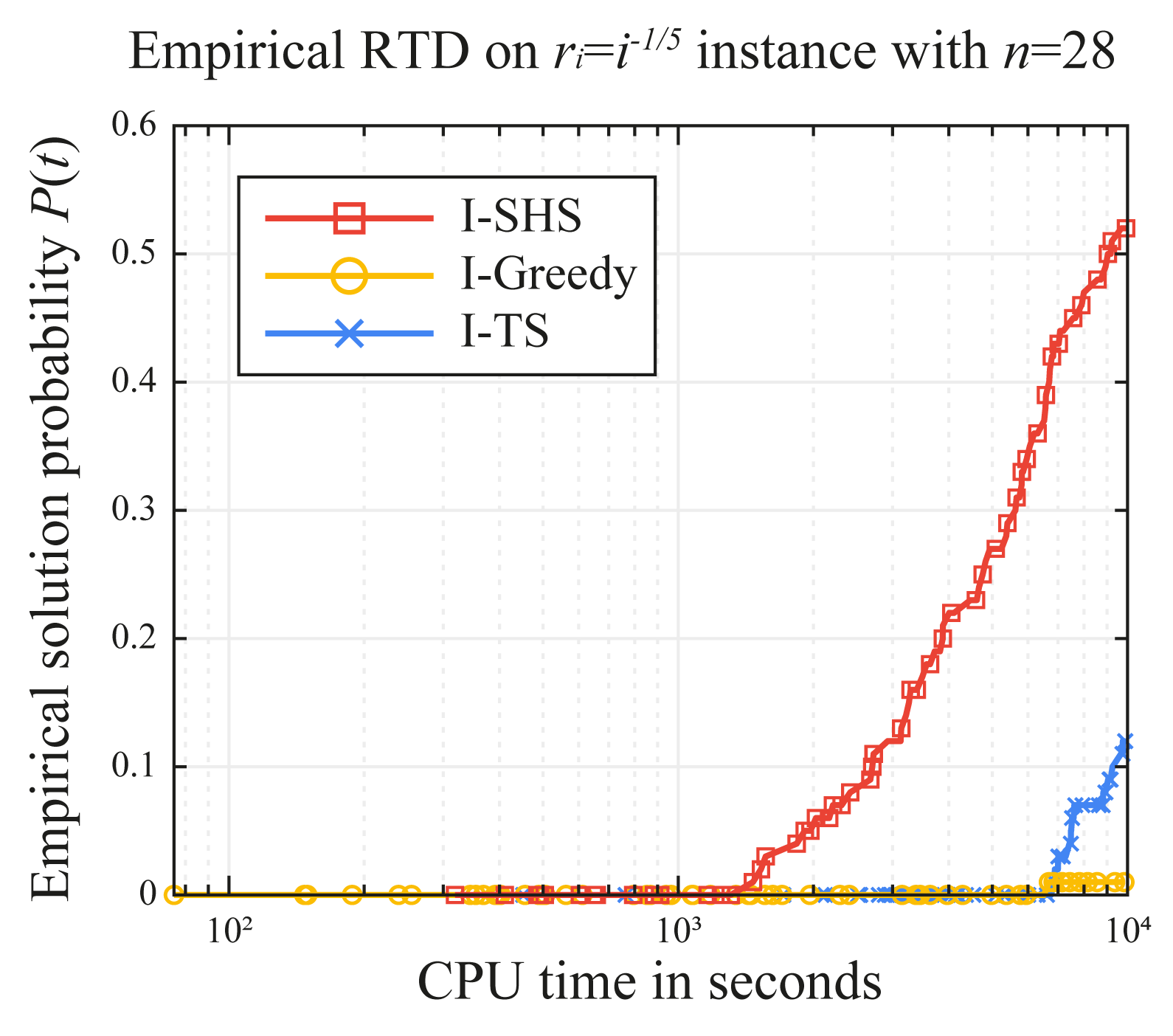}}
    \end{minipage}

    \caption{Empirical RTD of the proposed methods with different local search strategies on the representative instances.}
    \label{fig:cmp}
\end{figure}

To analyze and compare the behaviors of these algorithms, we employ the empirical Run-Time Distribution (RTD) for stochastic optimization methods~\citep{hoos2014empirical}. For a tested instance, the cumulative empirical RTD of a stochastic algorithm is a function $P(t)$ mapping the run time $t$ to the probability of obtaining the current best-known solution within time $t$. The function $P(t)$ is defined as follows:
\begin{align} 
    P(t) = \frac{\lvert \{ k : time(k) \leq t \} \rvert}{N},
\end{align}
where $time(k)$ indicates the running time of the $k$-th successful run to obtain the current best-known solution, and $N$ indicates the number of runs performed. The empirical RTD provides an efficient graphic representative for analyzing the behavior of stochastic optimization algorithms. The three algorithms, I-SHS, I-Greedy, and I-TS, were evaluated on six representative instances selected from each benchmark with the empirical RTD setting of $N = 100$. The experimental results are presented in Figure~\ref{fig:cmp}. 

From Figure~\ref{fig:cmp}, it is evident that I-SHS significantly outperforms the two variants, I-Greedy and I-TS, across the tested instances. 
Specifically, I-SHS is much more efficient than I-Greedy in five out of the six tested instances, and I-SHS 
shows comparable performance to 
I-Greedy on the $r_i = i^{-1/2}$ instance with $n = 23$. It demonstrates that the integration of greedy methods and the SHS core significantly enhances the algorithm performance. 
This underscores the potential and efficiency of our SHS core as a local search method.
Moreover, I-SHS exhibits much higher efficiency than I-TS in five out of the six test instances, 
I-SHS shows comparable performance to I-TS 
on the $r_i = i$ instance with $n=30$. It demonstrates that the proposed SHS core surpasses the traditional Tabu Search method in efficiency. 
Remarkably, on the $r_i = i^{-1/5}$ instance with $n=28$, the success rate of I-SHS surpasses 50\%, whereas I-TS achieves only over 10\%, and I-Greedy nearly approaches zero. 
These experimental findings 
underscore the exceptional performance of our proposed methods in solving PUCC, highlighting their applicability and generality across various benchmark instances.

\section{Conclusions} \label{sec:conclu}
In this study, we have addressed the classic and challenging Packing Unequal Circles in a Circle (PUCC) problem, renowned as one of the most popular and representative circle packing problems. 
Our study has introduced several efficient methodologies and achieved novel functions to tackle this problem, briefly recalled as follows:
1) The Adaptive Adjacency Maintenance (AAM) method achieves the feature that dynamically maintains the circle adjacency set during the layout optimization process to reduce unnecessary maintenance and computational overhead;
2) Two novel methods, named the vacancy detection method and the Voronoi-based locating method, are proposed to identify and quantify vacancies in a configuration properly and efficiently, which is essential for the \textit{insert} operation in solving PUCC;
3) The layout-graph transformation and hashing methods offer a unique approach of discerning isomorphism or similarity between different configurations;
4) The Solution-Hashing Search (SHS) heuristic can easily record explored configurations and determine whether the current configuration has been previously explored, avoiding duplicate exploration effectively.
By incorporating the abovementioned methods, we propose a stochastic optimization algorithm, called the Iterative Solution-Hashing Search (I-SHS) algorithm, for solving PUCC.
Notably, the ideas of our proposed methods are of a general nature. As such, they can be easily adapted to handle other packing problems. 


Extensive experiments demonstrate the remarkable superiority of  our I-SHS algorithm over existing state-of-the-art methods. I-SHS exhibits robust searching capabilities across instances featuring diverse radius distributions. Specifically, I-SHS improves the best-known results for 56 out of the 179 benchmark instances while concurrently matching the best-known results for the remaining instances.




\begin{appendices}
\section{Detailed Computational Results and Comparisons of PUCC Benchmarks} \label{append:detail}

Tables~\ref{tb:de_ins1}-\ref{tb:de_ins5} show, respectively, the detailed computational results and comparisons of our proposed algorithm (I-SHS) and the best-known results sourced from the Packomania website~\citep{Spechtweb} on six PUCC benchmarks.
The first column indicates the size of instances ($n$) (or the instance name for the NR benchmark). 
The second column shows the best-known results ($R^*$) sourced from the Packomania website. 
Columns 6-10 show the results of our algorithm where $R_{best}$, $R_{avg}$, and $R_{worst}$ show, respectively, the best result, the average result, and the worst result over multiple runs for each instance (20 runs for the $r_i = i$ benchmark and 10 runs for other benchmarks), $\Delta_{best}$, $\Delta_{avg}$, and $\Delta_{worst}$ show, respectively, the differences between $R_{best}$, $R_{avg}$, $R_{worst}$ and $R^*$ presented in scientific notation, i.e., $\Delta_{best} = R_{best} - R^*$, $\Delta_{avg} = R_{avg} - R^*$, $\Delta_{worst} = R_{worst} - R^*$, the success rate (SR) of obtaining the best result, and the average running time ($time(s)$) in seconds for each run of the algorithm to obtain its final result.
In addition, the last three rows of the tables summarize the number of instances in terms of $R_{best}$, $R_{avg}$, $R_{worst}$, where our algorithm obtained an improved, matched, or worse result compared with the best-known result.
For the $r_i = i$ benchmark instances, the matched results are indicated in bold compared with the best-known result in terms of $R_{best}$, $R_{avg}$, $R_{worst}$.
For other benchmark instances, the improved results are indicated in bold compared with the best-known result in terms of $R_{best}$, $R_{avg}$, $R_{worst}$.

\begin{table}[tb]
\centering
\caption{
Detailed computational results and comparison on the $r_i = i$ benchmark instances in the range of $5 \leq n \leq 35$.
}
\label{tb:de_ins1}

\resizebox{.75\columnwidth}{!}{
\begin{tabular}{lllllrrrrr}
\toprule
           &                      & \multicolumn{8}{l}{I-SHS (this study)}                                                                                                                                                       \\ \cline{3-10} 
$n$          & $R^*$ & $R_{best}$               & $R_{avg}$               & $R_{worst}$             & \multicolumn{1}{l}{$\Delta_{best}$} & \multicolumn{1}{l}{$\Delta_{avg}$} & \multicolumn{1}{l}{$\Delta_{worst}$} & SR    & $time(s)$ \\ \midrule
5          & 9.00139774           & \textbf{9.00139774}   & \textbf{9.00139774}  & \textbf{9.00139774}  & 0                           & 0                          & 0                            & 20/20 & 0.00    \\
6          & 11.05704039          & \textbf{11.05704039}  & \textbf{11.05704039} & \textbf{11.05704039} & 0                           & 0                          & 0                            & 20/20 & 0.01    \\
7          & 13.46211067          & \textbf{13.46211067}  & \textbf{13.46211067} & \textbf{13.46211067} & 0                           & 0                          & 0                            & 20/20 & 0.01    \\
8          & 16.22174667          & \textbf{16.22174667}  & \textbf{16.22174667} & \textbf{16.22174667} & 0                           & 0                          & 0                            & 20/20 & 0.02    \\
9          & 19.23319390          & \textbf{19.23319390}  & \textbf{19.23319390} & \textbf{19.23319390} & 0                           & 0                          & 0                            & 20/20 & 0.04    \\
10         & 22.00019301          & \textbf{22.00019301}  & \textbf{22.00019301} & \textbf{22.00019301} & 0                           & 0                          & 0                            & 20/20 & 0.02    \\
11         & 24.96063428          & \textbf{24.96063428}  & \textbf{24.96063428} & \textbf{24.96063428} & 0                           & 0                          & 0                            & 20/20 & 0.10    \\
12         & 28.37138943          & \textbf{28.37138943}  & \textbf{28.37138943} & \textbf{28.37138943} & 0                           & 0                          & 0                            & 20/20 & 0.19    \\
13         & 31.54586701          & \textbf{31.54586701}  & \textbf{31.54586701} & \textbf{31.54586701} & 0                           & 0                          & 0                            & 20/20 & 0.23    \\
14         & 35.09564714          & \textbf{35.09564714}  & \textbf{35.09564714} & \textbf{35.09564714} & 0                           & 0                          & 0                            & 20/20 & 0.76    \\
15         & 38.83799550          & \textbf{38.83799550}  & \textbf{38.83799550} & \textbf{38.83799550} & 0                           & 0                          & 0                            & 20/20 & 1.93    \\
16         & 42.45811643          & \textbf{42.45811643}  & \textbf{42.45811643} & \textbf{42.45811643} & 0                           & 0                          & 0                            & 20/20 & 13.43   \\
17         & 46.29134211          & \textbf{46.29134211}  & \textbf{46.29134211} & \textbf{46.29134211} & 0                           & 0                          & 0                            & 20/20 & 12.27   \\
18         & 50.11976262          & \textbf{50.11976262}  & \textbf{50.11976262} & \textbf{50.11976262} & 0                           & 0                          & 0                            & 20/20 & 5.37    \\
19         & 54.24029359          & \textbf{54.24029359}  & \textbf{54.24029359} & \textbf{54.24029359} & 0                           & 0                          & 0                            & 20/20 & 30.70   \\
20         & 58.40056747          & \textbf{58.40056747}  & \textbf{58.40056747} & \textbf{58.40056747} & 0                           & 0                          & 0                            & 20/20 & 2081.80 \\
21         & 62.55887709          & \textbf{62.55887709}  & \textbf{62.55887709} & \textbf{62.55887709} & 0                           & 0                          & 0                            & 20/20 & 56.72   \\
22         & 66.76028624          & \textbf{66.76028624}  & \textbf{66.76028624} & \textbf{66.76028624} & 0                           & 0                          & 0                            & 20/20 & 127.52  \\
23         & 71.19946160          & \textbf{71.19946160}  & \textbf{71.19946160} & \textbf{71.19946160} & 0                           & 0                          & 0                            & 20/20 & 1900.06 \\
24         & 75.74914258          & \textbf{75.74914258}  & \textbf{75.74914258} & \textbf{75.74914258} & 0                           & 0                          & 0                            & 20/20 & 643.55  \\
25         & 80.28586443          & \textbf{80.28586443}  & \textbf{80.28586443} & \textbf{80.28586443} & 0                           & 0                          & 0                            & 20/20 & 1596.54 \\
26         & 84.97819106          & \textbf{84.97819106}  & \textbf{84.97819106} & \textbf{84.97819106} & 0                           & 0                          & 0                            & 20/20 & 1417.22 \\
27         & 89.75096268          & \textbf{89.75096268}  & 89.75296818          & 89.79107273          & 0                           & 2.01E-03                   & 4.01E-02                     & 19/20 & 3903.17 \\
28         & 94.52587710          & \textbf{94.52587710}  & \textbf{94.52587710} & \textbf{94.52587710} & 0                           & 0                          & 0                            & 20/20 & 796.63  \\
29         & 99.48311156          & \textbf{99.48311156}  & 99.48895282          & 99.51231790          & 0                           & 5.84E-03                   & 2.92E-02                     & 16/20 & 4791.38 \\
30         & 104.54036376         & \textbf{104.54036376} & 104.54716843         & 104.57855508         & 0                           & 6.80E-03                   & 3.82E-02                     & 15/20 & 4148.65 \\
31         & 109.62924066         & \textbf{109.62924066} & 109.67261210         & 109.74451812         & 0                           & 4.34E-02                   & 1.15E-01                     & 6/20  & 4273.86 \\
32         & 114.79981466         & \textbf{114.79981466} & 114.83842698         & 114.91385096         & 0                           & 3.86E-02                   & 1.14E-01                     & 2/20  & 5014.04 \\
33         & 120.06565963         & \textbf{120.06565963} & 120.14171713         & 120.21732759         & 0                           & 7.61E-02                   & 1.52E-01                     & 4/20  & 5293.43 \\
34         & 125.36693920         & \textbf{125.36693920} & 125.49490944         & 125.58377406         & 0                           & 1.28E-01                   & 2.17E-01                     & 1/20  & 4257.74 \\
35         & 130.84907875         & \textbf{130.84907875} & 130.93428807         & 131.05260822         & 0                           & 8.52E-02                   & 2.04E-01                     & 2/20  & 5186.00 \\ \midrule
\# Improved &                      & 0                     & 0                    & 0                    & \multicolumn{1}{l}{}        & \multicolumn{1}{l}{}       & \multicolumn{1}{l}{}         &       &         \\
\# Matched  &                      & 31                    & 23                   & 23                   & \multicolumn{1}{l}{}        & \multicolumn{1}{l}{}       & \multicolumn{1}{l}{}         &       &         \\
\# Worse    &                      & 0                     & 8                    & 8                    & \multicolumn{1}{l}{}        & \multicolumn{1}{l}{}       & \multicolumn{1}{l}{}         &       &         \\ \bottomrule
\end{tabular}
}

\end{table}
\begin{table}[tb]
\centering
\caption{
Detailed computational results and comparison on the 24 NR benchmark instances.
}
\label{tb:de_NR}

\resizebox{.75\columnwidth}{!}{
\begin{tabular}{lllllrrrrr}
\toprule
           &                      & \multicolumn{8}{l}{I-SHS (this study)}                                                                                                                                                       \\ \cline{3-10} 
$n$          & $R^*$ & $R_{best}$               & $R_{avg}$               & $R_{worst}$             & \multicolumn{1}{l}{$\Delta_{best}$} & \multicolumn{1}{l}{$\Delta_{avg}$} & \multicolumn{1}{l}{$\Delta_{worst}$} & SR    & $time(s)$ \\ \midrule
NR10\_1    & 99.88507689          & 99.88507689           & 99.88507689           & 99.88507689  & 0                           & 0                          & 0                            & 10/10 & 0.04    \\
NR11\_1    & 60.70996138          & 60.70996138           & 60.70996138           & 60.70996138  & 0                           & 0                          & 0                            & 10/10 & 0.67    \\
NR12\_1    & 65.02442246          & 65.02442246           & 65.02442246           & 65.02442246  & 0                           & 0                          & 0                            & 10/10 & 0.70    \\
NR14\_1    & 113.55876291         & 113.55876291          & 113.55876291          & 113.55876291 & 0                           & 0                          & 0                            & 10/10 & 11.20   \\
NR15\_1    & 38.91138666          & 38.91138666           & 38.91138666           & 38.91138666  & 0                           & 0                          & 0                            & 10/10 & 2.29    \\
NR15\_2    & 38.83799550          & 38.83799550           & 38.83799550           & 38.83799550  & 0                           & 0                          & 0                            & 10/10 & 2.63    \\
NR16\_1    & 143.37978108         & 143.37978108          & 143.37978108          & 143.37978108 & 0                           & 0                          & 0                            & 10/10 & 71.72   \\
NR16\_2    & 127.69782537         & 127.69782537          & 127.69782537          & 127.69782537 & 0                           & 0                          & 0                            & 10/10 & 31.71   \\
NR17\_1    & 49.18730653          & 49.18730653           & 49.18730653           & 49.18730653  & 0                           & 0                          & 0                            & 10/10 & 45.51   \\
NR18\_1    & 196.98262400         & 196.98262400          & 196.98262400          & 196.98262400 & 0                           & 0                          & 0                            & 10/10 & 113.15  \\
NR20\_1    & 125.11775418         & 125.11775418          & 125.11775418          & 125.11775418 & 0                           & 0                          & 0                            & 10/10 & 16.25   \\
NR20\_2    & 121.78871660         & 121.78871660          & 121.78871660          & 121.78871660 & 0                           & 0                          & 0                            & 10/10 & 223.98  \\
NR21\_1    & 148.09678792         & 148.09678792          & 148.09678792          & 148.09678792 & 0                           & 0                          & 0                            & 10/10 & 293.77  \\
NR23\_1    & 174.34254220         & 174.34254220          & 174.38657308          & 174.43060397 & 0                           & 4.40E-02                   & 8.81E-02                     & 5/10  & 4528.68 \\
NR24\_1    & 137.75905206         & 137.75905206          & 137.75905206          & 137.75905206 & 0                           & 0                          & 0                            & 10/10 & 2023.08 \\
NR25\_1    & 188.71878994         & 188.71878994          & 188.74034084          & 188.77266719 & 0                           & 2.16E-02                   & 5.39E-02                     & 6/10  & 6458.62 \\
NR26\_1    & 244.57428028         & 244.57428028          & 244.57428028          & 244.57428028 & 0                           & 0                          & 0                            & 10/10 & 1360.38 \\
NR26\_2    & 300.26307937         & 300.26307937          & 300.26307937          & 300.26307937 & 0                           & 0                          & 0                            & 10/10 & 1728.43 \\
NR27\_1    & 220.65960596         & 220.65960596          & 220.97236951          & 221.12884403 & 0                           & 3.13E-01                   & 4.69E-01                     & 1/10  & 5215.47 \\
NR30\_1    & 177.25866811         & \textbf{177.17846105} & 177.41696810          & 177.52841637 & -8.02E-02                   & 1.58E-01                   & 2.70E-01                     & 1/10  & 4454.86 \\
NR30\_2    & 172.65018482         & \textbf{172.51354788} & 172.69807687          & 172.84784660 & -1.37E-01                   & 4.79E-02                   & 1.98E-01                     & 1/10  & 5281.77 \\
NR40\_1    & 352.40262684         & \textbf{352.23082161} & 352.46770385          & 352.65040551 & -1.72E-01                   & 6.51E-02                   & 2.48E-01                     & 1/10  & 5795.36 \\
NR50\_1    & 376.80638007         & \textbf{375.88541573} & \textbf{376.46731077} & 377.05732846 & -9.21E-01                   & -3.39E-01                  & 2.51E-01                     & 1/10  & 4726.23 \\
NR60\_1    & 514.83631921         & \textbf{514.09716814} & \textbf{514.64578784} & 515.39826769 & -7.39E-01                   & -1.91E-01                  & 5.62E-01                     & 1/10  & 7815.28 \\ \midrule
\# Improved &                      & 5                     & 2                     & 0            & \multicolumn{1}{l}{}        & \multicolumn{1}{l}{}       & \multicolumn{1}{l}{}         &       &         \\
\# Matched  &                      & 19                    & 16                    & 16           & \multicolumn{1}{l}{}        & \multicolumn{1}{l}{}       & \multicolumn{1}{l}{}         &       &         \\
\# Worse    &                      & 0                     & 6                     & 8            & \multicolumn{1}{l}{}        & \multicolumn{1}{l}{}       & \multicolumn{1}{l}{}         &       &         \\ \bottomrule
\end{tabular}
}

\end{table}
\begin{table}[tb]
\centering
\caption{
Detailed computational results and comparison on the $r_i = i^{-1/2}$ benchmark instances in the range of $5 \leq n \leq 35$.
}
\label{tb:de_ins3}

\resizebox{.75\columnwidth}{!}{
\begin{tabular}{lllllrrrrr}
\toprule
           &                      & \multicolumn{8}{l}{I-SHS (this study)}                                                                                                                                                       \\ \cline{3-10} 
$n$          & $R^*$ & $R_{best}$               & $R_{avg}$               & $R_{worst}$             & \multicolumn{1}{l}{$\Delta_{best}$} & \multicolumn{1}{l}{$\Delta_{avg}$} & \multicolumn{1}{l}{$\Delta_{worst}$} & SR    & $time(s)$ \\ \midrule
5          & 1.75155245           & 1.75155245          & 1.75155245          & 1.75155245          & 0                           & 0                          & 0                            & 10/10 & 0.01    \\
6          & 1.81007693           & 1.81007693          & 1.81007693          & 1.81007693          & 0                           & 0                          & 0                            & 10/10 & 0.01    \\
7          & 1.83872406           & 1.83872406          & 1.83872406          & 1.83872406          & 0                           & 0                          & 0                            & 10/10 & 0.02    \\
8          & 1.85840095           & 1.85840095          & 1.85840095          & 1.85840095          & 0                           & 0                          & 0                            & 10/10 & 0.01    \\
9          & 1.87881275           & 1.87881275          & 1.87881275          & 1.87881275          & 0                           & 0                          & 0                            & 10/10 & 0.06    \\
10         & 1.91343551           & 1.91343551          & 1.91343551          & 1.91343551          & 0                           & 0                          & 0                            & 10/10 & 0.57    \\
11         & 1.92918775           & 1.92918775          & 1.92918775          & 1.92918775          & 0                           & 0                          & 0                            & 10/10 & 0.57    \\
12         & 1.94982343           & 1.94982343          & 1.94982343          & 1.94982343          & 0                           & 0                          & 0                            & 10/10 & 1.38    \\
13         & 1.96523681           & 1.96523681          & 1.96523681          & 1.96523681          & 0                           & 0                          & 0                            & 10/10 & 1.92    \\
14         & 1.98024874           & 1.98024874          & 1.98024874          & 1.98024874          & 0                           & 0                          & 0                            & 10/10 & 15.55   \\
15         & 1.99270927           & 1.99270927          & 1.99270927          & 1.99270927          & 0                           & 0                          & 0                            & 10/10 & 35.51   \\
16         & 2.00458577           & 2.00458577          & 2.00458577          & 2.00458577          & 0                           & 0                          & 0                            & 10/10 & 27.59   \\
17         & 2.01525778           & 2.01525778          & 2.01525778          & 2.01525778          & 0                           & 0                          & 0                            & 10/10 & 30.57   \\
18         & 2.02814858           & 2.02814858          & 2.02814858          & 2.02814858          & 0                           & 0                          & 0                            & 10/10 & 63.38   \\
19         & 2.04199731           & 2.04199731          & 2.04199731          & 2.04199731          & 0                           & 0                          & 0                            & 10/10 & 541.85  \\
20         & 2.05144226           & \textbf{2.05144134} & \textbf{2.05144134} & \textbf{2.05144134} & -9.20E-07                   & -9.20E-07                  & -9.20E-07                    & 10/10 & 724.41  \\
21         & 2.06233110           & 2.06233110          & 2.06233110          & 2.06233110          & 0                           & 0                          & 0                            & 10/10 & 1568.79 \\
22         & 2.06796317           & 2.06796317          & 2.06796317          & 2.06796317          & 0                           & 0                          & 0                            & 10/10 & 622.06  \\
23         & 2.07977236           & 2.07977236          & 2.07986812          & 2.08025118          & 0                           & 9.58E-05                   & 4.79E-04                     & 8/10  & 4152.34 \\
24         & 2.09038316           & 2.09038316          & 2.09047590          & 2.09069230          & 0                           & 9.27E-05                   & 3.09E-04                     & 7/10  & 6347.90 \\
25         & 2.09752783           & 2.09752783          & 2.09828200          & 2.10010577          & 0                           & 7.54E-04                   & 2.58E-03                     & 5/10  & 4346.33 \\
26         & 2.10763017           & \textbf{2.10702271} & 2.10799866          & 2.11032448          & -6.07E-04                   & 3.68E-04                   & 2.69E-03                     & 5/10  & 4847.16 \\
27         & 2.11589049           & \textbf{2.11317010} & \textbf{2.11574662} & 2.11768298          & -2.72E-03                   & -1.44E-04                  & 1.79E-03                     & 2/10  & 4929.92 \\
28         & 2.12284917           & \textbf{2.12147566} & 2.12359093          & 2.12666393          & -1.37E-03                   & 7.42E-04                   & 3.81E-03                     & 1/10  & 4369.85 \\
29         & 2.13197913           & \textbf{2.12807595} & \textbf{2.13144729} & 2.13496476          & -3.90E-03                   & -5.32E-04                  & 2.99E-03                     & 1/10  & 6545.35 \\
30         & 2.13748392           & \textbf{2.13624146} & 2.13941024          & 2.14093792          & -1.24E-03                   & 1.93E-03                   & 3.45E-03                     & 1/10  & 6033.12 \\
31         & 2.14898330           & \textbf{2.14356970} & \textbf{2.14683660} & \textbf{2.14847470} & -5.41E-03                   & -2.15E-03                  & -5.09E-04                    & 1/10  & 5688.82 \\
32         & 2.15535415           & \textbf{2.15130900} & \textbf{2.15476643} & 2.15639006          & -4.05E-03                   & -5.88E-04                  & 1.04E-03                     & 1/10  & 5824.71 \\
33         & 2.16580368           & \textbf{2.15890144} & \textbf{2.15980076} & \textbf{2.16132212} & -6.90E-03                   & -6.00E-03                  & -4.48E-03                    & 1/10  & 4993.72 \\
34         & 2.17436411           & \textbf{2.16293568} & \textbf{2.16776129} & \textbf{2.17013877} & -1.14E-02                   & -6.60E-03                  & -4.23E-03                    & 1/10  & 6388.91 \\
35         & 2.17116210           & \textbf{2.16988756} & 2.17410842          & 2.17538000          & -1.27E-03                   & 2.95E-03                   & 4.22E-03                     & 1/10  & 5243.74 \\ \midrule
\# Improved &                      & 11                  & 7                   & 4                   & \multicolumn{1}{l}{}        & \multicolumn{1}{l}{}       & \multicolumn{1}{l}{}         &       &         \\
\# Matched  &                      & 20                  & 17                  & 17                  & \multicolumn{1}{l}{}        & \multicolumn{1}{l}{}       & \multicolumn{1}{l}{}         &       &         \\
\# Worse    &                      & 0                   & 7                   & 10                  & \multicolumn{1}{l}{}        & \multicolumn{1}{l}{}       & \multicolumn{1}{l}{}         &       &         \\ \bottomrule
\end{tabular}
}

\end{table}
\begin{table}[tb]
\centering
\caption{
Detailed computational results and comparison on the $r_i = i^{1/2}$ benchmark instances in the range of $5 \leq n \leq 35$.
}
\label{tb:de_ins2}

\resizebox{.75\columnwidth}{!}{
\begin{tabular}{lllllrrrrr}
\toprule
           &                      & \multicolumn{8}{l}{I-SHS (this study)}                                                                                                                                                       \\ \cline{3-10} 
$n$          & $R^*$ & $R_{best}$               & $R_{avg}$               & $R_{worst}$             & \multicolumn{1}{l}{$\Delta_{best}$} & \multicolumn{1}{l}{$\Delta_{avg}$} & \multicolumn{1}{l}{$\Delta_{worst}$} & SR    & $time(s)$ \\ \midrule
5          & 4.52148027           & 4.52148027           & 4.52148027           & 4.52148027           & 0                           & 0                          & 0                            & 10/10 & 0.00    \\
6          & 5.35096299           & 5.35096299           & 5.35096299           & 5.35096299           & 0                           & 0                          & 0                            & 10/10 & 0.00    \\
7          & 6.04937848           & 6.04937848           & 6.04937848           & 6.04937848           & 0                           & 0                          & 0                            & 10/10 & 0.01    \\
8          & 6.77426665           & 6.77426665           & 6.77426665           & 6.77426665           & 0                           & 0                          & 0                            & 10/10 & 0.02    \\
9          & 7.55900237           & 7.55900237           & 7.55900237           & 7.55900237           & 0                           & 0                          & 0                            & 10/10 & 0.12    \\
10         & 8.30346812           & 8.30346812           & 8.30346812           & 8.30346812           & 0                           & 0                          & 0                            & 10/10 & 0.07    \\
11         & 9.07212587           & 9.07212587           & 9.07212587           & 9.07212587           & 0                           & 0                          & 0                            & 10/10 & 0.23    \\
12         & 9.86532030           & 9.86532030           & 9.86532030           & 9.86532030           & 0                           & 0                          & 0                            & 10/10 & 0.97    \\
13         & 10.58832628          & 10.58832628          & 10.58832628          & 10.58832628          & 0                           & 0                          & 0                            & 10/10 & 1.64    \\
14         & 11.36497759          & 11.36497759          & 11.36497759          & 11.36497759          & 0                           & 0                          & 0                            & 10/10 & 2.28    \\
15         & 12.06692333          & 12.06692333          & 12.06692333          & 12.06692333          & 0                           & 0                          & 0                            & 10/10 & 3.17    \\
16         & 12.81931152          & 12.81931152          & 12.81931152          & 12.81931152          & 0                           & 0                          & 0                            & 10/10 & 12.29   \\
17         & 13.56954137          & 13.56954137          & 13.56954137          & 13.56954137          & 0                           & 0                          & 0                            & 10/10 & 41.90   \\
18         & 14.32166883          & 14.32166883          & 14.32166883          & 14.32166883          & 0                           & 0                          & 0                            & 10/10 & 672.46  \\
19         & 15.03535243          & 15.03535243          & 15.03535243          & 15.03535243          & 0                           & 0                          & 0                            & 10/10 & 28.68   \\
20         & 15.79127663          & 15.79127663          & 15.79127663          & 15.79127663          & 0                           & 0                          & 0                            & 10/10 & 24.28   \\
21         & 16.53963351          & 16.53963351          & 16.53963351          & 16.53963351          & 0                           & 0                          & 0                            & 10/10 & 394.46  \\
22         & 17.28558985          & 17.28558985          & 17.28558985          & 17.28558985          & 0                           & 0                          & 0                            & 10/10 & 151.43  \\
23         & 18.04998520          & 18.04998520          & 18.04998520          & 18.04998520          & 0                           & 0                          & 0                            & 10/10 & 955.72  \\
24         & 18.78807360          & 18.78807360          & 18.78807360          & 18.78807360          & 0                           & 0                          & 0                            & 10/10 & 1323.27 \\
25         & 19.54468228          & \textbf{19.52687708} & \textbf{19.52771930} & \textbf{19.53529929} & -1.78E-02                   & -1.70E-02                  & -9.38E-03                    & 9/10  & 3613.19 \\
26         & 20.29026179          & \textbf{20.27823046} & \textbf{20.27876167} & \textbf{20.28002429} & -1.20E-02                   & -1.15E-02                  & -1.02E-02                    & 5/10  & 3091.49 \\
27         & 21.04950183          & \textbf{21.02971196} & \textbf{21.03309266} & \textbf{21.03557141} & -1.98E-02                   & -1.64E-02                  & -1.39E-02                    & 1/10  & 5021.10 \\
28         & 21.79712526          & \textbf{21.76057988} & \textbf{21.77440758} & \textbf{21.77707831} & -3.65E-02                   & -2.27E-02                  & -2.00E-02                    & 4/10  & 4412.49 \\
29         & 22.54735751          & \textbf{22.49836633} & \textbf{22.51347222} & \textbf{22.53777417} & -4.90E-02                   & -3.39E-02                  & -9.58E-03                    & 2/10  & 4804.19 \\
30         & 23.25868018          & \textbf{23.24258773} & \textbf{23.24950759} & \textbf{23.25685873} & -1.61E-02                   & -9.17E-03                  & -1.82E-03                    & 2/10  & 5651.52 \\
31         & 24.03727192          & \textbf{23.99738366} & \textbf{24.00997257} & \textbf{24.02602933} & -3.99E-02                   & -2.73E-02                  & -1.12E-02                    & 4/10  & 5295.84 \\
32         & 24.78200903          & \textbf{24.74151096} & \textbf{24.75531148} & \textbf{24.77298015} & -4.05E-02                   & -2.67E-02                  & -9.03E-03                    & 2/10  & 4349.66 \\
33         & 25.54486051          & \textbf{25.47851512} & \textbf{25.50823919} & \textbf{25.52382729} & -6.63E-02                   & -3.66E-02                  & -2.10E-02                    & 1/10  & 4535.92 \\
34         & 26.30459206          & \textbf{26.23123612} & \textbf{26.25362520} & \textbf{26.26690361} & -7.34E-02                   & -5.10E-02                  & -3.77E-02                    & 1/10  & 5895.22 \\
35         & 27.03382244          & \textbf{26.97763983} & \textbf{27.00203009} & \textbf{27.01150944} & -5.62E-02                   & -3.18E-02                  & -2.23E-02                    & 1/10  & 2539.64 \\ \midrule
\# Improved &                      & 11                   & 11                   & 11                   & \multicolumn{1}{l}{}        & \multicolumn{1}{l}{}       & \multicolumn{1}{l}{}         &       &         \\
\# Matched  &                      & 20                   & 20                   & 20                   & \multicolumn{1}{l}{}        & \multicolumn{1}{l}{}       & \multicolumn{1}{l}{}         &       &         \\
\# Worse    &                      & 0                    & 0                    & 0                    & \multicolumn{1}{l}{}        & \multicolumn{1}{l}{}       & \multicolumn{1}{l}{}         &       &         \\ \bottomrule
\end{tabular}
}

\end{table}
\begin{table}[tb]
\centering
\caption{
Detailed computational results and comparison on the $r_i = i^{-2/3}$ benchmark instances in the range of $5 \leq n \leq 35$.
}
\label{tb:de_ins4}

\resizebox{.75\columnwidth}{!}{
\begin{tabular}{lllllrrrrr}
\toprule
           &                      & \multicolumn{8}{l}{I-SHS (this study)}                                                                                                                                                       \\ \cline{3-10} 
$n$          & $R^*$ & $R_{best}$               & $R_{avg}$               & $R_{worst}$             & \multicolumn{1}{l}{$\Delta_{best}$} & \multicolumn{1}{l}{$\Delta_{avg}$} & \multicolumn{1}{l}{$\Delta_{worst}$} & SR    & $time(s)$ \\ \midrule
5          & 1.62996052           & 1.62996052          & 1.62996052          & 1.62996052          & 0                           & 0                          & 0                            & 10/10 & 0.01    \\
6          & 1.62996052           & 1.62996052          & 1.62996052          & 1.62996052          & 0                           & 0                          & 0                            & 10/10 & 0.01    \\
7          & 1.62997277           & 1.62997277          & 1.62997277          & 1.62997277          & 0                           & 0                          & 0                            & 10/10 & 0.02    \\
8          & 1.63148407           & 1.63148407          & 1.63148407          & 1.63148407          & 0                           & 0                          & 0                            & 10/10 & 0.08    \\
9          & 1.63786399           & 1.63786399          & 1.63786399          & 1.63786399          & 0                           & 0                          & 0                            & 10/10 & 0.07    \\
10         & 1.64695723           & 1.64695723          & 1.64695723          & 1.64695723          & 0                           & 0                          & 0                            & 10/10 & 0.34    \\
11         & 1.65031382           & 1.65031382          & 1.65031382          & 1.65031382          & 0                           & 0                          & 0                            & 10/10 & 0.23    \\
12         & 1.65676025           & 1.65676025          & 1.65676025          & 1.65676025          & 0                           & 0                          & 0                            & 10/10 & 1.58    \\
13         & 1.66277803           & 1.66277803          & 1.66277803          & 1.66277803          & 0                           & 0                          & 0                            & 10/10 & 0.99    \\
14         & 1.67018352           & 1.67018352          & 1.67018352          & 1.67018352          & 0                           & 0                          & 0                            & 10/10 & 3.65    \\
15         & 1.67300105           & 1.67300105          & 1.67300105          & 1.67300105          & 0                           & 0                          & 0                            & 10/10 & 4.53    \\
16         & 1.67963181           & 1.67963181          & 1.67963181          & 1.67963181          & 0                           & 0                          & 0                            & 10/10 & 34.00   \\
17         & 1.68384097           & 1.68384097          & 1.68384097          & 1.68384097          & 0                           & 0                          & 0                            & 10/10 & 51.39   \\
18         & 1.68602524           & 1.68602524          & 1.68602524          & 1.68602524          & 0                           & 0                          & 0                            & 10/10 & 75.43   \\
19         & 1.68953260           & 1.68953260          & 1.68953260          & 1.68953260          & 0                           & 0                          & 0                            & 10/10 & 176.03  \\
20         & 1.69644772           & \textbf{1.69279776} & \textbf{1.69279776} & \textbf{1.69279776} & -3.65E-03                   & -3.65E-03                  & -3.65E-03                    & 10/10 & 637.58  \\
21         & 1.69894302           & \textbf{1.69473728} & \textbf{1.69473728} & \textbf{1.69473728} & -4.21E-03                   & -4.21E-03                  & -4.21E-03                    & 10/10 & 1405.80 \\
22         & 1.70066803           & \textbf{1.69838000} & \textbf{1.69838000} & \textbf{1.69838000} & -2.29E-03                   & -2.29E-03                  & -2.29E-03                    & 10/10 & 2530.81 \\
23         & 1.70393378           & \textbf{1.70223864} & \textbf{1.70225978} & \textbf{1.70245013} & -1.70E-03                   & -1.67E-03                  & -1.48E-03                    & 9/10  & 4382.03 \\
24         & 1.70538770           & \textbf{1.70420700} & \textbf{1.70425668} & \textbf{1.70434172} & -1.18E-03                   & -1.13E-03                  & -0.001046                    & 6/10  & 3660.89 \\
25         & 1.71233042           & \textbf{1.70631982} & \textbf{1.70712281} & \textbf{1.70861343} & -6.01E-03                   & -5.21E-03                  & -3.72E-03                    & 5/10  & 3756.70 \\
26         & 1.71459329           & \textbf{1.70969312} & \textbf{1.71030108} & \textbf{1.71124697} & -4.90E-03                   & -4.29E-03                  & -3.35E-03                    & 3/10  & 4675.17 \\
27         & 1.71898563           & \textbf{1.71211430} & \textbf{1.71261504} & \textbf{1.71404416} & -6.87E-03                   & -6.37E-03                  & -4.94E-03                    & 2/10  & 5060.78 \\
28         & 1.72190904           & \textbf{1.71437832} & \textbf{1.71544246} & \textbf{1.71593815} & -7.53E-03                   & -6.47E-03                  & -5.97E-03                    & 1/10  & 5382.01 \\
29         & 1.72242777           & \textbf{1.71658642} & \textbf{1.71809375} & \textbf{1.71923871} & -5.84E-03                   & -4.33E-03                  & -3.19E-03                    & 2/10  & 5322.33 \\
30         & 1.72504075           & \textbf{1.71816387} & \textbf{1.72057373} & \textbf{1.72148070} & -6.88E-03                   & -4.47E-03                  & -3.56E-03                    & 1/10  & 5268.07 \\
31         & 1.72537598           & \textbf{1.72046887} & \textbf{1.72260839} & \textbf{1.72359974} & -4.91E-03                   & -2.77E-03                  & -1.78E-03                    & 1/10  & 5054.68 \\
32         & 1.72708467           & \textbf{1.72253372} & \textbf{1.72468851} & \textbf{1.72587996} & -4.55E-03                   & -2.40E-03                  & -1.20E-03                    & 1/10  & 5840.75 \\
33         & 1.73189080           & \textbf{1.72382222} & \textbf{1.72613734} & \textbf{1.72822282} & -8.07E-03                   & -5.75E-03                  & -3.67E-03                    & 1/10  & 4188.38 \\
34         & 1.73298460           & \textbf{1.72519586} & \textbf{1.72836344} & \textbf{1.72964263} & -7.79E-03                   & -4.62E-03                  & -3.34E-03                    & 1/10  & 4844.66 \\
35         & 1.73378808           & \textbf{1.72867585} & \textbf{1.73013070} & \textbf{1.73168510} & -5.11E-03                   & -3.66E-03                  & -2.10E-03                    & 1/10  & 4636.03 \\ \midrule
\# Improved &                      & 16                  & 16                  & 16                  & \multicolumn{1}{l}{}        & \multicolumn{1}{l}{}       & \multicolumn{1}{l}{}         &       &         \\
\# Matched  &                      & 15                  & 15                  & 15                  & \multicolumn{1}{l}{}        & \multicolumn{1}{l}{}       & \multicolumn{1}{l}{}         &       &         \\
\# Worse    &                      & 0                   & 0                   & 0                   & \multicolumn{1}{l}{}        & \multicolumn{1}{l}{}       & \multicolumn{1}{l}{}         &       &         \\ \bottomrule
\end{tabular}
}

\end{table}
\begin{table}[tb]
\centering
\caption{
Detailed computational results and comparison on the $r_i = i^{-1/5}$ benchmark instances in the range of $5 \leq n \leq 35$.
}
\label{tb:de_ins5}

\resizebox{.75\columnwidth}{!}{
\begin{tabular}{lllllrrrrr}
\toprule
           &                      & \multicolumn{8}{l}{I-SHS (this study)}                                                                                                                                                       \\ \cline{3-10} 
$n$          & $R^*$ & $R_{best}$               & $R_{avg}$               & $R_{worst}$             & \multicolumn{1}{l}{$\Delta_{best}$} & \multicolumn{1}{l}{$\Delta_{avg}$} & \multicolumn{1}{l}{$\Delta_{worst}$} & SR    & $time(s)$ \\ \midrule
5          & 2.24461584           & 2.24461584          & 2.24461584          & 2.24461584          & 0                           & 0                          & 0                            & 10/10 & 0.00    \\
6          & 2.38798638           & 2.38798638          & 2.38798638          & 2.38798638          & 0                           & 0                          & 0                            & 10/10 & 0.02    \\
7          & 2.42262334           & 2.42262334          & 2.42262334          & 2.42262334          & 0                           & 0                          & 0                            & 10/10 & 0.04    \\
8          & 2.52382090           & \textbf{2.52381953} & \textbf{2.52381953} & \textbf{2.52381953} & -1.37E-06                   & -1.37E-06                  & -1.37E-06                    & 10/10 & 3.86    \\
9          & 2.63002657           & 2.63002657          & 2.63002657          & 2.63002657          & 0                           & 0                          & 0                            & 10/10 & 0.07    \\
10         & 2.71578482           & 2.71578482          & 2.71578482          & 2.71578482          & 0                           & 0                          & 0                            & 10/10 & 5.57    \\
11         & 2.76746630           & 2.76746630          & 2.76746630          & 2.76746630          & 0                           & 0                          & 0                            & 10/10 & 8.38    \\
12         & 2.82901840           & 2.82901840          & 2.82901840          & 2.82901840          & 0                           & 0                          & 0                            & 10/10 & 0.23    \\
13         & 2.92391778           & 2.92391778          & 2.92391778          & 2.92391778          & 0                           & 0                          & 0                            & 10/10 & 14.51   \\
14         & 2.98478396           & 2.98478396          & 2.98478396          & 2.98478396          & 0                           & 0                          & 0                            & 10/10 & 35.02   \\
15         & 3.04049205           & 3.04049205          & 3.04049205          & 3.04049205          & 0                           & 0                          & 0                            & 10/10 & 1.35    \\
16         & 3.09641443           & 3.09641443          & 3.09641443          & 3.09641443          & 0                           & 0                          & 0                            & 10/10 & 5.30    \\
17         & 3.15212915           & 3.15212915          & 3.15212915          & 3.15212915          & 0                           & 0                          & 0                            & 10/10 & 90.60   \\
18         & 3.21482378           & 3.21482378          & 3.21482378          & 3.21482378          & 0                           & 0                          & 0                            & 10/10 & 72.09   \\
19         & 3.25933278           & 3.25933278          & 3.25933278          & 3.25933278          & 0                           & 0                          & 0                            & 10/10 & 38.06   \\
20         & 3.31183101           & 3.31183101          & 3.31183101          & 3.31183101          & 0                           & 0                          & 0                            & 10/10 & 156.98  \\
21         & 3.36159380           & 3.36159380          & 3.36159380          & 3.36159380          & 0                           & 0                          & 0                            & 10/10 & 45.54   \\
22         & 3.41145125           & 3.41145125          & 3.41145125          & 3.41145125          & 0                           & 0                          & 0                            & 10/10 & 689.79  \\
23         & 3.45013735           & 3.45013735          & 3.45013735          & 3.45013735          & 0                           & 0                          & 0                            & 10/10 & 385.53  \\
24         & 3.50251027           & \textbf{3.49921018} & \textbf{3.49921018} & \textbf{3.49921018} & -3.30E-03                   & -3.30E-03                  & -3.30E-03                    & 10/10 & 1090.83 \\
25         & 3.54524557           & \textbf{3.54088547} & \textbf{3.54088547} & \textbf{3.54088547} & -4.36E-03                   & -4.36E-03                  & -4.36E-03                    & 10/10 & 2927.83 \\
26         & 3.59113378           & \textbf{3.58686982} & \textbf{3.58686982} & \textbf{3.58686982} & -4.26E-03                   & -4.26E-03                  & -4.26E-03                    & 10/10 & 2167.83 \\
27         & 3.63268719           & \textbf{3.62413712} & \textbf{3.62416144} & \textbf{3.62438041} & -8.55E-03                   & -8.53E-03                  & -8.31E-03                    & 9/10  & 5784.57 \\
28         & 3.67394734           & \textbf{3.66722675} & \textbf{3.66760263} & \textbf{3.66841087} & -6.72E-03                   & -6.34E-03                  & -5.54E-03                    & 6/10  & 5601.51 \\
29         & 3.70961474           & \textbf{3.70434027} & \textbf{3.70434027} & \textbf{3.70434027} & -5.27E-03                   & -5.27E-03                  & -5.27E-03                    & 10/10 & 3949.97 \\
30         & 3.75302053           & \textbf{3.74670554} & \textbf{3.74714608} & \textbf{3.74768010} & -6.31E-03                   & -5.87E-03                  & -5.34E-03                    & 1/10  & 6646.12 \\
31         & 3.79463689           & \textbf{3.78299544} & \textbf{3.78541974} & \textbf{3.78667499} & -1.16E-02                   & -9.22E-03                  & -7.96E-03                    & 1/10  & 3836.33 \\
32         & 3.82636068           & \textbf{3.81794567} & \textbf{3.81940565} & \textbf{3.82085799} & -8.42E-03                   & -6.96E-03                  & -5.50E-03                    & 2/10  & 4459.13 \\
33         & 3.86602448           & \textbf{3.85326849} & \textbf{3.85557648} & \textbf{3.85725232} & -1.28E-02                   & -1.04E-02                  & -8.77E-03                    & 1/10  & 6053.48 \\
34         & 3.90172754           & \textbf{3.88952299} & \textbf{3.89129944} & \textbf{3.89261602} & -1.22E-02                   & -1.04E-02                  & -9.11E-03                    & 1/10  & 5267.17 \\
35         & 3.93668876           & \textbf{3.92408008} & \textbf{3.92680599} & \textbf{3.92934621} & -1.26E-02                   & -9.88E-03                  & -7.34E-03                    & 1/10  & 3452.73 \\ \midrule
\# Improved &                      & 13                  & 13                  & 13                  & \multicolumn{1}{l}{}        & \multicolumn{1}{l}{}       & \multicolumn{1}{l}{}         &       &         \\
\# Matched  &                      & 18                  & 18                  & 18                  & \multicolumn{1}{l}{}        & \multicolumn{1}{l}{}       & \multicolumn{1}{l}{}         &       &         \\
\# Worse    &                      & 0                   & 0                   & 0                   & \multicolumn{1}{l}{}        & \multicolumn{1}{l}{}       & \multicolumn{1}{l}{}         &       &         \\ \bottomrule
\end{tabular}
}

\end{table}

\end{appendices}













\bibliographystyle{cas-model2-names}

\bibliography{mybibfile}


\end{document}